\documentclass[twocolumn]{aastex63}

\usepackage{amsmath}

\submitjournal{\apj}

\shorttitle{Deuterium fractionation of methanol}
\shortauthors{Sakai et al.}

\begin{document}

\title{Digging into the Interior of Hot Cores with ALMA (DIHCA). V. Deuterium Fractionation of Methanol}

\correspondingauthor{Takeshi Sakai}
\email{takeshi.sakai@uec.ac.jp}

\author[0000-0003-4521-7492]{Takeshi Sakai}
\affiliation{Graduate School of Informatics and Engineering, The University of Electro-Communications, Chofu, Tokyo 182-8585, Japan}
\affiliation{Institute for Advanced Science, The University of Electro-Communications, Chofu, Tokyo 182-8585, Japan}
\affiliation{Center for Space Science and Radio Engineering, The University of Electro-Communications, Chofu, Tokyo 182-8585, Japan}

\author{Nobuhito Shiomura}
\affiliation{Graduate School of Informatics and Engineering, The University of Electro-Communications, Chofu, Tokyo 182-8585, Japan}

\author[0000-0002-7125-7685]{Patricio Sanhueza}
\affiliation{Department of Earth and Planetary Sciences, Institute of Science Tokyo, Meguro, Tokyo, 152-8551, Japan}
\affiliation{National Astronomical Observatory of Japan, National Institutes of Natural Sciences, 2-21-1 Osawa, Mitaka, Tokyo 181-8588, Japan}

\author[0000-0002-2026-8157]{Kenji Furuya}
\affiliation{RIKEN Cluster for Pioneering Research, Wako, Saitama 351-0198, Japan}
\affiliation{Department of Astronomy, School of Science, The University of Tokyo, 7-3-1 Hongo, Bunkyo, Tokyo 113-0033, Japan}

\author[0000-0002-8250-6827]{Fernando A. Olguin}
\affiliation{Yukawa Institute for Theoretical Physics, Kyoto University, Kyoto, 606-8502, Japan}
\affiliation{National Astronomical Observatory of Japan, National Institutes of Natural Sciences, 2-21-1 Osawa, Mitaka, Tokyo 181-8588, Japan}
\affiliation{Institute of Astronomy and Department of Physics, National Tsing Hua University, Hsinchu 30013, Taiwan}

\author[0000-0002-8149-8546]{Ken'ichi Tatematsu}
\affiliation{Nobeyama Radio Observatory, National Astronomical Observatory of Japan, National Institutes of Natural Sciences, 462-2 Nobeyama, Minamimaki, Minamisaku, Nagano 384-1305, Japan}
\affiliation{Department of Astronomical Science, The Graduate University for Advanced Studies, SOKENDAI, 2-21-1 Osawa, Mitaka, Tokyo 181-8588, Japan}

\author[0000-0003-3283-6884]{Yuri Aikawa}
\affiliation{Department of Astronomy, School of Science, The University of Tokyo, 7-3-1 Hongo, Bunkyo, Tokyo 113-0033, Japan}

\author[0000-0003-4402-6475]{Kotomi Taniguchi}
\affiliation{National Astronomical Observatory of Japan, National Institutes of Natural Sciences, 2-21-1 Osawa, Mitaka, Tokyo 181-8588, Japan}

\author[0000-0002-9774-1846]{Huei-Ru Vivien Chen}
\affiliation{Institute of Astronomy and Department of Physics, National Tsing Hua University, Hsinchu 30013, Taiwan}

\author[0000-0002-6752-6061]{Kaho Morii}
\affiliation {Department of Astronomy, Graduate School of Science, The University of Tokyo, 7-3-1 Hongo, Bunkyo-ku, Tokyo 113-0033, Japan}
\affiliation {National Astronomical Observatory of Japan, National Institutes of Natural Sciences, 2-21-1 Osawa, Mitaka, Tokyo 181-8588, Japan}

\author[0000-0001-5431-2294]{Fumitaka Nakamura}
\affiliation{National Astronomical Observatory of Japan, National Institutes of Natural Sciences, 2-21-1 Osawa, Mitaka, Tokyo 181-8588, Japan}
\affiliation{Department of Astronomical Science, The Graduate University for Advanced Studies, SOKENDAI, 2-21-1 Osawa, Mitaka, Tokyo 181-8588, Japan}
\affiliation{Department of Astronomy, School of Science, The University of Tokyo, 7-3-1 Hongo, Bunkyo, Tokyo 113-0033, Japan}

\author[0000-0003-1275-5251]{Shanghuo Li}
\affiliation{Korea Astronomy and Space Science Institute, 776 Daedeokdae-ro, Yuseong-gu, Daejeon 34055, Republic of Korea}

\author[0000-0003-2619-9305]{Xing Lu}
\affiliation{Shanghai Astronomical Observatory, Chinese Academy of Sciences, 80 Nandan Road, Shanghai 200030, People’s Republic of China}

\author[0000-0003-2384-6589]{Qizhou Zhang}
\affiliation{Center for Astrophysics, Harvard \& Smithsonian, 60 Garden Street, Cambridge, MA 02138, USA}

\author[0000-0003-1659-095X]{Tomoya Hirota}
\affiliation{Mizusawa VLBI Observatory, National Astronomical Observatory of Japan, 2-12 Hoshigaoka, Mizusawa, Oshu, Iwate 023-0861, Japan}
\affiliation{Department of Astronomical Science, The Graduate University for Advanced Studies, SOKENDAI, 2-21-1 Osawa, Mitaka, Tokyo 181-8588, Japan}

\author[0000-0001-7080-2808]{Kousuke Ishihara}
\affiliation {Department of Astronomical Science, The Graduate University for Advanced Studies, SOKENDAI, 2-21-1 Osawa, Mitaka, Tokyo 181-8588, Japan}
\affiliation {National Astronomical Observatory of Japan, National Institutes of Natural Sciences, 2-21-1 Osawa, Mitaka, Tokyo 181-8588, Japan}

\author{Hongda Ke}
\affiliation{Graduate School of Informatics and Engineering, The University of Electro-Communications, Chofu, Tokyo 182-8585, Japan}

\author[0000-0002-3297-4497]{Nami Sakai}
\affiliation{Star and Planet Formation Laboratory, RIKEN Cluster for Pioneering Research, Wako, Saitama 351-0198, Japan}

\author[0000-0002-9865-0970]{Satoshi Yamamoto}
\affiliation{SOKENDAI (The Graduate University for Advanced Studies), Shonan Village, Hayama, Kanagawa 240-0193, Japan}




\begin{abstract}

We have observed the $^{13}$CH$_3$OH $5_1-4_1$ A$^+$, $^{13}$CH$_3$OH $14_1-13_2$ A$^-$, and CH$_2$DOH $8_{2,6}-8_{1,7}$ $e_0$ lines toward 24 high-mass star-forming regions by using Atacama Large Millimeter/submillimeter Array (ALMA) with an angular resolution of about 0$^{\prime\prime}$.3. This resolution corresponds to a linear scale of 400-1600 au, allowing us to resolve individual cores properly. We detected the $^{13}$CH$_3$OH and CH$_2$DOH emission near the continuum peaks in many of these regions. From the two $^{13}$CH$_3$OH lines, we calculated the temperature toward the $^{13}$CH$_3$OH peaks, and confirm that the emission traces hot ($>$100 K) regions. The $N$(CH$_2$DOH)/$N$($^{12}$CH$_3$OH) ratio in the observed high-mass star-forming regions is found to be lower than that in low-mass star-forming regions. We have found no correlation between the $N$(CH$_2$DOH)/$N$($^{13}$CH$_3$OH) or $N$(CH$_2$DOH)/$N$($^{12}$CH$_3$OH) ratios and either temperatures or distance to the sources, and have also found a source-to-source variation in these ratios. Our model calculations predict that the $N$(CH$_2$DOH)/$N$($^{12}$CH$_3$OH) ratio in hot cores depends on the duration of the cold phase; the shorter the cold phase, the lower the deuterium fractionation in the hot cores. We have suggested that the lower $N$(CH$_2$DOH)/$N$($^{12}$CH$_3$OH) ratio in high-mass star-forming regions compared to that in low-mass star-forming regions is due to the shorter duration of the cold phase and that the diversity in the $N$(CH$_2$DOH)/$N$($^{12}$CH$_3$OH) ratio in high-mass star-forming regions is due to the diversity in the length of the cold prestellar phase, and not the time that the objects have been in the hot core phase.

\end{abstract}



\section{Introduction} \label{sec:intro}

In cold molecular clouds, the ratio of deuterated molecules to their non-deuterated counterparts is significantly higher than the cosmic D/H abundance ratio ($\sim$10$^{-5}$) (e.g. Caselli et al. 1999). This enhancement, known as deuterium fractionation, results from the following exothermic reaction:
\begin{equation}
{\rm H}_3^+ + {\rm HD} \rightarrow {\rm H}_2{\rm D}^+ + {\rm H}_2 + 230\ {\rm K},
\end{equation}
where the backward reaction is inefficient in cold conditions, leading to an increased H$_2$D$^+$/H$_3^+$ ratio. Since the H$_3^+$ and H$_2$D$^+$ ions are the parent molecules of many molecular species, a high H$_2$D$^+$/H$_3^+$ ratio facilitates a high deuterium fractionation of their daughter molecules, such as N$_2$D$^+$/N$_2$H$^+$ (e.g. Emprechtinger et al. 2009).

Since the rate coefficient of the above backward reaction depends on temperature, the deuterium fractionation of molecules is influenced by gas temperature. In addition, the CO depletion onto dust grains leads to an increase in the H$_2$D$^+$/H$_3^+$ ratio, since CO is the main destructor of H$_2$D$^+$ \footnote{Note that the destruction of H$_3^+$ by CO does not affect the H$_2$D$^+$/H$_3^+$ ratio because H$_2$D$^+$ is a daughter molecule of H$_3^+$.}. This depletion is more pronounced in cold ($<$25 K) and dense ($>$10$^{6}$ cm$^{-3}$) regions. Thus, cold and dense regions, such as prestellar dense cores, exhibit significant deuterium fractionation. 
Furthermore, the duration of a dense and cold phase also plays a role in deuterium fractionation. For example, younger dense cores show less fractionation, as reported by Hirota et al. (2001). Chemical model calculations of the time dependence of deuterium fractionation have also been carried out and confirm this trend (e.g., Roberts \& Millar 2000; Taquet et al. 2012; Kong et al. 2016).

As gas temperature rises due to star formation, depleted CO molecules are released into the gas phase, and the H$_2$D$^+$/H$_3^+$ ratio decreases.
Consequently, the deuterium fractionation in molecules starts to decrease after the onset of star formation. The timescale of this decrease may vary among molecular species, depending on the destruction timescale of each molecular species; the deuterium fractionation of ionic molecules such as N$_2$D$^+$/N$_2$H$^+$ decreases quickly (e.g. Chen et al. 2010, 2011), while the deuterium fractionation of neutral molecules such as DNC/HNC takes a longer time to decrease (Sakai et al. 2015).
However, it is still not well understood how the deuterium fractionation of molecules varies after the onset of star formation and what affects the deuterium fractionation levels in hot regions around protostars. In this paper, we focus on the deuterium fractionation of methanol (CH$_3$OH) toward high-mass star-forming cores.

Several studies have been conducted on the deuterium fractionation of methanol in high-mass star-forming regions (Charnley et al. 1997; Peng et al. 2012; Fontani et al. 2015; Belloche et al. 2016; B{\o}gelund et al. 2018; van der Walt et al. 2021; van Gelder et al. 2022; Baek et al. 2022). van Gelder et al. (2022) investigated the deuterium fractionation of methanol in 99 objects based on the data from the ALMA Evolutionary Study of High Mass Protocluster Formation in the Galaxy (ALMAGAL) survey and determined the D/H ratio of methanol in 38 of those objects. They found that the deuterium fractionation of methanol in high-mass star-forming regions is lower than that in low-mass star-forming regions. 
B{\o}gelund et al. (2018) also reported the low deuterium fractionation of methanol in NGC6334I using the ALMA data. Peng et al. (2012) revealed the distribution of the deuterated methanol toward the Orion BN/KL region, and they found that the deuterium fractionation of methanol is not significantly different over the observed regions (CH$_2$DOH/CH$_3$OH$\sim$(1.1-8.8)$\times$10$^{-4}$), except for Orion KL ($<$2.2$\times$10$^{-4}$).
Despite these studies, the extent of variation in the deuterium fractionation of methanol across different high-mass star-forming regions remains unclear. In this paper, we explore the diversity of the deuterium fractionation level of methanol among high-mass star-forming regions in order to answer the question: What kind of information can the deuterium fractionation of methanol in high-mass star-forming regions deliver to us?

For this purpose, we used the data from the DIHCA (Digging into the Interior of Hot Cores with ALMA) project. The DIHCA project aims at observing high-mass star-forming regions with ALMA in order to elucidate the fragmentation process of molecular clouds associated with high-mass star formation. Several papers have been published on the results of the DIHCA project (Olguin et al. 2021, 2022, 2023; Taniguchi et al. 2023; Li et al. 2024; Ishihara et al. 2024). Ishihara et al. (2024) reported on the distribution of dense cores based on the continuum data, and Taniguchi et al. (2023) reported on the chemical composition of hot cores. In this study, we present the distributions of $^{13}$CH$_3$OH and CH$_2$DOH for 24 fields observed in the DIHCA project and investigate how much the deuterium fractionation of methanol differs among the 24 sources. In addition, we examine the distribution of the deuterium fractionation in each individual region and discuss the relationship between star formation activity and the deuterium fractionation of methanol.

\section{Observations} \label{sec:obs}

In the DIHCA project, bright ($>$ 0.1 Jy at 230 GHz) targets that have been studied with interferometers were selected from the literature, and 30 high-mass star-forming clumps were then observed (see Ishihara et al. 2024). In this paper, we used the data of 24 targets where the $^{13}$CH$_3$OH emission line was detected. The distances range from 1.30 kpc to 5.26 kpc. The properties of the selected targets are listed in Table 1.

The observations were performed with ALMA during Cycles 4, 5, and 6 (Project ID: 2016.1.01036.S; 2017.1.00237.S; PI: P. Sanhueza).
The observations were conducted in the single pointing mode, and the data used in this paper were obtained with the single array configuration (C-5 or C-6) of the 12 m array.
In the DIHCA project, four spectral windows (216.9-218.7 GHz, 219.0-221.0 GHz, 231.0-233.0 GHz, and 233.5-235.5 GHz) were observed. Although many methanol emission lines are detected in the observed frequency range, we make no use of the $^{12}$CH$_3$OH lines in this work, because they are thought to be optically thick.
In this paper, we use the $^{13}$CH$_3$OH and CH$_2$DOH lines, which are likely to be optically thin.  In particular, we have selected 3 transitions that are unlikely to be contaminated by other molecular lines: $^{13}$CH$_3$OH $5_1-4_1$ A$^+$, $^{13}$CH$_3$OH $14_1-13_2$ A$^-$, and CH$_2$DOH $8_{2,6}-8_{1,7}$ $e_0$. 
The molecular line parameters used are given in Table 2.

Data calibrations were carried out by using the Common Astronomy Software Applications (CASA) software package versions 4.7.0, 4.7.2, 5.1.1-5, 5.4.0-70, and 5.6.1-8. The observational data were self-calibrated, and we adopted the procedure for selecting emission-free channels for continuum subtraction that is described in the Appendix of Olguin et al. (2022). 
According to the ALMA technical handbook, the accuracy of the absolute flux calibration for ALMA Band 6 data is estimated to be 10\%.
For imaging of the data cube, we use an automatic masking script YCLEAN that makes use of tclean in CASA (Contreras et al. 2018). For cleaning, we used the Briggs weighting with a robust parameter of 0.5. The angular resolution of the images is typically $\sim$0$^{\prime\prime}$.3, which corresponds to $\sim$0.0019 pc ($\sim$400 au) at a distance of 1.3 kpc and $\sim$0.0077 pc ($\sim$1600 au) at a distance of 5.26 kpc. The beam sizes are listed in Table 1. The 1$\sigma$ rms noise level of the spectral data ranges from 0.002 Jy beam$^{-1}$ to 0.031 Jy beam$^{-1}$ at a velocity resolution of 0.67 km s$^{-1}$ for $^{13}$CH$_3$OH $14_1-13_2$ A$^-$, and from 0.002 Jy beam$^{-1}$ to 0.019 Jy beam$^{-1}$ at a velocity resolution of 0.62 km s$^{-1}$ for $^{13}$CH$_3$OH $5_1-4_1$ A and CH$_2$DOH $8_{2,6}-8_{1,7}$ $e_0$.

\section{Results} \label{sec:results}

\subsection{Integrated Intensity Maps} \label{subsec:integ}

Figure 1 shows the integrated intensity maps for $^{13}$CH$_3$OH $5_1-4_1$ A$^+$, $^{13}$CH$_3$OH $14_1-13_2$ A$^-$, and CH$_2$DOH $8_{2,6}-8_{1,7}$ $e_0$. 
For the integrated intensity maps, we integrated over a velocity range of 10 km s$^{-1}$; specific values are given in the captions of Figure 1.
The errors shown in the captions of Figure 1 were calculated based on the rms noise in emission-free areas of the maps. Since the purpose of these maps is to investigate spatial distributions within each region, the errors do not account for the uncertainty in absolute flux calibration.
As seen in Figure 1, the $^{13}$CH$_3$OH emission is detected near continuum peaks in many sources. 
Thus, the $^{13}$CH$_3$OH emission is likely to trace the dense regions around the protostars.
Since we use the $^{13}$CH$_3$OH data instead of the $^{12}$CH$_3$OH data, outflow shocked regions with low H$_2$ density and/or lower methanol column density are not seen in the maps.

In Figure 1, we selected the peaks of $^{13}$CH$_3$OH and labelled them as P1, P2, P3, etc. For sources with many peaks, such as G351.77 and NGC6334I, we chose an appropriate number of peaks based on the intensity of the $^{13}$CH$_3$OH $14_1-13_2$ A$^-$ transitions, starting with the strongest.
In some regions, the $^{13}$CH$_3$OH emission appears unassociated with the continuum peaks. For example, a $^{13}$CH$_3$OH peak (P2) does not coincide with any of the continuum peaks in G5.89-0.37. It is known that G5.89-0.37 has a compact HII region and a large outflow (e.g., Su et al. 2012; Zapata et al. 2019; Fern{\'a}ndez-L{\'o}pez et al.2021), so that G5.89-0.37 P2 could trace a shocked region with a high H$_2$ density and/or high methanol column density.
We also note that even the $^{13}$CH$_3$OH emission detected near the continuum peaks might also trace shocked regions by outflows. In order to determine whether the emission traces regions heated by radiation from protostars or shocked regions near the protostars, further high-resolution observations are required.

The $^{13}$CH$_3$OH emission is relatively widespread in G351.77-0.54 and NGC6334I, which could be partly due to these two sources being the closest among the observed sources.
In G351.77-0.54 and NGC6334I, the $^{13}$CH$_3$OH emission is weakened toward some peaks of the continuum emission. This suggests the absorption due to the bright continuum or a decrease of the CH$_3$OH abundance near the protostars.

It is also found that continuum sources do not always reveal $^{13}$CH$_3$OH emission. The fraction of such peaks varies by region, but G333.46-0.16 and G336.01-0.82 appear to have a higher fraction.
The continuum sources without $^{13}$CH$_3$OH emission may be in either an earlier (i.e. starless or very early stages of star formation) or later evolutionary stage (i.e. ultra-compact HII regions) and may also be low-mass cluster members. 
According to Ishihara et al. (2024), there are 45 cores with a mass of $>$10 $M_\odot$ in the regions reported in this paper, and we found that 18 of these cores exhibit the $^{13}$CH$_3$OH emission.
The cores without $^{13}$CH$_3$OH emission are not investigated in the present work.

In the sources where the CH$_2$DOH emission is detected, the distributions of $^{13}$CH$_3$OH and CH$_2$DOH are found to be similar to each other. However, slight discrepancies in peak positions between $^{13}$CH$_3$OH and CH$_2$DOH are observed in some sources (e.g., G335.579 and G335.78), which will be discussed in Section \ref{sec:discuss}. 

\subsection{Spectra} \label{subsec:spec}

 Figures 2 and 3 show the spectral line profiles of a single pixel at the peak positions of $^{13}$CH$_3$OH. 
The line widths are found to be broad ($\sim$5 km s$^{-1}$) for most of them. In addition, non-Gaussian spectral profiles are seen in some sources (e.g. G35.13, G351.77), suggesting that these sources have a few velocity components in the line of sight.
Toward some sources (e.g. G335.579 P1), a strong spectral line is seen about 5 km s$^{-1}$ red-shifted relative to the CH$_2$DOH line (see Figure 2). This spectral line could be the line of acetone (CH$_3$COCH$_3$ 10$_{8,3}$-9$_{5,4}$ EE, 234.4681493 GHz, $E_u$$\sim$47 K). In G351.77 (Figure 3), an emission line can be seen around -23 km s$^{-1}$, which is likely to be a line of methyl formate (HCOOCH$_3$ 19$_{9,10}$-18$_{9,9}$, 234.486395 GHz, $E_u$$\sim$166 K).

In Figures 2 and 3, it is also found that the relative intensity of CH$_2$DOH to $^{13}$CH$_3$OH varies between sources. For example, the intensities of the two $^{13}$CH$_3$OH lines are comparable between G29.96 P1 and G34.43 P1, while the intensity of CH$_2$DOH is significantly higher in G34.43 P1 than in G29.96 P1 (Figure 2). The difference between $^{13}$CH$_3$OH and CH$_2$DOH will be discussed in Section \ref{sec:discuss}.

We calculate the integrated intensities of each spectral line within a $\pm$2.5 km s$^{-1}$ range from the peak velocity of $^{13}$CH$_3$OH. 
Since the acetone line may partially overlap with the spectral lines of CH$_2$DOH in some sources, we set the integration range to minimize its contamination. 
Although the $\pm$2.5 km s$^{-1}$ range does not cover the entire velocity range of the lines, we chose this range because the main purpose is to examine the CH$_2$DOH/CH$_3$OH ratio. 
The results are listed in Table 3. The errors in the integrated intensities described in the table were derived by combining the 1$\sigma$ rms noise level of the spectra with the 10\% uncertainty from the absolute flux calibration.

Figure 4 shows a correlation plot of the integrated intensities. The integrated intensities of the two $^{13}$CH$_3$OH lines are found to be strongly correlated with a correlation coefficient of 0.86, while the integrated intensities of $^{13}$CH$_3$OH and CH$_2$DOH are moderately correlated with a correlation coefficient of 0.65.

\section{Analysis} \label{subsec:analysis}

\subsection{Temperature} \label{subsec:temp}

Assuming local thermodynamic equilibrium (LTE) and optically thin emission, we estimate the rotation temperature of $^{13}$CH$_3$OH toward the $^{13}$CH$_3$OH peaks from the integrated intensities of the observed $^{13}$CH$_3$OH lines (Table 3). 
We assume the optically thin condition as a first (minimum order)  approximation, because $^{13}$CH$_3$OH is generally about 2.5\% of the abundance of $^{12}$CH$_3$OH.
It is difficult to confirm that these lines are truly optically thin from the data used in this paper, because we only analyze two lines. 

The rotation temperature ($T_{rot}$) can be derived using the following equation,
\begin{equation}
T_{\rm rot} = \frac{E_{u1}-E_{u2}}{k \ln(\frac{\nu_1 \mu_1^2 S_1}{\nu_2 \mu_2^2 S_2} \frac{W_2}{W_1})}, \label{eq:temp}
\end{equation}
where $E_{u}$ is the upper state energy, $\nu$ is the rest frequency, $k$ is the Boltzmann constant, $\mu$ is the dipole moment, $S$ is the line strength, and $W$ is the integrated intensity in K km s$^{-1}$. The subscript numbers 1 and 2 refer to the $^{13}$CH$_3$OH $14_1-13_2$ A$^-$ and $^{13}$CH$_3$OH $5_1-4_1$ A$^+$ lines, respectively. 

The integrated intensity in K km s$^{-1}$ is calculated from the following equation:
\begin{equation}
W = \frac{c^2}{2k\nu^2 \Omega_B} \int I_{\nu} dV,
\end{equation} 
where $c$ is the light speed, $I_{\nu}$ is the flux density, and $\Omega_{\rm B}$ is a solid angle of the beam. 
The error was calculated from the 1$\sigma$ rms noise level of the integrated intensities (Table 3). 
Since the rotation temperature was derived from the ratio of integrated intensities of two spectral lines observed simultaneously, the uncertainty due to absolute flux calibration is negligible and therefore not included in the error estimation of the rotation temperature. The calculated temperatures are listed in Table 4. 
The derived temperatures are mostly above 100 K. The temperatures derived here are comparable to those obtained from CH$_3$CN by Taniguchi et al. (2023). However, this is not an exact comparison because the positions for the temperatures derived from CH$_3$CN and those determined in this study are not exactly the same.

\subsection{Column Densities} \label{subsec:col}

Under the assumptions of the LTE and optically thin conditions, as in the case deriving $T_{rot}$, we calculated the column densities ($N$) using the following equation:
\begin{equation}
N=\frac{3hQ}{8\pi^3 \mu^2 S } \frac{\exp\left(\frac{E_u}{kT_{\rm rot}}\right)}{\exp\left(\frac{h \nu}{kT_{\rm rot}}\right)-1} \frac{W}{J(T_{\rm rot})-J(T_{\rm B})}, \label{eq:col}
\end{equation}
where $h$ is the Planck constant, $Q$ is the partition function, $T_{\rm B}$ is the temperature of the cosmic background radiation (2.73 K), and $J$ is the Planck function for the given temperature and frequency as:
\begin{equation}
J(T) = \frac{\frac{h\nu}{k}}{\exp\left(\frac{h\nu}{kT}\right)-1}.
\end{equation}
The parameters used are listed in Table 2.

The partition function $Q$ is evaluated from the data in the Cologne Database for Molecular Spectroscopy (CDMS; M{\"u}ller et al. 2005) and the Jet Propulsion Laboratory molecular spectroscopy database (JPL; Pickett et al. 1998) for $^{13}$CH$_3$OH and CH$_2$DOH, respectively. We fit the CDMS or JPL data to the function of $Q$($T$) = $a$$\times$$T^b$, where $a$ and $b$ are fitting parameters. For $^{13}$CH$_3$OH, ($a$, $b$)=(0.35711, 1.78288). For CH$_2$DOH, ($a$, $b$)=(1.33986, 1.63442).

For $^{13}$CH$_3$OH, the column density was calculated using the parameters of the $5_1-4_1$ A$^+$ emission line. The rotation temperature used in these calculations was derived in Section 4.1. We assume that the rotation temperatures for CH$_2$DOH and $^{13}$CH$_3$OH are identical. Errors in the column density were derived from the 1$\sigma$ rms noise level of the observed integrated intensities (Table 3), the uncertainty in the temperature estimate (Table 4), and the uncertainty of the absolute flux calibration (10\%). It should be noted that the derived column densities do not reflect the total column densities along the line of sight, because the integrated intensities are calculated over a range of $\pm$2.5 km s$^{-1}$.

We also calculate the column density of $^{12}$CH$_3$OH, with assuming the $^{12}$C/$^{13}$C ratio. The $^{12}$C/$^{13}$C ratio was estimated using the relationship with the distance from the galactic center ($D_{\rm GC}$) (Yan et al. 2019): 
\begin{equation}
{\rm ^{12}C/^{13}C} = (5.08 \pm 1.10) D_{\rm GC} + (11.86 \pm 6.60). \label{eq:cratio}
\end{equation}
The values used are the same as those used in Taniguchi et al. (2023). Errors in the column density of $^{12}$CH$_3$OH includes the uncertainty of $^{12}$C/$^{13}$C, in addition to the uncertainty of the column density of $^{13}$CH$_3$OH. The results are listed in Table 4.
The $N$(CH$_2$DOH)/$N$($^{13}$CH$_3$OH) and $N$(CH$_2$DOH)/$N$($^{12}$CH$_3$OH) ratios are also listed in Table 4.

\section{Discussion} \label{sec:discuss}

\subsection{$N${\rm (CH}$_2${\rm DOH)}/$N${\rm (CH}$_3${\rm OH)} Ratio} \label{subsec:ch2doh}

In this section, we compare the $N$(CH$_2$DOH)/$N$($^{13}$CH$_3$OH) and $N$(CH$_2$DOH)/$N$($^{12}$CH$_3$OH) ratios in our observed regions with those in other regions.  
Figure 5a shows a plot between the rotation temperature and the $N$(CH$_2$DOH)/$N$($^{13}$CH$_3$OH) ratio of the $^{13}$CH$_3$OH peaks.
In Figure 5a, we found that there is no apparent correlation between the $N$(CH$_2$DOH)/$N$($^{13}$CH$_3$OH) ratio and the rotation temperature. While the rotation temperatures of many sources are around 100-200 K, the $N$(CH$_2$DOH)/$N$($^{13}$CH$_3$OH) ratio is spread over a range of more than an order of magnitude. 
This result suggests that some factors other than the current temperature influence the degree of the deuterium fractionation of methanol (see Section 5.3).

Figure 5b shows a plot between the rotation temperature and the $N$(CH$_2$DOH)/$N$($^{12}$CH$_3$OH) ratio.
In Figure 5b, a range of the corresponding ratio for high-mass protostars observed by van Gelder et al. (2022) and that for low-mass protostars (for references on low-mass protostars, see van Gelder et al. 2022) are also shown.
Although van Gelder et al. (2022) consider a D/H ratio that takes into account the statistical weight, which is one third of the column density ratio, we use only the column density ratio or abundance ratio in this paper. This choice does not affect the discussion.

In Figure 5b, the ratios for our observed high-mass star-forming regions are comparable to or slightly higher than those of the high-mass protostars reported by van Gelder et al. (2022). The low $N$(CH$_2$DOH)/$N$($^{12}$CH$_3$OH) ratios were observed toward NGC6334I (see also Table 4), which are consistent with the results of B{\o}gelund et al. (2018) and van Gelder et al. (2022).
It is also clear that the values for our observed high-mass star-forming regions are lower than those for low-mass protostars.
Thus, all the studies consistently show that the $N$(CH$_2$DOH)/$N$($^{12}$CH$_3$OH) ratio in high-mass star-forming regions is clearly different from that in low-mass star-forming regions.

In Figure 5b, there is also no correlation between the $N$(CH$_2$DOH)/$N$($^{12}$CH$_3$OH) ratio and the rotation temperature. Due to the large uncertainty in the $^{12}$C/$^{13}$C ratio, the $N$(CH$_2$DOH)/$N$($^{12}$CH$_3$OH) ratio appear to fall mostly within the range of 10$^{-3}$ to 10$^{-2}$. 
However, we note here that the source-to-source variation in the $^{12}$CH$_3$OH/$^{13}$CH$_3$OH ratio caused by chemical reactions is not significant according to chemical model calculations (Ichimura et al. 2024). Therefore, the source-to-source variation in the $N$(CH$_2$DOH)/$N$($^{13}$CH$_3$OH) ratio should similarly be observed in the $N$(CH$_2$DOH)/$N$($^{12}$CH$_3$OH) ratio, if there are no significant differences in the $^{12}$C/$^{13}$C ratio. Since many of the observed sources are located in the inner Galaxy, with galactic center distances of 4–7 kpc, the variation in the $^{12}$C/$^{13}$C ratio could be smaller than the uncertainties estimated from Equation 6.
Thus, it is possible that the $N$(CH$_2$DOH)/$N$($^{12}$CH$_3$OH) ratio also exhibits source-to-source variation, similar to the variation observed in the $N$(CH$_2$DOH)/$N$($^{13}$CH$_3$OH) ratio, although additional observations to confirm the differences in the $^{12}$C/$^{13}$C ratio are required.

In Figures 5c and 5d, we show plots of the distance against the $N$(CH$_2$DOH)/$N$($^{13}$CH$_3$OH) ratio and $N$(CH$_2$DOH)/$N$($^{12}$CH$_3$OH) ratio, respectively. There is no correlation between the distance and the $N$(CH$_2$DOH)/$N$($^{13}$CH$_3$OH) ratio or the $N$(CH$_2$DOH)/$N$($^{12}$CH$_3$OH) ratio, indicating that the spatial resolution of the observations is not responsible for this variation.

Figures 5e and 5f show plots of the dust temperature of the clumps (Urquhart et al. 2018) against the $N$(CH$_2$DOH)/$N$($^{13}$CH$_3$OH) ratio and $N$(CH$_2$DOH)/$N$($^{12}$CH$_3$OH) ratio, respectively.
Although the rotation temperature derived from the $^{13}$CH$_3$OH emission reflects the temperature of the dense and hot regions around protostars, the dust temperature reported by Urquhart et al. (2018) rather reflects the temperature of the outer envelopes of the embedded dense cores. In fact, the sources with dust temperatures below 17 K (G11.1-0.12, G14.22-0.50S, and G333.12-0.56) are associated with infrared dark clouds (which contain the earliest stages of high-mass star formation; Sanhueza et al. 2012, 2019, Morii et al. 2023), according to the Spitzer images (see Figure 1 of Ishihara et al. 2024).
In Figures 5e and 5f, the $N$(CH$_2$DOH)/$N$($^{13}$CH$_3$OH) ratio and $N$(CH$_2$DOH)/$N$($^{12}$CH$_3$OH) ratio is different from source to source in the dust temperature range of 20-30 K. This indicates that the current clump-scale temperature alone cannot explain the variation of the $N$(CH$_2$DOH)/$N$($^{13}$CH$_3$OH) or $N$(CH$_2$DOH)/$N$($^{12}$CH$_3$OH) ratios, either. Another important result in Figures 5e and 5f is that there are no sources with low $N$(CH$_2$DOH)/$N$($^{13}$CH$_3$OH) ratio or $N$(CH$_2$DOH)/$N$($^{12}$CH$_3$OH) ratios for sources with low dust temperatures ($<$20 K). This point will be discussed later.

\subsection{Spatial Distribution of the N{\rm (CH}$_2${\rm DOH)/}N{\rm (CH}$_3${\rm OH)} Ratio} \label{subsec:dist}

In this section, we examine the spatial distribution of the $N$(CH$_2$DOH)/$N$($^{12}$CH$_3$OH) ratio in eight selected regions, where the intensity of $^{13}$CH$_3$OH is bright, higher than 0.2 Jy beam$^{-1}$ and the distance is less than 3.5 kpc.
To create the integrated intensity maps, we first evaluate the peak velocity of $^{13}$CH$_3$OH $5_1-4_1$ A$^+$ toward each pixel. Then, we set the integrated velocity range to $\pm$2.5 km s$^{-1}$ from the peak velocity of $^{13}$CH$_3$OH $5_1-4_1$ A$^+$ for all of the three lines toward each pixel.
Using the derived integrated intensity maps, we created temperature maps using Equation \ref{eq:temp}.
In the calculations of the rotation temperature, we only used data from the positions where the $^{13}$CH$_3$OH $14_1-13_2$ A$^-$ intensity is above the $3\sigma$ noise level.
Using the temperature map, Equation 4, and Equation 6, we made maps of the $N$(CH$_2$DOH)/$N$($^{12}$CH$_3$OH) ratio.
In the calculations of the $N$(CH$_2$DOH)/$N$($^{12}$CH$_3$OH) ratio, we only used data from the positions where the intensities of all the lines, including the CH$_2$DOH line, are above the $3\sigma$ noise level.
The results are shown in Figure 6.

Comparing the maps of the $N$(CH$_2$DOH)/$N$($^{12}$CH$_3$OH) ratio and the rotation temperature among the regions, it is clear that the $N$(CH$_2$DOH)/$N$($^{12}$CH$_3$OH) ratios vary over the entire region, even if the rotation temperature does not vary significantly. Furthermore, there is no significant correlation between the $N$(CH$_2$DOH)/$N$($^{12}$CH$_3$OH) ratio and the rotation temperature. For example, the $N$(CH$_2$DOH)/$N$($^{12}$CH$_3$OH) ratio is different between G34.43+0.24 ($\sim$0.02) and IRAS18089-1732 ($\sim$0.002), although the temperature is similar ($\sim$200 K). This difference is significant even if we take the uncertainty of $^{12}$C/$^{13}$C ratio into account.
In addition, the $N$(CH$_2$DOH)/$N$($^{12}$CH$_3$OH) ratio in G335.579-0.272 is comparable to that in G35.20-00.74 ($\sim$0.01), while the temperature of G335.579-0.272 ($\sim$300 K) is much higher than that of G35.20-00.74 ($\sim$140 K). These results confirm that the degree of deuterium fractionation of methanol is independent of the present gas temperature.

We also find a variation in the $N$(CH$_2$DOH)/$N$($^{12}$CH$_3$OH) ratio even within each region as well as no clear correlation between $N$(CH$_2$DOH)/$N$($^{12}$CH$_3$OH) and the temperature within each region. In particular, the $N$(CH$_2$DOH)/$N$($^{12}$CH$_3$OH) ratio in G335.78+00.17 differs significantly between the eastern and western parts of the continuum peak, and the temperature distribution is not correlated with the $N$(CH$_2$DOH)/$N$($^{12}$CH$_3$OH) ratio. Since the difference could not be due to the difference in the temperature, it is likely that regions with high and low deuterium fractionation exist in this region. This implies that the factors affecting the $N$(CH$_2$DOH)/$N$($^{12}$CH$_3$OH) ratio can vary even on a relatively small scale. High angular resolution observations are necessary for further detailed examination of such small-scale variation.

\subsection{Origins of the Diversity of $N${\rm (CH}$_2${\rm DOH)}/$N${\rm (CH}$_3${\rm OH)} ratio} \label{subsec:div}

We have suggested that the $N$(CH$_2$DOH)/$N$($^{13}$CH$_3$OH) and $N$(CH$_2$DOH)/$N$($^{12}$CH$_3$OH) ratios is independent on the temperature at present. In this section, we discuss the origins of the diversity of the $N$(CH$_2$DOH)/$N$($^{12}$CH$_3$OH) ratio. Figure 7 shows a diagram of the formation mechanism of CH$_3$OH and CH$_2$DOH. The main reaction pathways on grain surfaces suggested by the experiments in the laboratory (Watanabe et al. 2006; Watanabe \& Kouchi 2008; Hidaka et al. 2009) are represented by bold arrows. CH$_3$OH is mainly formed on dust grains through hydrogenation reactions of CO, and CH$_2$DOH is also formed on dust grains by deuteration of CH$_2$OH. Therefore, the $N$(CH$_2$DOH)/$N$($^{12}$CH$_3$OH) ratio is expected to depend on the atomic D/H ratio on the dust grains.
The atomic D/H ratio on dust grains depends on that in the gas phase. As shown in Figure 7, D and H atoms in the gas phase are mainly produced by dissociative recombination of H$_3^+$ or H$_2$D$^+$ with electrons. Thus, the atomic D/H ratio in the gas phase would depend on the H$_2$D$^+$/H$_3^+$ ratio. Since the H$_2$D$^+$/H$_3^+$ ratio is enhanced in cold and dense regions, as mentioned in Section 1, the $N$(CH$_2$DOH)/$N$($^{12}$CH$_3$OH) ratio on dust grains is expected to increase in cold dense cores.

After the onset of star formation, CH$_2$DOH and CH$_3$OH on dust grains are released into the gas phase due to heating by star formation activity.
In the case of CH$_3$OD, it could be destroyed on the grain surface before sublimation (around 80 K) through the reaction CH$_3$OD + H$_2$O$\rightarrow$HDO + CH$_3$OH (e.g., Ratajczak et al. 2009), while CH$_2$DOH does not react with H$_2$O and simply sublimates.
Thus, the $N$(CH$_2$DOH)/$N$($^{12}$CH$_3$OH) ratio around protostars can reflect the ratio on dust grains, and the diversity of the $N$(CH$_2$DOH)/$N$($^{12}$CH$_3$OH) ratio in the gas around protostars can be considered to better depict the condition of the cold phase experienced before sublimation.

To verify the hypothesis presented above, we reanalyze the chemical model calculations reported by Sakai et al. (2022). For the formation reactions of CH$_3$OH and CH$_2$DOH on grain surface, we adopt the same reactions and the rate coefficients as in Taquet et al. (2012) (see also Figure 1 in Taquet et al.).
We compute the temporal changes of the chemical composition during two phases: the prestellar phase and the warm-up phase. During the prestellar phase, we simulate collapsing cores with a constant temperature of 10 K.
In the calculations, we assume three different collapsing speeds; $Model$ 1: one free fall time, $Model$ 2: three times slower than the free fall time, and $Model$ 3: ten times slower than the free fall time. 
In all models, the initial and final volume densities are set to 3$\times$10$^{3}$ cm$^{-3}$ and 10$^{7}$ cm$^{-3}$, respectively.
For the warm-up phase, we simulate a core with a constant density of 10$^7$ cm$^{-3}$. The temperature of the core increases from 10 to 200 K on a timescale of 10$^4$ yr, a period chosen as an estimate based on the age range of high-mass outflows, which typically ranges from 10$^4$ to 10$^5$ years (Beuther et al. 2002).
We assume that the temperature changes in proportion to the square of time (Garrod \& Herbst 2006).
The initial chemical composition of the warm-up phase is set to the chemical composition at the end of the prestellar phase, at which the density is 10$^7$ cm$^{-3}$. More detailed information on the chemical model calculations is described in Sakai et al. (2022).

The model calculation results of the CH$_2$DOH/CH$_3$OH abundance ratio and the abundances are shown in Figure 8 and Figure 9, respectively. For the results of the prestellar phase (Figures 8a, 9a, and 9c), we plot the density on the horizontal axis and the CH$_2$DOH/CH$_3$OH ratio on the vertical axis. In Figures 9a and 9c, the abundances of CH$_3$OH and CH$_2$DOH on the grain surface is much higher than those in the gas phase.
Thus, it is most likely that CH$_3$OH and CH$_2$DOH are mostly formed on the grain surface in the prestellar phase. In Figure 8a, the CH$_2$DOH/CH$_3$OH ratio appears to increase with density and collapsing speed. In particular, the CH$_2$DOH/CH$_3$OH ratio on grain surface is higher in the model with longer collapse timescales; the difference in CH$_2$DOH/CH$_3$OH is more than one order of magnitude.

For the results of the warm-up phase (Figures 8b, 9b, and 9d), we plot the temperature on the horizontal axis and the CH$_2$DOH/CH$_3$OH ratio on the vertical axis.
In Figure 8b, it appears that the CH$_2$DOH/CH$_3$OH ratio in the gas phase decreases with increasing temperature from 10 to 90 K. In contrast, the ratio on the grain surface remains unchanged over the whole temperature range. As shown in Figures 9b and 8d, both CH$_3$OH and CH$_2$DOH begin to sublimate from dust grains at around 90 K. 
Since the derived temperatures are mostly above 100 K, it is likely that the observed CH$_3$OH and CH$_2$DOH are those sublimated from the dust grains.
Above the temperature of 90 K, the CH$_2$DOH/CH$_3$OH ratio in the gas-phase becomes comparable to that on dust grains in the prestellar phase (Figure 8b). Consequently, the CH$_2$DOH/CH$_3$OH ratio in hot cores depends on the ratio on the grain surface before the onset of star formation, and the CH$_2$DOH/CH$_3$OH ratio in the hot cores can differ by more than an order of magnitude depending on the collapsing time scale.
This indicates that the CH$_2$DOH/CH$_3$OH ratio in hot cores depends on the duration of the cold prestellar phase; if the duration of the cold phase is long, the CH$_2$DOH/CH$_3$OH ratio in hot cores would be high. This is confirmed by Figure 10, which is a plot of time versus the CH$_2$DOH/CH$_3$OH ratio in the prestellar phase. 
We should note that the CH$_2$DOH/CH$_3$OH ratio in shocked regions is also expected to depend on the duration of the cold phase, because the CH$_2$DOH/CH$_3$OH ratio in shocked regions caused by outflows is likely to reflect the values on the dust grains before the shock occurred.
Therefore, the lower N(CH$_2$DOH)/N($^{12}$CH$_3$OH) ratio in high-mass star-forming regions compared to low-mass star-forming regions may suggest that the collapsing timescale in high-mass star-forming regions is shorter.

In Figure 8, it should also be noted that the final CH$_2$DOH/CH$_3$OH ratio in the gas phase from the model calculations is higher than the observed values in the high-mass star-forming regions, although the $N$(CH$_2$DOH)/$N$(CH$_3$OH) ratio in some of our observed sources is comparable to the results from the chemical model. 
One possible explanation for the observed low $N$(CH$_2$DOH)/$N$($^{12}$CH$_3$OH) ratio is that the model calculations assume a constant density, whereas there are density structures within the observed regions.
If density structures were the primary cause, a trend dependent on resolution would be expected. However, as shown in Figure 5, there is no correlation between the resolution and the $N$(CH$_2$DOH)/$N$($^{12}$CH$_3$OH) ratio, suggesting that the impact of this effect is minimal.
Several other factors could contribute to the observed low $N$(CH$_2$DOH)/$N$($^{12}$CH$_3$OH) ratios, which are discussed below.

Another possibility is that some destruction processes of CH$_3$OH and CH$_2$DOH, which are not included in the model calculations, could occur in hot cores. For example, the destruction processes due to strong UV radiation are not involved in the model. However, since both CH$_2$DOH and CH$_3$OH are destroyed at the almost same rate by UV radiation and they are not formed in the gas phase, the CH$_2$DOH/CH$_3$OH ratio is likely to be unaffected by the UV radiation.
In addition, the self-shielding effects are also expected to be negligible, because the CH$_3$OH column densities are calculated from the $^{13}$CH$_3$OH data, instead of the $^{12}$CH$_3$OH data.
Furthermore, the CH$_3$OH abundance is expected to decrease in more evolved hot cores, because CH$_3$OH reacts with H$_3^+$ to form CH$_3$OH$_2^+$, which is subsequently destroyed by reactions with electrons (Garrod \& Herbst 2006). The timescale for the CH$_3$OH destruction is estimated to be 10$^5$ years (or 10$^4$ years) when the ionization rate is $\sim$1$\times$10$^{-17}$ s$^{-1}$ (or 1$\times$10$^{-16}$ s$^{-1}$) (e.g., Nomura \& Millar 2004). In fact, the peaks of the $^{13}$CH$_3$OH emission are observed to be clearly offset from the continuum peaks in some sources, such as G351.77 and W33A (Figure 6), indicating that the envelope is mainly observed in these regions.
Thus, in evolved hot cores, the deuterium fractionation of CH$_3$OH near the embedded stars is not reflected in the observed results, although we expect that it is comparable to that of the envelope.

Moreover, since molecular destruction due to shocks caused by outflows is likely to occur at the same rates for both CH$_3$OH and CH$_2$DOH, this effect is also considered to have no impact on the CH$_2$DOH/CH$_3$OH ratio. Thus, the destruction processes are unlikely to affect the deuterium fractionation of methanol, implying that the deuterium fractionation of methanol is independent of the age of a hot core.

Another explanation is that the dense cores had been heated by surrounding star formation activity, which would have shortened the duration of their cold phase and resulted in a lower CH$_2$DOH/CH$_3$OH ratio, as pointed out by van Gelder et al. (2022).
As mentioned in Section \ref{subsec:ch2doh}, none of the sources with dust temperature below 20 K exhibit low $N$(CH$_2$DOH)/$N$($^{12}$CH$_3$OH) ratios (Figure 5), indicating that these sources are not likely to be heated by the surroundings.
Furthermore, as mentioned in Section \ref{subsec:dist}, the $N$(CH$_2$DOH)/$N$($^{12}$CH$_3$OH) ratios in G34.43+0.24 and IRAS18089-1732 differs significantly, despite their similar temperatures. According to the Spitzer images (see Figure 1 of Ishihara et al. 2024), G34.43+0.24 is associated with an infrared dark cloud (Rathborne et al. 2005; Sanhueza et al. 2010; Sakai et al. 2013), in contrast to the IRAS18089-1732 case (Sanhueza et al. 2021).
These results seem consistent with the idea that heating from surrounding star formation activity contributes to lower $N$(CH$_2$DOH)/$N$($^{12}$CH$_3$OH) ratios.
It is important to note here that, even if the dense cores with low $N$(CH$_2$DOH)/$N$(CH$_3$OH) ratios were heated by surrounding star formation activity, they must have experienced a period of low temperature, because methanol cannot form on grain surfaces unless the temperature drops below the sublimation point of CO ($\sim$20 K). 

Lastly, the assumed initial ortho/para (o/p) ratio of the H$_2$ molecule, which is 10$^{-3}$, may differ from the actual value, because the deuterium fractionation level depends on the o/p ratio of H$_2$ molecule (Sipil{\"a} et al. 2013; Furuya et al. 2015; Kong et al. 2016; Goodson et al. 2016); the higher the o/p ratio, the lower the deuterium fractionation ratio. If so, the initial o/p ratio is expected to be higher than 10$^{-3}$. However, Furuya et al. (2015) suggest that the o/p ratio reaches 10$^{-3}$ relatively early in the molecular cloud formation (around Av$\sim$2), which is also suggested by Lupi et al. (2021). Since our calculations start from a density of 3$\times$10$^3$ cm$^{-3}$, a higher o/p ratio may not be expected at this stage. It should also be noted that a reliable measurement of the initial o/p ratio is extremely challenging, making it impossible to rule out this possibility at this moment.

In summary, we suggest that the chemical composition in hot cores or shocked regions is influenced by the physical condition of their starless phase. The lower N(CH$_2$DOH)/N($^{12}$CH$_3$OH) ratio in high-mass star-forming regions compared to in low-mass star-forming regions can be attributed to the shorter duration of the cold phase in high-mass star-forming regions than that in low-mass star-forming regions.
The diversity in the N(CH$_2$DOH)/N($^{12}$CH$_3$OH) ratio can be attributed to differences in the duration of the cold starless phase, which could result from differences in the collapsing timescale or the heating by surrounding star formation activity. Furthermore, the effects of the differences in the initial ortho-to-para (o/p) ratio may also be considered.

\section{Summary} \label{subsec:summary}

We summarize the results below.

\begin{itemize} 

\item Observations of $^{13}$CH$_3$OH and CH$_2$DOH in 24 high-mass star-forming regions were conducted using ALMA. The $^{13}$CH$_3$OH and CH$_2$DOH lines were detected near the peaks of the continuum emission, and the $^{13}$CH$_3$OH emission was detected toward 18 of 45 cores with a mass of $>$10 $M_\odot$.
The rotation temperatures estimated from the $^{13}$CH$_3$OH lines are mostly above 100 K. Thus, the $^{13}$CH$_3$OH and CH$_2$DOH emission is likely to trace the hot and dense regions around the protostars.

\item The $N$(CH$_2$DOH)/$N$($^{12}$CH$_3$OH) ratio in the observed high-mass star-forming regions is found to be lower than that in the low-mass star-forming regions and tends to be similar to previously observed values in high-mass star-forming regions. 

\item No correlation was found between the $N$(CH$_2$DOH)/$N$($^{13}$CH$_3$OH) or $N$(CH$_2$DOH)/$N$($^{12}$CH$_3$OH) ratios and either the rotation temperature or the source distance. We also found a source-to-source variation of more than an order of magnitude in the observed $N$(CH$_2$DOH)/$N$($^{13}$CH$_3$OH) ratio. Although there is a relatively large uncertainties of the $^{12}$C/$^{13}$C ratio, the variation of the $N$(CH$_2$DOH)/$N$($^{12}$CH$_3$OH) ratio was also observed among sources.
This suggests that the $N$(CH$_2$DOH)/$N$($^{13}$CH$_3$OH) or $N$(CH$_2$DOH)/$N$($^{12}$CH$_3$OH) ratios do not depend on the temperature at present.

\item Model calculations confirmed that the CH$_2$DOH/CH$_3$OH ratio in hot cores depends on the duration time of the cold phase; the shorter the cold phase, the lower the deuterium fractionation in the hot cores. The lower N(CH$_2$DOH)/N($^{12}$CH$_3$OH) ratio in high-mass star-forming regions compared to low-mass star-forming regions suggests that the duration of the cold phase in high-mass star-forming regions is shorter than that in low-mass star-forming regions. This may be due to the collapsing timescale being shorter in high-mass star-forming regions.

\item The diversity in the $N$(CH$_2$DOH)/$N$($^{12}$CH$_3$OH) ratio across the observed high-mass star-forming regions may be attributed to differences in the duration of the cold starless phase, potentially caused by variations in the collapsing timescale or heating effects from nearby star formation activity. In addition, the influence of an initial ortho-to-para (o/p) ratio may also be an influencing factor.

\end{itemize}

\acknowledgments
We would like to thank the anonymous referee for her/his thoughtful comments, which have improved the quality of this paper.
This paper makes use of the following ALMA data: ADS/JAO.ALMA\#2016.1.01036.S and 2017.1.00237.S. ALMA is a partnership of ESO (representing its member states), NSF (USA) and NINS (Japan), together with NRC (Canada), MOST and ASIAA (Taiwan), and KASI (Republic of Korea), in cooperation with the Republic of Chile. The Joint ALMA Observatory is operated by ESO, AUI/NRAO and NAOJ. 
P.S. was partially supported by a Grant-in-Aid for Scientific Research (KAKENHI Number JP22H01271 and JP23H01221) of JSPS.
PS was supported by Yoshinori Ohsumi Fund (Yoshinori Ohsumi Award for Fundamental Research).
K.T. is supported by JSPS KAKENHI grant Nos. JP20K14523, 21H01142, 24K17096, and 24H00252. 
X.L. acknowledges support from the National Key R\&D Program of China (No. 2022YFA1603101), the Strategic Priority Research Program of the Chinese Academy of Sciences (CAS) Grant No. XDB0800300, the National Natural Science Foundation of China (NSFC) through grant Nos. 12273090 and 12322305, the Natural Science Foundation of Shanghai (No. 23ZR1482100), and the CAS “Light of West China” Program No. xbzg-zdsys-202212.
This study is supported by KAKENHI (JP20K04025, JP20H05645, JP20H05845, JP20H05847, and JP18H05222). 

%

\vspace{5mm}
\facilities{ALMA}


\software{CASA \citep{2007ASPC..376..127M}}
\software{yclean \citep{2018ApJ...861...14C}}



\clearpage








\clearpage

\begin{figure*}
\figurenum{1}
\epsscale{1.0}
\plotone{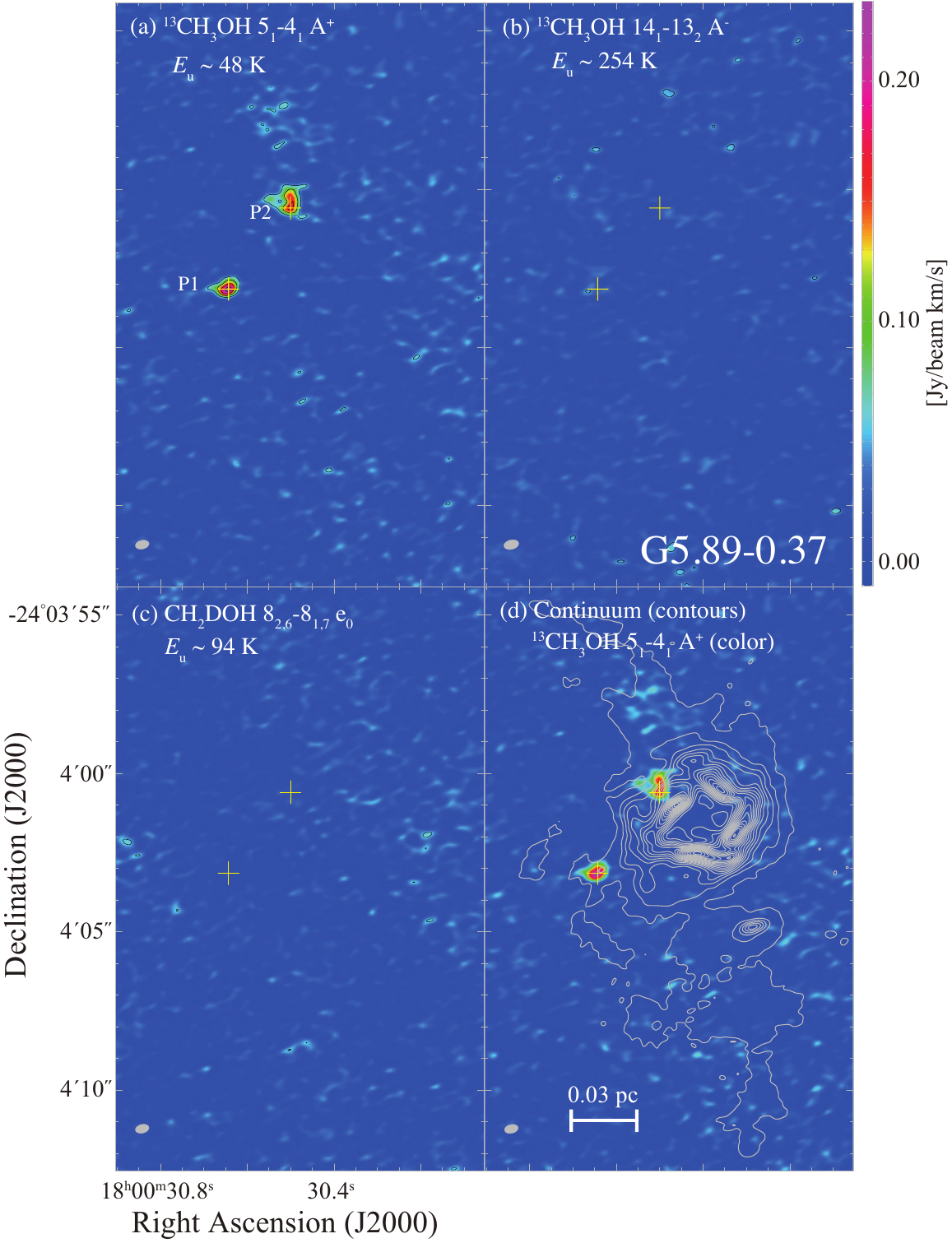}
\caption{Integrated intensity maps of $^{13}$CH$_3$OH $5_1-4_1$ A$^+$ (a), $^{13}$CH$_3$OH $14_1-13_2$ A$^-$ (b), and CH$_2$DOH $8_{2,6}-8_{1,7}$ $e_0$ (c) toward G5.89-00.37. The integrated velocity range is from 5 km s$^{-1}$ to 15 km s$^{-1}$. Contour levels start from 3$\sigma$ and increase in steps of 1$\sigma$ for (a), (b), and (c) [(a) $1\sigma=18$ mJy beam$^{-1}$ km s$^{-1}$, (b) $1\sigma=15$ mJy beam$^{-1}$ km s$^{-1}$, (c) $1\sigma=18$ mJy beam$^{-1}$ km s$^{-1}$]. (d) the continuum image (contours) overlaid on $^{13}$CH$_3$OH $14_1-13_2$ A$^-$ integrated intensity map (color) . For (d), contour levels start from 3$\sigma$ and increase in steps of 20$\sigma$ [$1\sigma=0.2$ mJy beam$^{-1}$]. Cross marks indicate the positions of the $^{13}$CH$_3$OH peak. The same color scale is used in (a)-(d). The synthesized beam is shown at the bottom left of each panel. 
\label{fig:fig1}}
\end{figure*}

\clearpage

\begin{figure*}
\figurenum{1}
\epsscale{1.0}
\plotone{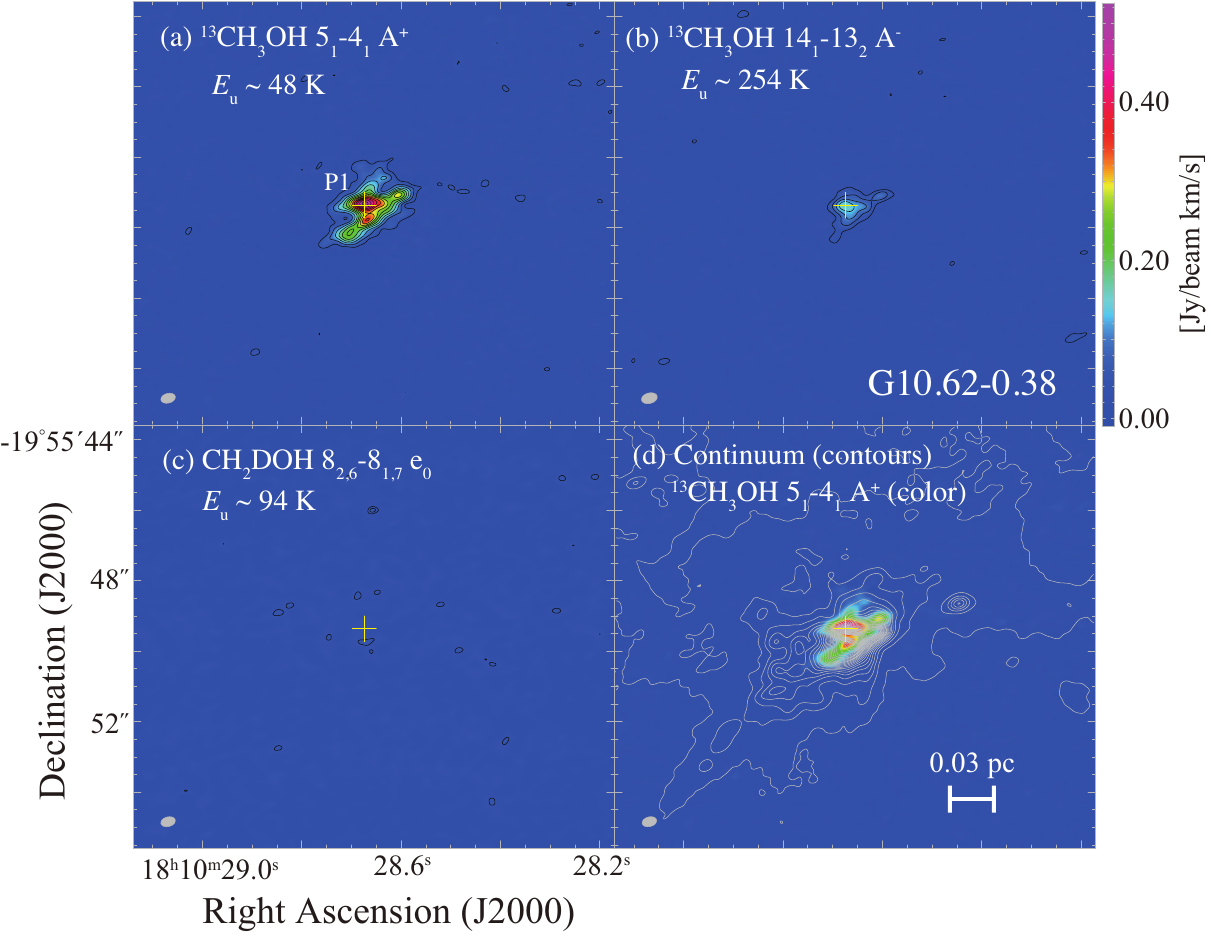}
\caption{Integrated intensity maps of $^{13}$CH$_3$OH $5_1-4_1$ A$^+$ (a), $^{13}$CH$_3$OH $14_1-13_2$ A$^-$ (b), and CH$_2$DOH $8_{2,6}-8_{1,7}$ $e_0$ (c) toward G10.62-00.38. The integrated velocity range is from -5 km s$^{-1}$ to 5 km s$^{-1}$. Contour levels start from 3$\sigma$ and increase in steps of 1$\sigma$ for (a), (b), and (c) [(a) $1\sigma=13$ mJy beam$^{-1}$ km s$^{-1}$, (b) $1\sigma=11$ mJy beam$^{-1}$ km s$^{-1}$, (c) $1\sigma=13$ mJy beam$^{-1}$ km s$^{-1}$]. (d) the continuum image (contours) overlaid on $^{13}$CH$_3$OH $14_1-13_2$ A$^-$ integrated intensity map (color) . For (d), contour levels start from 3$\sigma$ and increase in steps of 20$\sigma$ [$1\sigma=0.4$ mJy beam$^{-1}$]. Cross marks indicate the positions of the $^{13}$CH$_3$OH peak. The same color scale is used in (a)-(d). The synthesized beam is shown at the bottom left of each panel.}
\end{figure*}

\clearpage

\begin{figure*}
\figurenum{1}
\epsscale{1.0}
\plotone{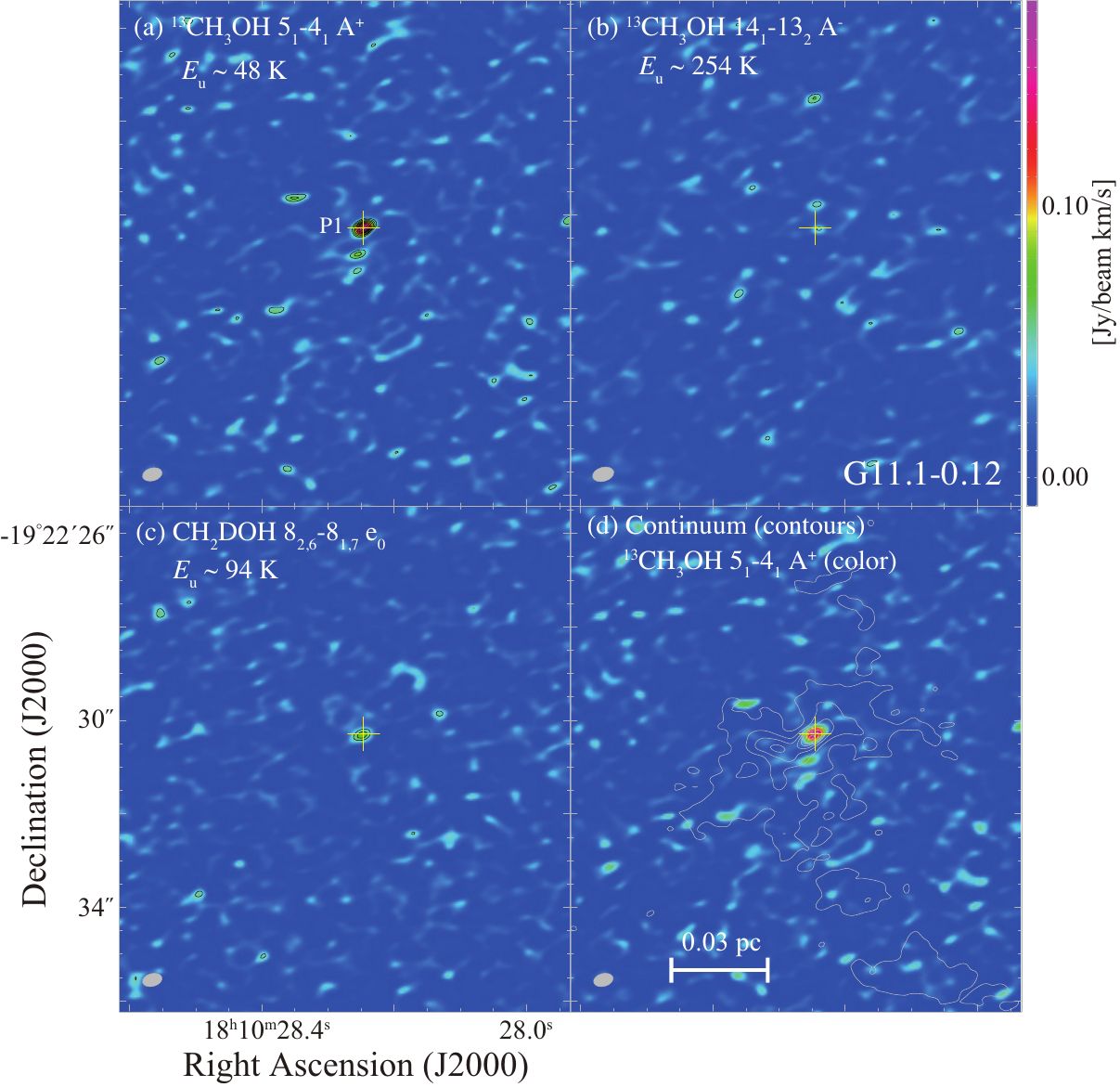}
\caption{Integrated intensity maps of $^{13}$CH$_3$OH $5_1-4_1$ A$^+$ (a), $^{13}$CH$_3$OH $14_1-13_2$ A$^-$ (b), and CH$_2$DOH $8_{2,6}-8_{1,7}$ $e_0$ (c) toward G11.1-00.12. The integrated velocity range is from 25 km s$^{-1}$ to 35 km s$^{-1}$. Contour levels start from 3$\sigma$ and increase in steps of 1$\sigma$ for (a), (b), and (c) [(a) $1\sigma=16$ mJy beam$^{-1}$ km s$^{-1}$, (b) $1\sigma=15$ mJy beam$^{-1}$ km s$^{-1}$, (c) $1\sigma=16$ mJy beam$^{-1}$ km s$^{-1}$]. (d) the continuum image (contours) overlaid on $^{13}$CH$_3$OH $14_1-13_2$ A$^-$ integrated intensity map (color) . For (d), contour levels start from 3$\sigma$ and increase in steps of 10$\sigma$ [$1\sigma=0.1$ mJy beam$^{-1}$]. Cross marks indicate the positions of the $^{13}$CH$_3$OH peak. The same color scale is used in (a)-(d). The synthesized beam is shown at the bottom left of each panel.}
\end{figure*}

\clearpage

\begin{figure*}
\figurenum{1}
\epsscale{1.0}
\plotone{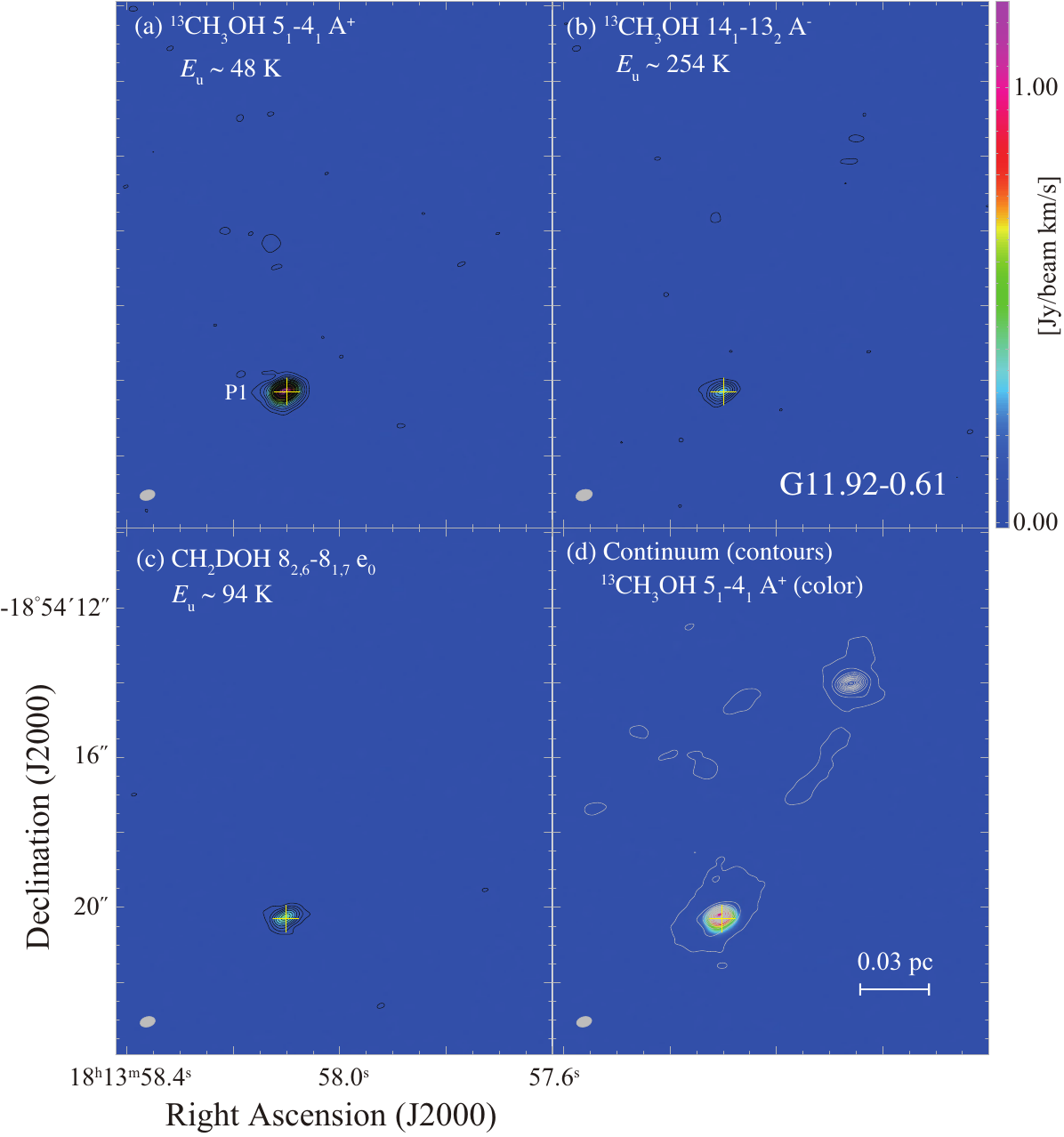}
\caption{Integrated intensity maps of $^{13}$CH$_3$OH $5_1-4_1$ A$^+$ (a), $^{13}$CH$_3$OH $14_1-13_2$ A$^-$ (b), and CH$_2$DOH $8_{2,6}-8_{1,7}$ $e_0$ (c) toward G11.92-00.61. The integrated velocity range is from 30 km s$^{-1}$ to 40 km s$^{-1}$. Contour levels start from 3$\sigma$ and increase in steps of 1$\sigma$ for (a), (b), and (c) [(a) $1\sigma=18$ mJy beam$^{-1}$ km s$^{-1}$, (b) $1\sigma=15$ mJy beam$^{-1}$ km s$^{-1}$, (c) $1\sigma=18$ mJy beam$^{-1}$ km s$^{-1}$]. (d) the continuum image (contours) overlaid on $^{13}$CH$_3$OH $14_1-13_2$ A$^-$ integrated intensity map (color) . For (d), contour levels start from 3$\sigma$ and increase in steps of 10$\sigma$ [$1\sigma=0.4$ mJy beam$^{-1}$]. Cross marks indicate the positions of the $^{13}$CH$_3$OH peak. The same color scale is used in (a)-(d). The synthesized beam is shown at the bottom left of each panel.}
\end{figure*}

\clearpage

\begin{figure*}
\figurenum{1}
\epsscale{1.0}
\plotone{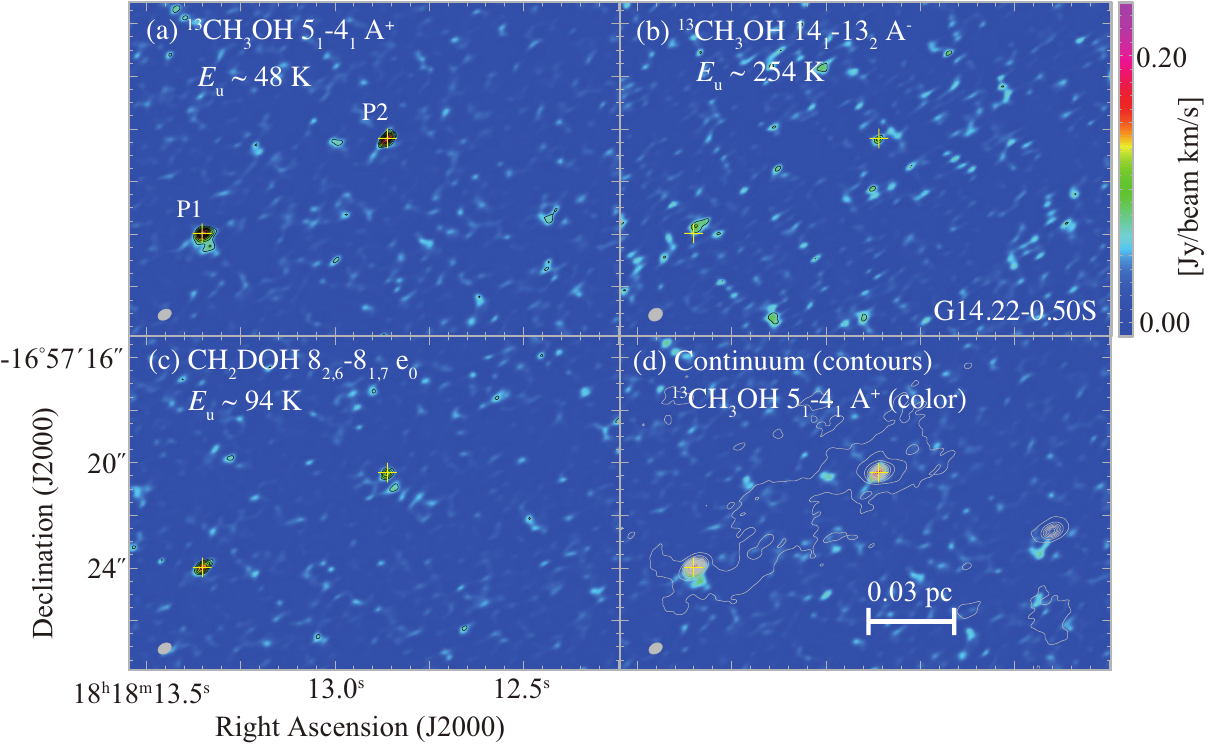}
\caption{Integrated intensity maps of $^{13}$CH$_3$OH $5_1-4_1$ A$^+$ (a), $^{13}$CH$_3$OH $14_1-13_2$ A$^-$ (b), and CH$_2$DOH $8_{2,6}-8_{1,7}$ $e_0$ (c) toward G14.22-00.50S. The integrated velocity range is from 15 km s$^{-1}$ to 25 km s$^{-1}$. Contour levels start from 3$\sigma$ and increase in steps of 1$\sigma$ for (a), (b), and (c) [(a) $1\sigma=20$ mJy beam$^{-1}$ km s$^{-1}$, (b) $1\sigma=23$ mJy beam$^{-1}$ km s$^{-1}$, (c) $1\sigma=20$ mJy beam$^{-1}$ km s$^{-1}$]. (d) the continuum image (contours) overlaid on $^{13}$CH$_3$OH $14_1-13_2$ A$^-$ integrated intensity map (color) . For (d), contour levels start from 3$\sigma$ and increase in steps of 10$\sigma$ [$1\sigma=0.2$ mJy beam$^{-1}$]. Cross marks indicate the positions of the $^{13}$CH$_3$OH peak. The same color scale is used in (a)-(d). The synthesized beam is shown at the bottom left of each panel.}
\end{figure*}

\clearpage

\begin{figure*}
\figurenum{1}
\epsscale{1.0}
\plotone{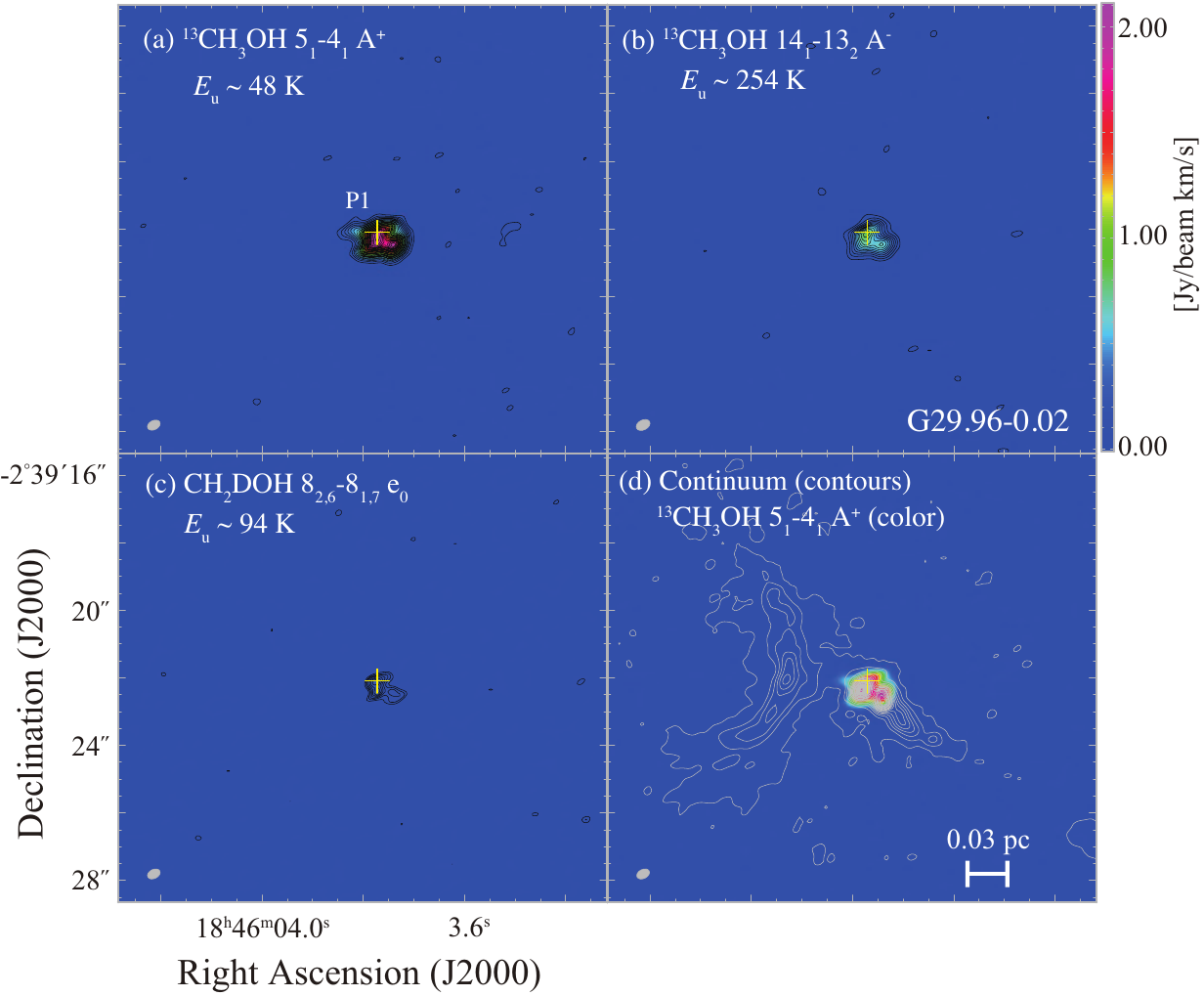}
\caption{Integrated intensity maps of $^{13}$CH$_3$OH $5_1-4_1$ A$^+$ (a), $^{13}$CH$_3$OH $14_1-13_2$ A$^-$ (b), and CH$_2$DOH $8_{2,6}-8_{1,7}$ $e_0$ (c) toward G29.96-00.02. The integrated velocity range is from 92.5 km s$^{-1}$ to 102.5 km s$^{-1}$. Contour levels start from 3$\sigma$ and increase in steps of 1$\sigma$ for (a), (b), and (c) [(a) $1\sigma=27$ mJy beam$^{-1}$ km s$^{-1}$, (b) $1\sigma=24$ mJy beam$^{-1}$ km s$^{-1}$, (c) $1\sigma=27$ mJy beam$^{-1}$ km s$^{-1}$]. (d) the continuum image (contours) overlaid on $^{13}$CH$_3$OH $14_1-13_2$ A$^-$ integrated intensity map (color) . For (d), contour levels start from 3$\sigma$ and increase in steps of 10$\sigma$ [$1\sigma=0.4$ mJy beam$^{-1}$]. Cross marks indicate the positions of the $^{13}$CH$_3$OH peak. The same color scale is used in (a)-(d). The synthesized beam is shown at the bottom left of each panel.}
\end{figure*}

\clearpage

\begin{figure*}
\figurenum{1}
\epsscale{1.0}
\plotone{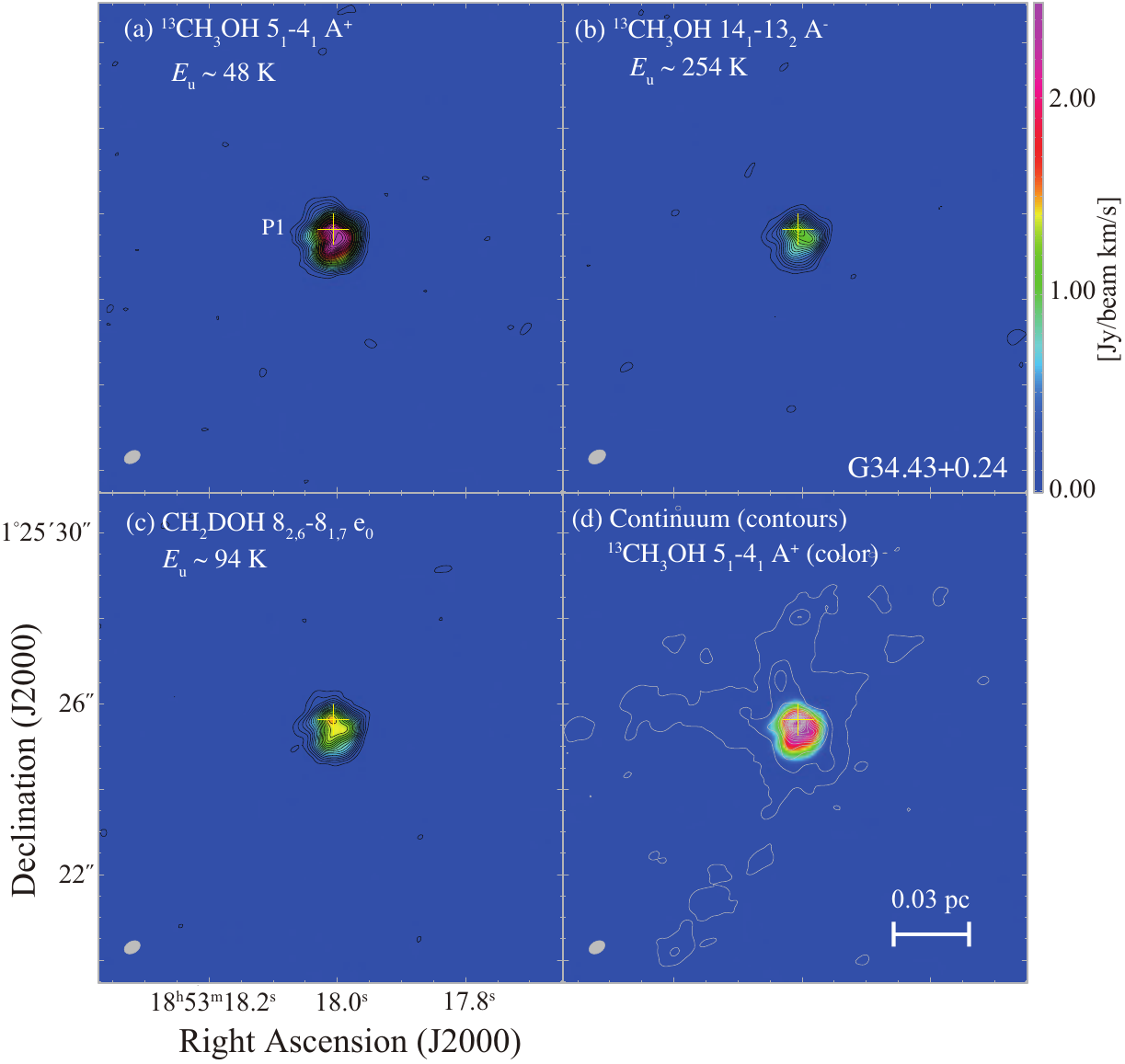}
\caption{Integrated intensity maps of $^{13}$CH$_3$OH $5_1-4_1$ A$^+$ (a), $^{13}$CH$_3$OH $14_1-13_2$ A$^-$ (b), and CH$_2$DOH $8_{2,6}-8_{1,7}$ $e_0$ (c) toward G34.43+00.24. The integrated velocity range is from 52.5 km s$^{-1}$ to 62.5 km s$^{-1}$. Contour levels start from 3$\sigma$ and increase in steps of 1$\sigma$ for (a), (b), and (c) [(a) $1\sigma=28$ mJy beam$^{-1}$ km s$^{-1}$, (b) $1\sigma=24$ mJy beam$^{-1}$ km s$^{-1}$, (c) $1\sigma=28$ mJy beam$^{-1}$ km s$^{-1}$]. (d) the continuum image (contours) overlaid on $^{13}$CH$_3$OH $14_1-13_2$ A$^-$ integrated intensity map (color) . For (d), contour levels start from 3$\sigma$ and increase in steps of 20$\sigma$ [$1\sigma=0.5$ mJy beam$^{-1}$]. Cross marks indicate the positions of the $^{13}$CH$_3$OH peak. The same color scale is used in (a)-(d). The synthesized beam is shown at the bottom left of each panel.}
\end{figure*}

\clearpage

\begin{figure*}
\figurenum{1}
\epsscale{1.0}
\plotone{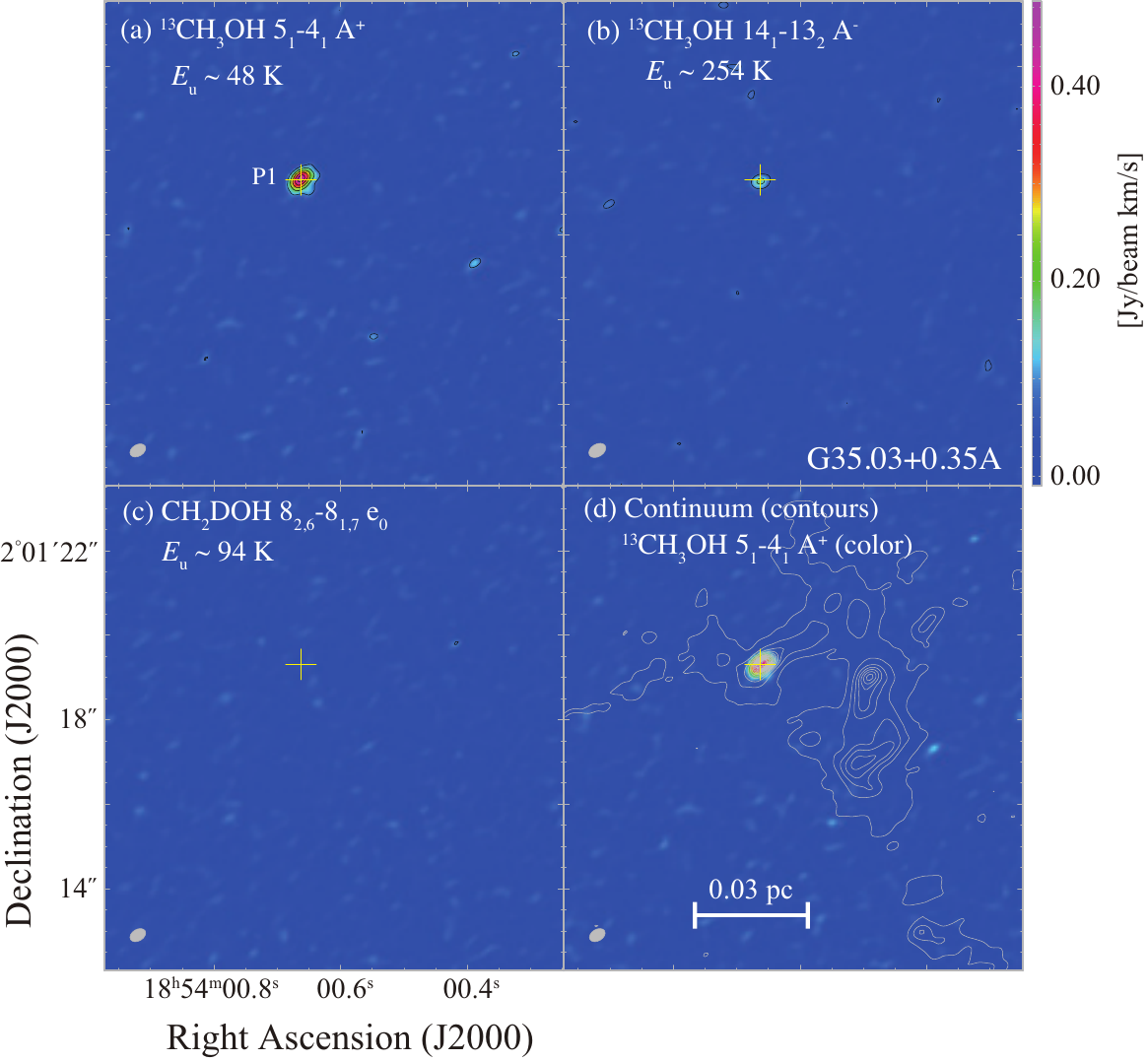}
\caption{Integrated intensity maps of $^{13}$CH$_3$OH $5_1-4_1$ A$^+$ (a), $^{13}$CH$_3$OH $14_1-13_2$ A$^-$ (b), and CH$_2$DOH $8_{2,6}-8_{1,7}$ $e_0$ (c) toward G35.03+00.35A. The integrated velocity range is from 40 km s$^{-1}$ to 50 km s$^{-1}$. Contour levels start from 3$\sigma$ and increase in steps of 1$\sigma$ for (a), (b), and (c) [(a) $1\sigma=29$ mJy beam$^{-1}$ km s$^{-1}$, (b) $1\sigma=24$ mJy beam$^{-1}$ km s$^{-1}$, (c) $1\sigma=29$ mJy beam$^{-1}$ km s$^{-1}$]. (d) the continuum image (contours) overlaid on $^{13}$CH$_3$OH $14_1-13_2$ A$^-$ integrated intensity map (color) . For (d), contour levels start from 3$\sigma$ and increase in steps of 10$\sigma$ [$1\sigma=0.2$ mJy beam$^{-1}$]. Cross marks indicate the positions of the $^{13}$CH$_3$OH peak. The same color scale is used in (a)-(d). The synthesized beam is shown at the bottom left of each panel.}
\end{figure*}

\clearpage

\begin{figure*}
\figurenum{1}
\epsscale{1.0}
\plotone{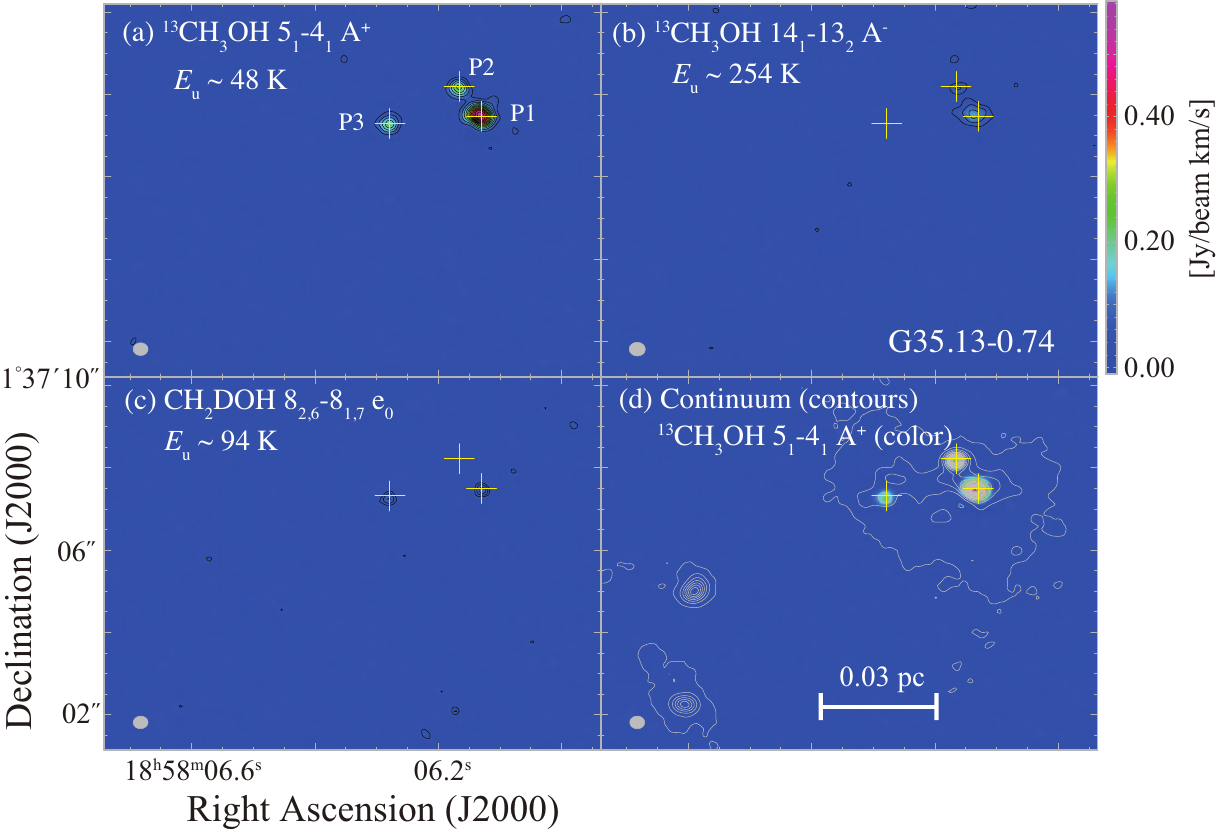}
\caption{Integrated intensity maps of $^{13}$CH$_3$OH $5_1-4_1$ A$^+$ (a), $^{13}$CH$_3$OH $14_1-13_2$ A$^-$ (b), and CH$_2$DOH $8_{2,6}-8_{1,7}$ $e_0$ (c) toward G35.13-00.74. The integrated velocity range is from 30 km s$^{-1}$ to 40 km s$^{-1}$. Contour levels start from 3$\sigma$ and increase in steps of 1$\sigma$ for (a), (b), and (c) [(a) $1\sigma=15$ mJy beam$^{-1}$ km s$^{-1}$, (b) $1\sigma=13$ mJy beam$^{-1}$ km s$^{-1}$, (c) $1\sigma=15$ mJy beam$^{-1}$ km s$^{-1}$]. (d) the continuum image (contours) overlaid on $^{13}$CH$_3$OH $14_1-13_2$ A$^-$ integrated intensity map (color) . For (d), contour levels start from 3$\sigma$ and increase in steps of 10$\sigma$ [$1\sigma=0.2$ mJy beam$^{-1}$]. Cross marks indicate the positions of the $^{13}$CH$_3$OH peak. The same color scale is used in (a)-(d). The synthesized beam is shown at the bottom left of each panel.}
\end{figure*}

\clearpage

\begin{figure*}
\figurenum{1}
\epsscale{1.0}
\plotone{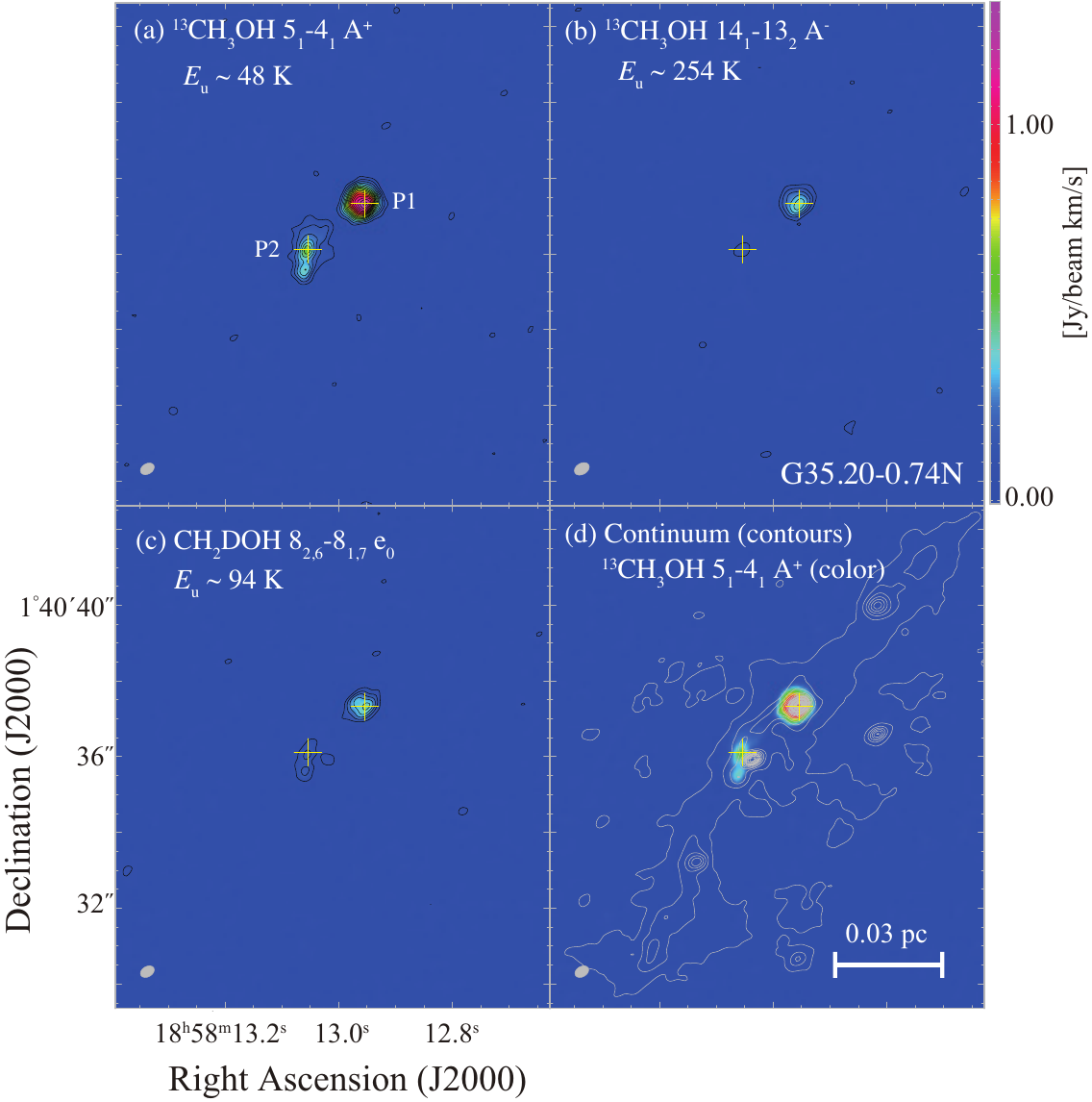}
\caption{Integrated intensity maps of $^{13}$CH$_3$OH $5_1-4_1$ A$^+$ (a), $^{13}$CH$_3$OH $14_1-13_2$ A$^-$ (b), and CH$_2$DOH $8_{2,6}-8_{1,7}$ $e_0$ (c) toward G35.20-00.74N. The integrated velocity range is from 25 km s$^{-1}$ to 35 km s$^{-1}$. Contour levels start from 3$\sigma$ and increase in steps of 1$\sigma$ for (a), (b), and (c) [(a) $1\sigma=27$ mJy beam$^{-1}$ km s$^{-1}$, (b) $1\sigma=25$ mJy beam$^{-1}$ km s$^{-1}$, (c) $1\sigma=27$ mJy beam$^{-1}$ km s$^{-1}$]. (d) the continuum image (contours) overlaid on $^{13}$CH$_3$OH $14_1-13_2$ A$^-$ integrated intensity map (color) . For (d), contour levels start from 3$\sigma$ and increase in steps of 10$\sigma$ [$1\sigma=0.3$ mJy beam$^{-1}$]. Cross marks indicate the positions of the $^{13}$CH$_3$OH peak. The same color scale is used in (a)-(d). The synthesized beam is shown at the bottom left of each panel.}
\end{figure*}

\clearpage

\begin{figure*}
\figurenum{1}
\epsscale{1.0}
\plotone{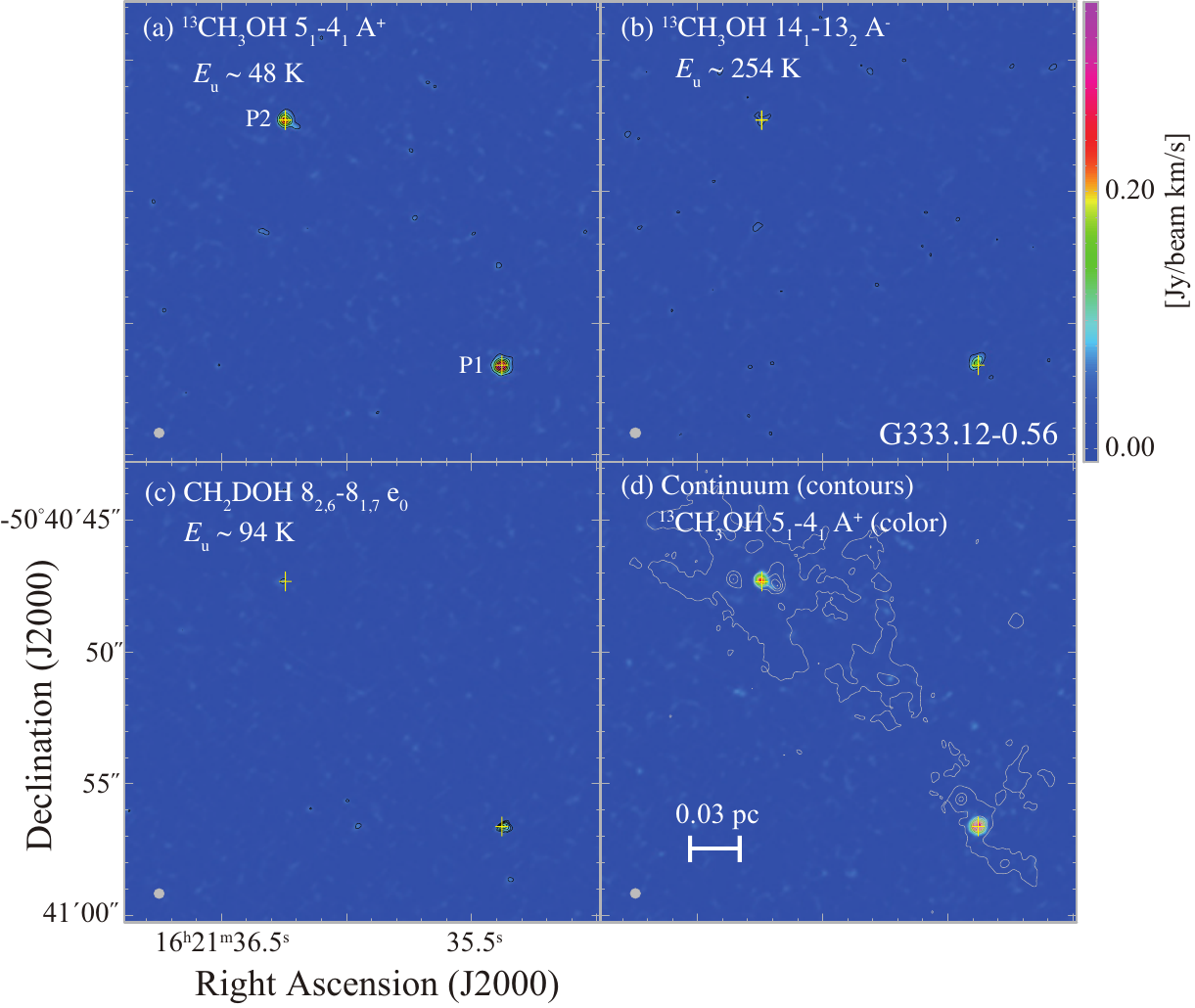}
\caption{Integrated intensity maps of $^{13}$CH$_3$OH $5_1-4_1$ A$^+$ (a), $^{13}$CH$_3$OH $14_1-13_2$ A$^-$ (b), and CH$_2$DOH $8_{2,6}-8_{1,7}$ $e_0$ (c) toward G333.12-00.56. The integrated velocity range is from -62.5 km s$^{-1}$ to -52.5 km s$^{-1}$. Contour levels start from 3$\sigma$ and increase in steps of 1$\sigma$ for (a), (b), and (c) [(a) $1\sigma=18$ mJy beam$^{-1}$ km s$^{-1}$, (b) $1\sigma=15$ mJy beam$^{-1}$ km s$^{-1}$, (c) $1\sigma=18$ mJy beam$^{-1}$ km s$^{-1}$]. (d) the continuum image (contours) overlaid on $^{13}$CH$_3$OH $14_1-13_2$ A$^-$ integrated intensity map (color) . For (d), contour levels start from 3$\sigma$ and increase in steps of 20$\sigma$ [$1\sigma=0.2$ mJy beam$^{-1}$]. Cross marks indicate the positions of the $^{13}$CH$_3$OH peak. The same color scale is used in (a)-(d). The synthesized beam is shown at the bottom left of each panel.}
\end{figure*}

\clearpage

\begin{figure*}
\figurenum{1}
\epsscale{1.0}
\plotone{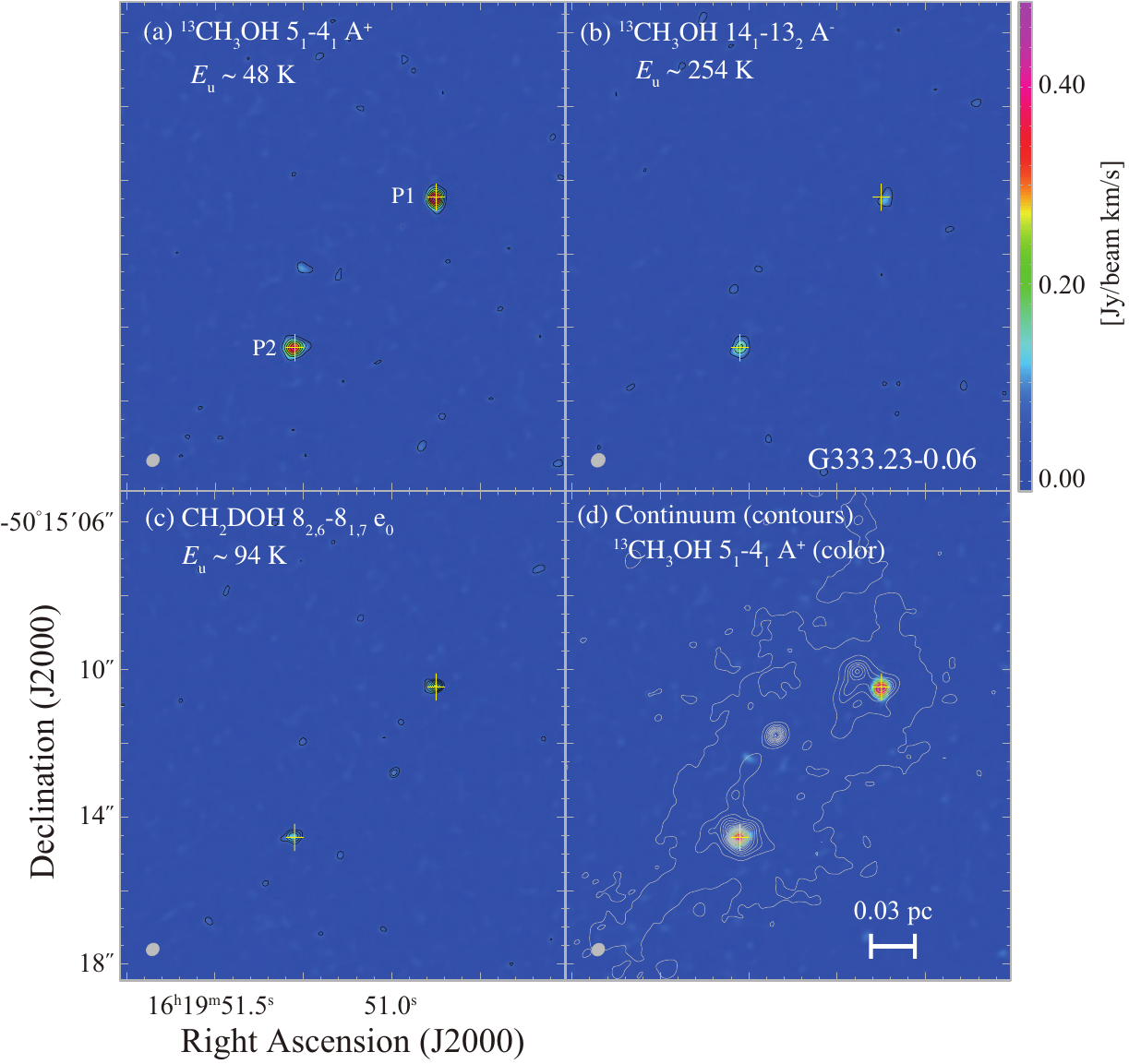}
\caption{Integrated intensity maps of $^{13}$CH$_3$OH $5_1-4_1$ A$^+$ (a), $^{13}$CH$_3$OH $14_1-13_2$ A$^-$ (b), and CH$_2$DOH $8_{2,6}-8_{1,7}$ $e_0$ (c) toward G333.23-00.06. The integrated velocity range is from -90 km s$^{-1}$ to -80 km s$^{-1}$. Contour levels start from 3$\sigma$ and increase in steps of 1$\sigma$ for (a), (b), and (c) [(a) $1\sigma=21$ mJy beam$^{-1}$ km s$^{-1}$, (b) $1\sigma=20$ mJy beam$^{-1}$ km s$^{-1}$, (c) $1\sigma=21$ mJy beam$^{-1}$ km s$^{-1}$]. (d) the continuum image (contours) overlaid on $^{13}$CH$_3$OH $14_1-13_2$ A$^-$ integrated intensity map (color) . For (d), contour levels start from 3$\sigma$ and increase in steps of 20$\sigma$ [$1\sigma=0.2$ mJy beam$^{-1}$]. Cross marks indicate the positions of the $^{13}$CH$_3$OH peak. The same color scale is used in (a)-(d). The synthesized beam is shown at the bottom left of each panel.}
\end{figure*}

\clearpage

\begin{figure*}
\figurenum{1}
\epsscale{1.0}
\plotone{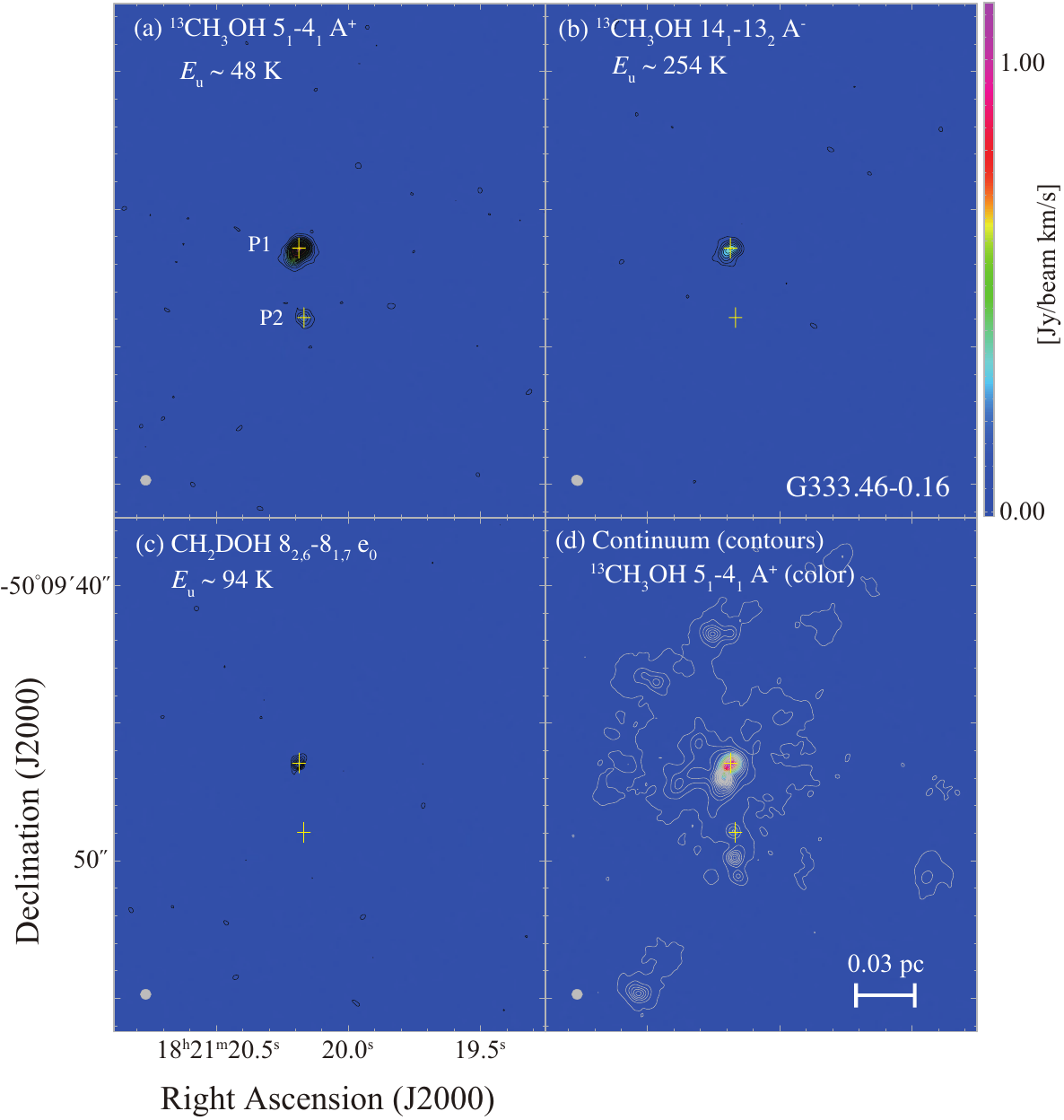}
\caption{Integrated intensity maps of $^{13}$CH$_3$OH $5_1-4_1$ A$^+$ (a), $^{13}$CH$_3$OH $14_1-13_2$ A$^-$ (b), and CH$_2$DOH $8_{2,6}-8_{1,7}$ $e_0$ (c) toward G333.46-00.16. The integrated velocity range is from -47.5 km s$^{-1}$ to -37.5 km s$^{-1}$. Contour levels start from 3$\sigma$ and increase in steps of 1$\sigma$ for (a), (b), and (c) [(a) $1\sigma=16$ mJy beam$^{-1}$ km s$^{-1}$, (b) $1\sigma=16$ mJy beam$^{-1}$ km s$^{-1}$, (c) $1\sigma=16$ mJy beam$^{-1}$ km s$^{-1}$]. (d) the continuum image (contours) overlaid on $^{13}$CH$_3$OH $14_1-13_2$ A$^-$ integrated intensity map (color) . For (d), contour levels start from 3$\sigma$ and increase in steps of 10$\sigma$ [$1\sigma=0.2$ mJy beam$^{-1}$]. Cross marks indicate the positions of the $^{13}$CH$_3$OH peak. The same color scale is used in (a)-(d). The synthesized beam is shown at the bottom left of each panel.}
\end{figure*}

\clearpage

\begin{figure*}
\figurenum{1}
\epsscale{1.0}
\plotone{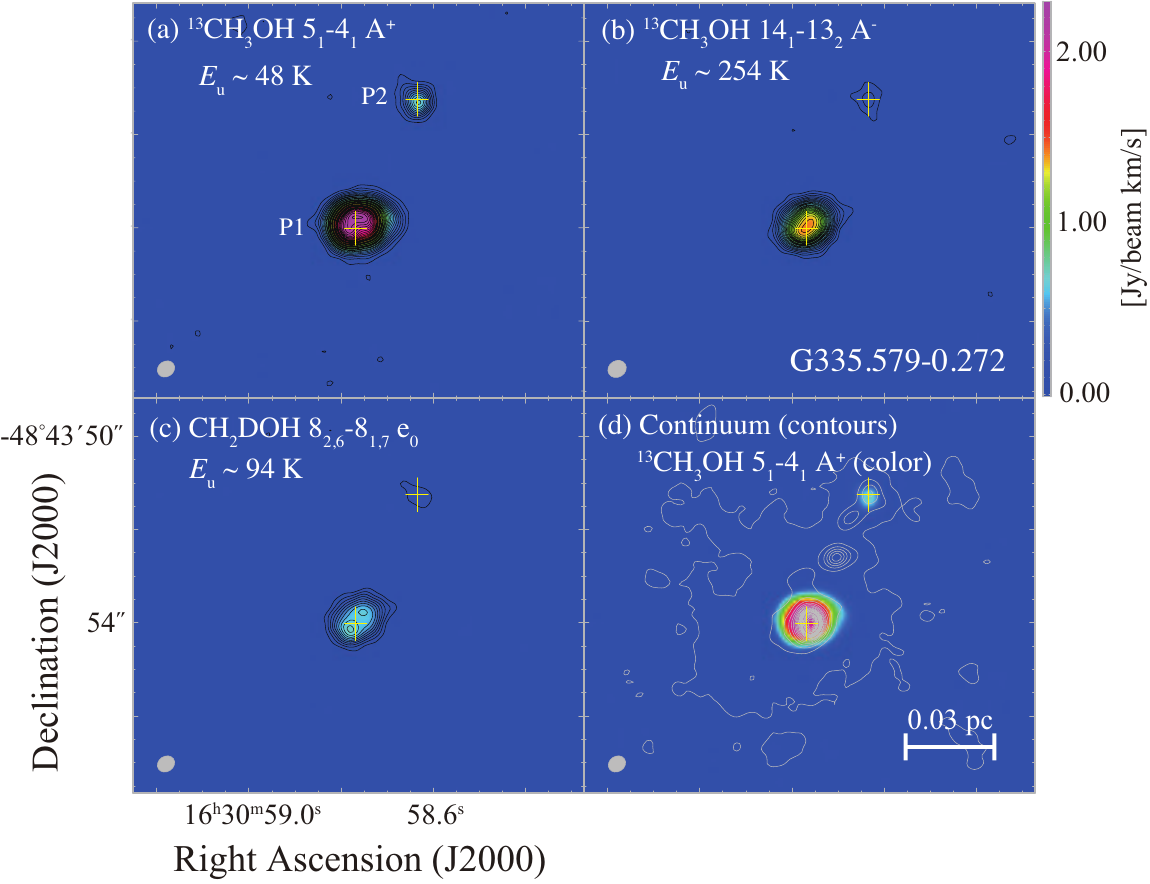}
\caption{Integrated intensity maps of $^{13}$CH$_3$OH $5_1-4_1$ A$^+$ (a), $^{13}$CH$_3$OH $14_1-13_2$ A$^-$ (b), and CH$_2$DOH $8_{2,6}-8_{1,7}$ $e_0$ (c) toward G335.579-00.272. The integrated velocity range is from -52.5 km s$^{-1}$ to -42.5 km s$^{-1}$. Contour levels start from 3$\sigma$ and increase in steps of 1$\sigma$ for (a), (b), and (c) [(a) $1\sigma=23$ mJy beam$^{-1}$ km s$^{-1}$, (b) $1\sigma=21$ mJy beam$^{-1}$ km s$^{-1}$, (c) $1\sigma=23$ mJy beam$^{-1}$ km s$^{-1}$]. (d) the continuum image (contours) overlaid on $^{13}$CH$_3$OH $14_1-13_2$ A$^-$ integrated intensity map (color) . For (d), contour levels start from 3$\sigma$ and increase in steps of 20$\sigma$ [$1\sigma=0.5$ mJy beam$^{-1}$]. Cross marks indicate the positions of the $^{13}$CH$_3$OH peak. The same color scale is used in (a)-(d). The synthesized beam is shown at the bottom left of each panel.}
\end{figure*}

\clearpage

\begin{figure*}
\figurenum{1}
\epsscale{1.0}
\plotone{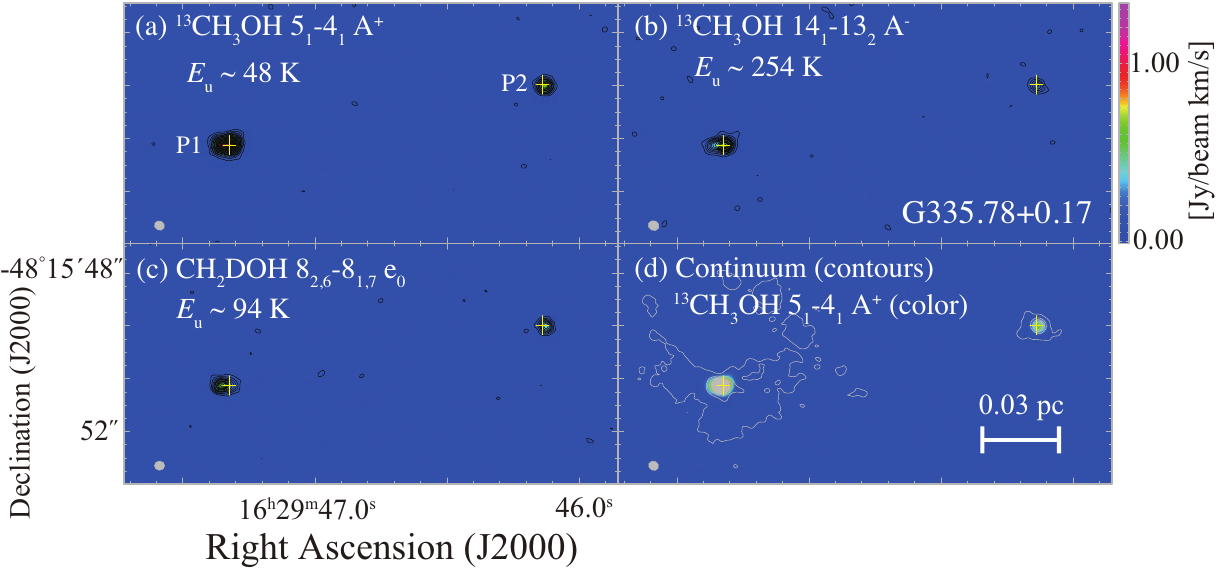}
\caption{Integrated intensity maps of $^{13}$CH$_3$OH $5_1-4_1$ A$^+$ (a), $^{13}$CH$_3$OH $14_1-13_2$ A$^-$ (b), and CH$_2$DOH $8_{2,6}-8_{1,7}$ $e_0$ (c) toward G335.78+00.17. The integrated velocity range is from -55 km s$^{-1}$ to -45 km s$^{-1}$. Contour levels start from 3$\sigma$ and increase in steps of 1$\sigma$ for (a), (b), and (c) [(a) $1\sigma=17$ mJy beam$^{-1}$ km s$^{-1}$, (b) $1\sigma=15$ mJy beam$^{-1}$ km s$^{-1}$, (c) $1\sigma=17$ mJy beam$^{-1}$ km s$^{-1}$]. (d) the continuum image (contours) overlaid on $^{13}$CH$_3$OH $14_1-13_2$ A$^-$ integrated intensity map (color) . For (d), contour levels start from 3$\sigma$ and increase in steps of 10$\sigma$ [$1\sigma=0.5$ mJy beam$^{-1}$]. Cross marks indicate the positions of the $^{13}$CH$_3$OH peak. The same color scale is used in (a)-(d). The synthesized beam is shown at the bottom left of each panel.}
\end{figure*}

\clearpage

\begin{figure*}
\figurenum{1}
\epsscale{1.0}
\plotone{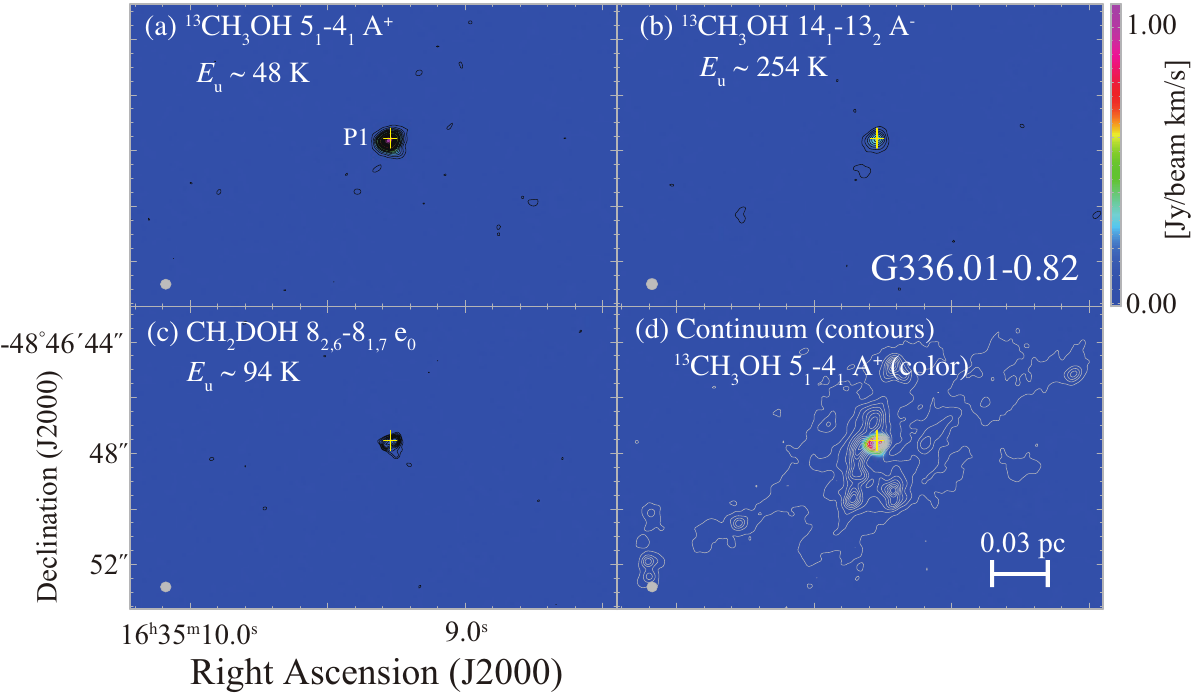}
\caption{Integrated intensity maps of $^{13}$CH$_3$OH $5_1-4_1$ A$^+$ (a), $^{13}$CH$_3$OH $14_1-13_2$ A$^-$ (b), and CH$_2$DOH $8_{2,6}-8_{1,7}$ $e_0$ (c) toward G336.01-00.82. The integrated velocity range is from -52.5 km s$^{-1}$ to -42.5 km s$^{-1}$. Contour levels start from 3$\sigma$ and increase in steps of 1$\sigma$ for (a), (b), and (c) [(a) $1\sigma=18$ mJy beam$^{-1}$ km s$^{-1}$, (b) $1\sigma=17$ mJy beam$^{-1}$ km s$^{-1}$, (c) $1\sigma=18$ mJy beam$^{-1}$ km s$^{-1}$]. (d) the continuum image (contours) overlaid on $^{13}$CH$_3$OH $14_1-13_2$ A$^-$ integrated intensity map (color) . For (d), contour levels start from 3$\sigma$ and increase in steps of 10$\sigma$ [$1\sigma=0.2$ mJy beam$^{-1}$]. Cross marks indicate the positions of the $^{13}$CH$_3$OH peak. The same color scale is used in (a)-(d). The synthesized beam is shown at the bottom left of each panel.}
\end{figure*}

\clearpage

\begin{figure*}
\figurenum{1}
\epsscale{1.0}
\plotone{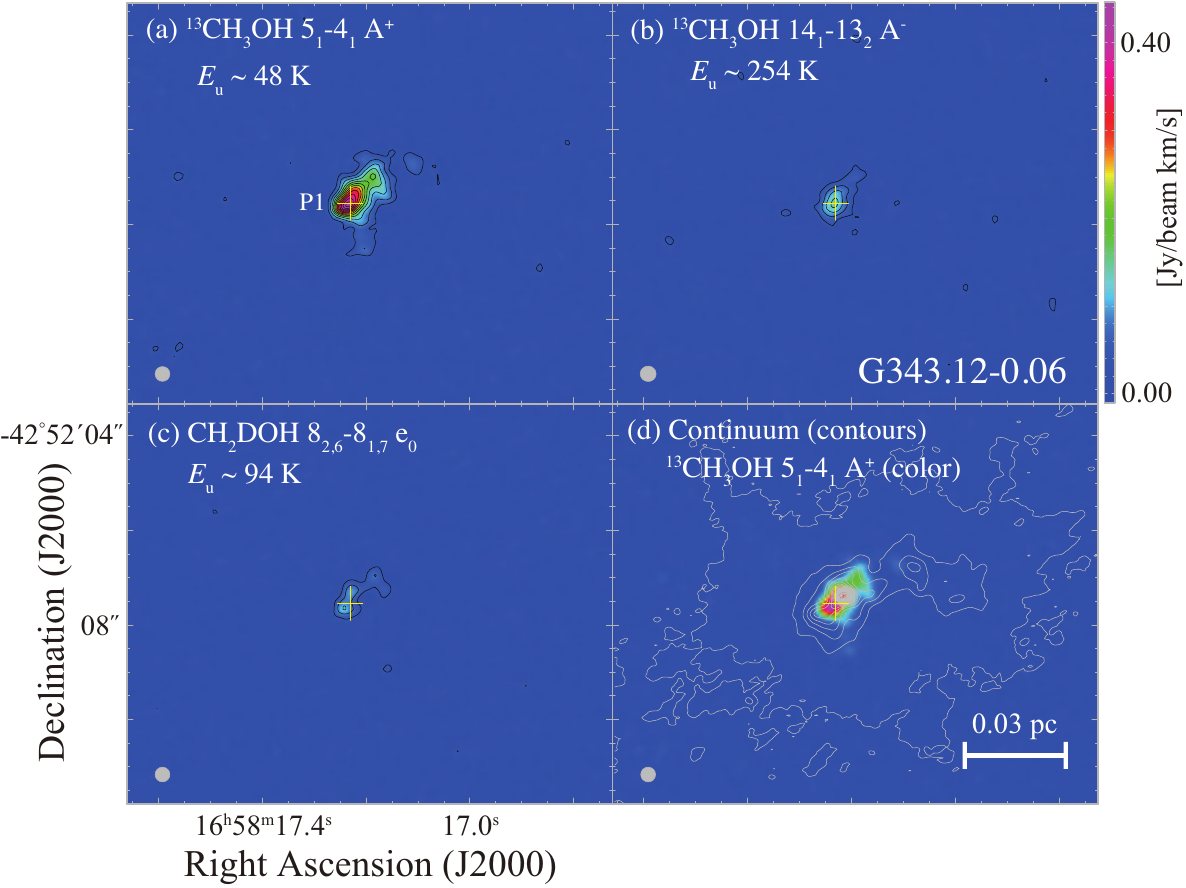}
\caption{Integrated intensity maps of $^{13}$CH$_3$OH $5_1-4_1$ A$^+$ (a), $^{13}$CH$_3$OH $14_1-13_2$ A$^-$ (b), and CH$_2$DOH $8_{2,6}-8_{1,7}$ $e_0$ (c) toward G343.12-0.06. The integrated velocity range is from -40 km s$^{-1}$ to -30 km s$^{-1}$. Contour levels start from 3$\sigma$ and increase in steps of 1$\sigma$ for (a), (b), and (c) [(a) $1\sigma=12$ mJy beam$^{-1}$ km s$^{-1}$, (b) $1\sigma=11$ mJy beam$^{-1}$ km s$^{-1}$, (c) $1\sigma=12$ mJy beam$^{-1}$ km s$^{-1}$]. (d) the continuum image (contours) overlaid on $^{13}$CH$_3$OH $14_1-13_2$ A$^-$ integrated intensity map (color) . For (d), contour levels start from 3$\sigma$ and increase in steps of 10$\sigma$ [$1\sigma=0.2$ mJy beam$^{-1}$]. Cross marks indicate the positions of the $^{13}$CH$_3$OH peak. The same color scale is used in (a)-(d). The synthesized beam is shown at the bottom left of each panel.}
\end{figure*}

\clearpage

\begin{figure*}
\figurenum{1}
\epsscale{1.0}
\plotone{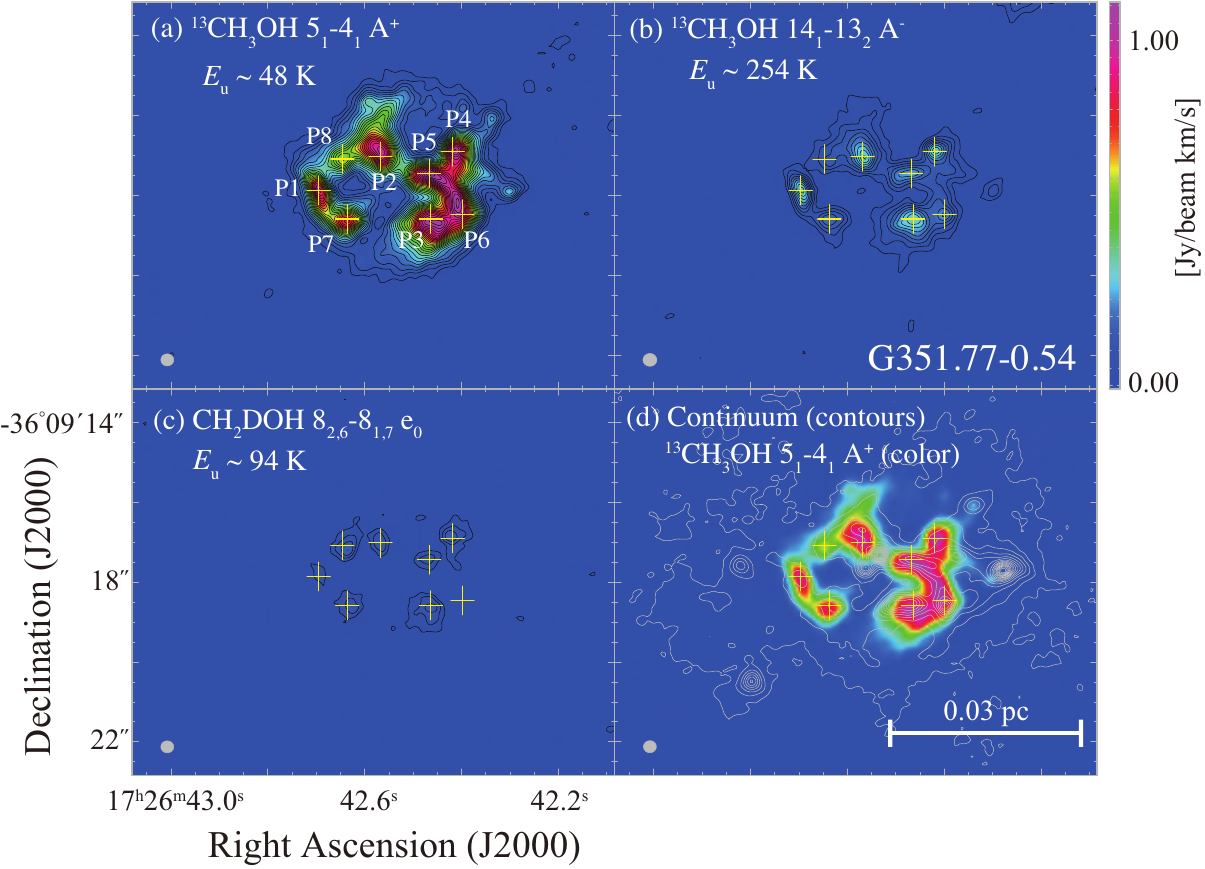}
\caption{Integrated intensity maps of $^{13}$CH$_3$OH $5_1-4_1$ A$^+$ (a), $^{13}$CH$_3$OH $14_1-13_2$ A$^-$ (b), and CH$_2$DOH $8_{2,6}-8_{1,7}$ $e_0$ (c) toward G351.77-00.54. The integrated velocity range is from -7.5 km s$^{-1}$ to 2.5 km s$^{-1}$. Contour levels start from 3$\sigma$ and increase in steps of 1$\sigma$ for (a), (b), and (c) [(a) $1\sigma=14$ mJy beam$^{-1}$ km s$^{-1}$, (b) $1\sigma=13$ mJy beam$^{-1}$ km s$^{-1}$, (c) $1\sigma=14$ mJy beam$^{-1}$ km s$^{-1}$]. (d) the continuum image (contours) overlaid on $^{13}$CH$_3$OH $14_1-13_2$ A$^-$ integrated intensity map (color) . For (d), contour levels start from 3$\sigma$ and increase in steps of 10$\sigma$ [$1\sigma=0.9$ mJy beam$^{-1}$]. Cross marks indicate the positions of the $^{13}$CH$_3$OH peak. The same color scale is used in (a)-(d). The synthesized beam is shown at the bottom left of each panel.}
\end{figure*}

\clearpage

\begin{figure*}
\figurenum{1}
\epsscale{1.0}
\plotone{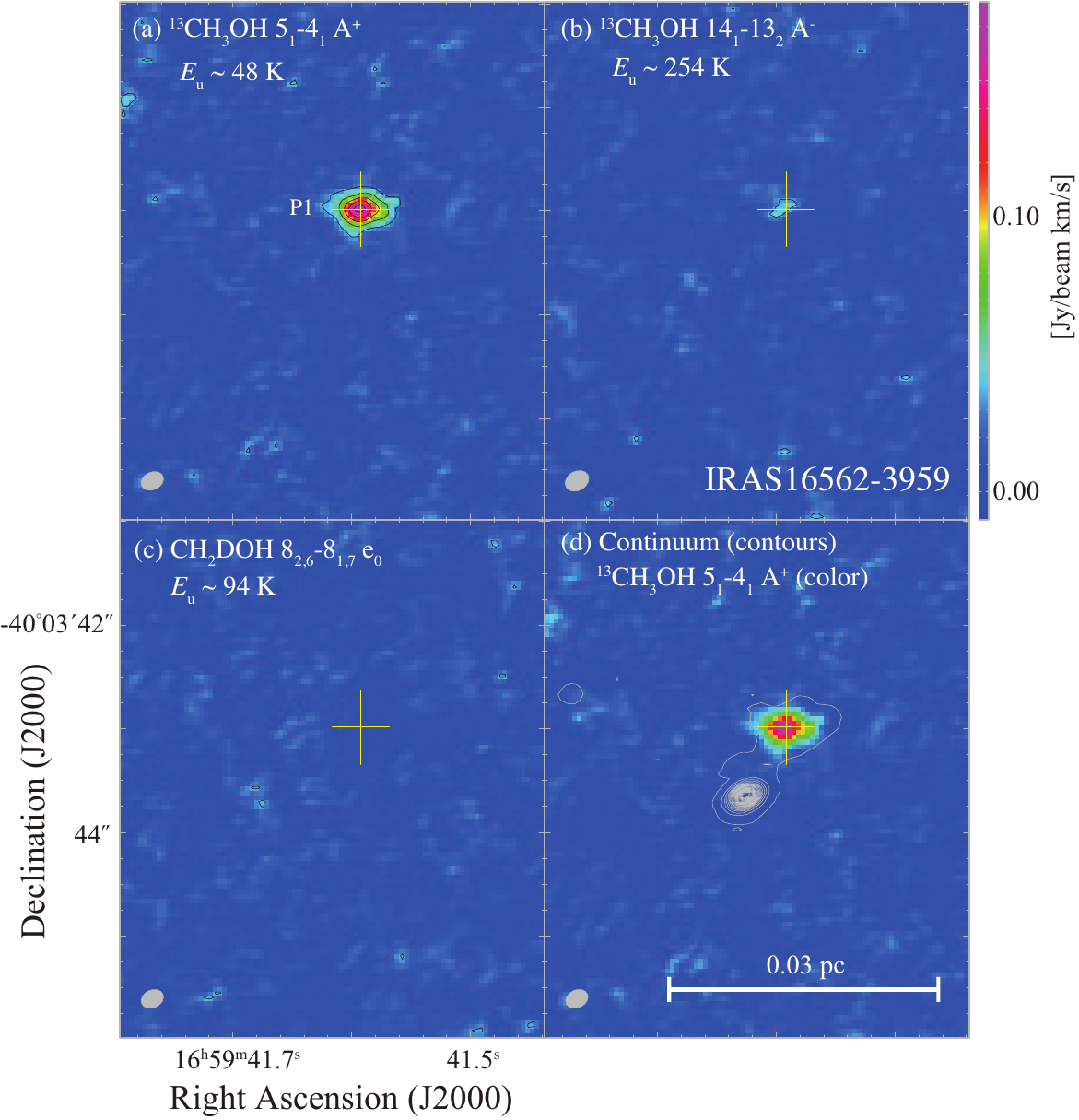}
\caption{Integrated intensity maps of $^{13}$CH$_3$OH $5_1-4_1$ A$^+$ (a), $^{13}$CH$_3$OH $14_1-13_2$ A$^-$ (b), and CH$_2$DOH $8_{2,6}-8_{1,7}$ $e_0$ (c) toward IRAS 16562-3959. The integrated velocity range is from -20 km s$^{-1}$ to -10 km s$^{-1}$. Contour levels start from 3$\sigma$ and increase in steps of 1$\sigma$ for (a), (b), and (c) [(a) $1\sigma=11$ mJy beam$^{-1}$ km s$^{-1}$, (b) $1\sigma=11$ mJy beam$^{-1}$ km s$^{-1}$, (c) $1\sigma=11$ mJy beam$^{-1}$ km s$^{-1}$]. (d) the continuum image (contours) overlaid on $^{13}$CH$_3$OH $14_1-13_2$ A$^-$ integrated intensity map (color) . For (d), contour levels start from 3$\sigma$ and increase in steps of 20$\sigma$ [$1\sigma=0.4$ mJy beam$^{-1}$]. Cross marks indicate the positions of the $^{13}$CH$_3$OH peak. The same color scale is used in (a)-(d). The synthesized beam is shown at the bottom left of each panel.}
\end{figure*}

\clearpage

\begin{figure*}
\figurenum{1}
\epsscale{1.0}
\plotone{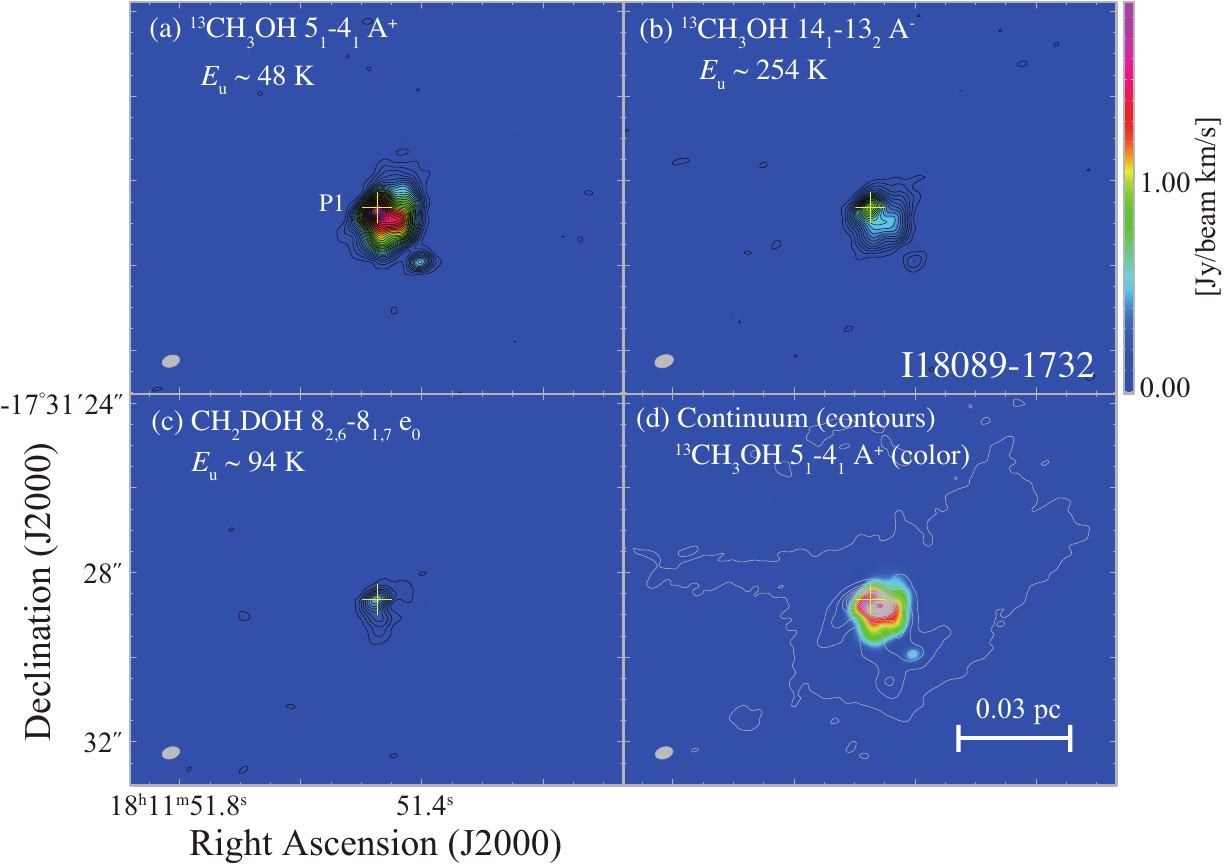}
\caption{Integrated intensity maps of $^{13}$CH$_3$OH $5_1-4_1$ A$^+$ (a), $^{13}$CH$_3$OH $14_1-13_2$ A$^-$ (b), and CH$_2$DOH $8_{2,6}-8_{1,7}$ $e_0$ (c) toward IRAS18089-1732. The integrated velocity range is from 27.5 km s$^{-1}$ to 37.5 km s$^{-1}$. Contour levels start from 3$\sigma$ and increase in steps of 1$\sigma$ for (a), (b), and (c) [(a) $1\sigma=17$ mJy beam$^{-1}$ km s$^{-1}$, (b) $1\sigma=15$ mJy beam$^{-1}$ km s$^{-1}$, (c) $1\sigma=17$ mJy beam$^{-1}$ km s$^{-1}$]. (d) the continuum image (contours) overlaid on $^{13}$CH$_3$OH $14_1-13_2$ A$^-$ integrated intensity map (color) . For (d), contour levels start from 3$\sigma$ and increase in steps of 20$\sigma$ [$1\sigma=0.3$ mJy beam$^{-1}$]. Cross marks indicate the positions of the $^{13}$CH$_3$OH peak. The same color scale is used in (a)-(d). The synthesized beam is shown at the bottom left of each panel.}
\end{figure*}

\clearpage

\begin{figure*}
\figurenum{1}
\epsscale{1.0}
\plotone{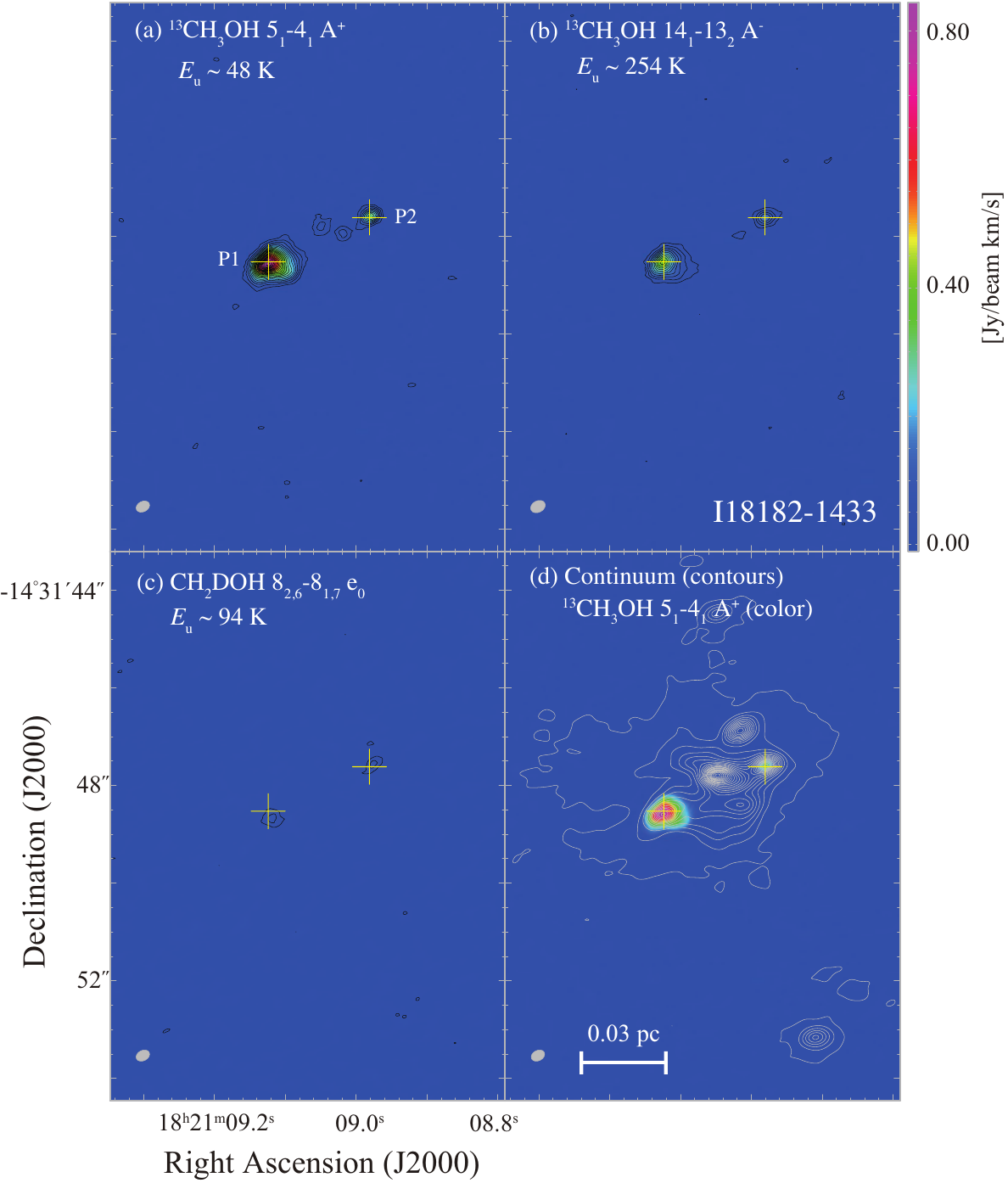}
\caption{Integrated intensity maps of $^{13}$CH$_3$OH $5_1-4_1$ A$^+$ (a), $^{13}$CH$_3$OH $14_1-13_2$ A$^-$ (b), and CH$_2$DOH $8_{2,6}-8_{1,7}$ $e_0$ (c) toward IRAS18182-1433. The integrated velocity range is from 57.5 km s$^{-1}$ to 67.5 km s$^{-1}$. Contour levels start from 3$\sigma$ and increase in steps of 1$\sigma$ for (a), (b), and (c) [(a) $1\sigma=12$ mJy beam$^{-1}$ km s$^{-1}$, (b) $1\sigma=11$ mJy beam$^{-1}$ km s$^{-1}$, (c) $1\sigma=12$ mJy beam$^{-1}$ km s$^{-1}$]. (d) the continuum image (contours) overlaid on $^{13}$CH$_3$OH $14_1-13_2$ A$^-$ integrated intensity map (color) . For (d), contour levels start from 3$\sigma$ and increase in steps of 10$\sigma$ [$1\sigma=0.2$ mJy beam$^{-1}$]. Cross marks indicate the positions of the $^{13}$CH$_3$OH peak. The same color scale is used in (a)-(d). The synthesized beam is shown at the bottom left of each panel.}
\end{figure*}

\clearpage

\begin{figure*}
\figurenum{1}
\epsscale{1.0}
\plotone{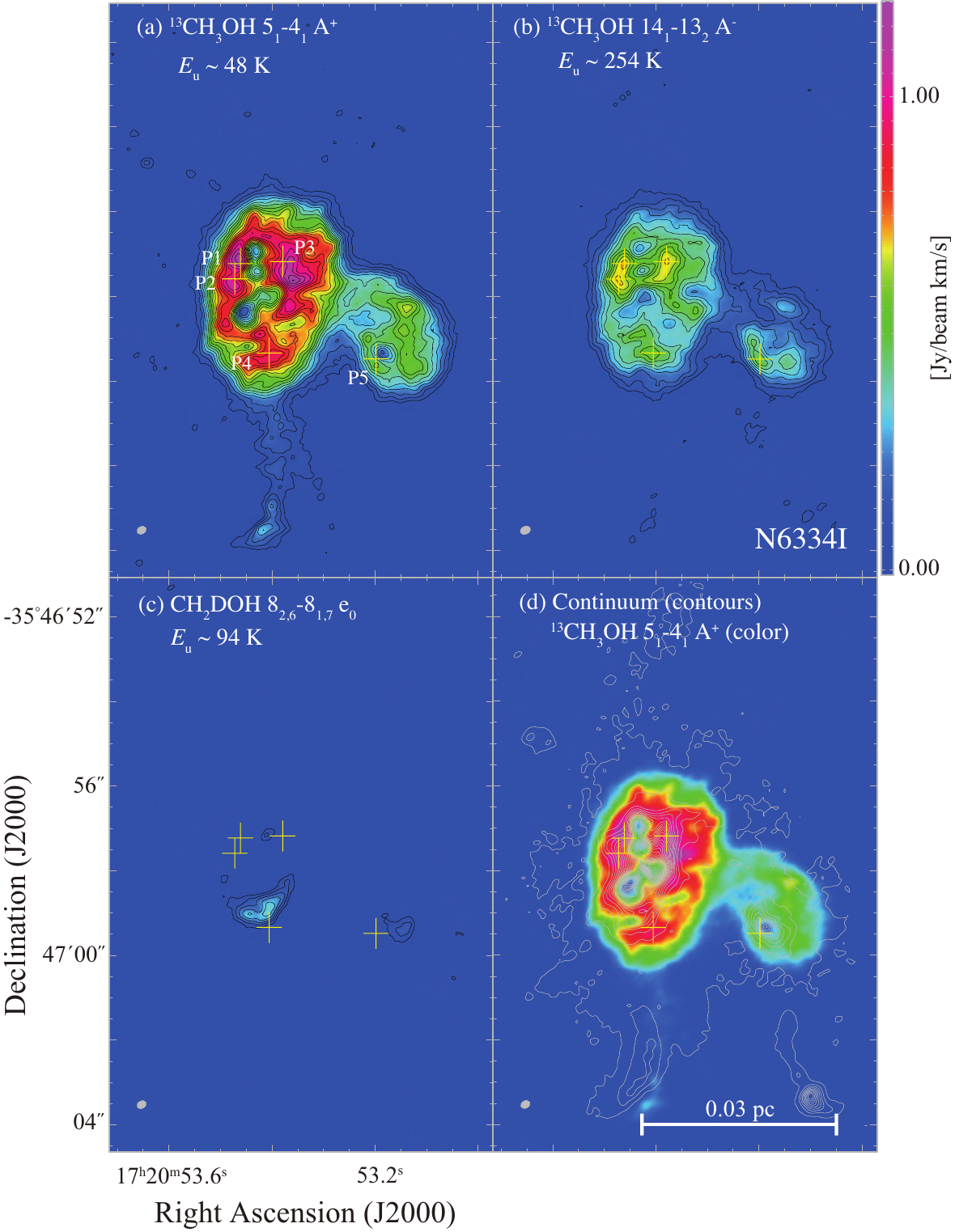}
\caption{Integrated intensity maps of $^{13}$CH$_3$OH $5_1-4_1$ A$^+$ (a), $^{13}$CH$_3$OH $14_1-13_2$ A$^-$ (b), and CH$_2$DOH $8_{2,6}-8_{1,7}$ $e_0$ (c) toward N6334I. The integrated velocity range is from -10 km s$^{-1}$ to 0 km s$^{-1}$. Contour levels start from 3$\sigma$ and increase in steps of 1$\sigma$ for (a), (b), and (c) [(a) $1\sigma=21$ mJy beam$^{-1}$ km s$^{-1}$, (b) $1\sigma=21$ mJy beam$^{-1}$ km s$^{-1}$, (c) $1\sigma=21$ mJy beam$^{-1}$ km s$^{-1}$]. (d) the continuum image (contours) overlaid on $^{13}$CH$_3$OH $14_1-13_2$ A$^-$ integrated intensity map (color) . For (d), contour levels start from 3$\sigma$ and increase in steps of 10$\sigma$ [$1\sigma=1$ mJy beam$^{-1}$]. Cross marks indicate the positions of the $^{13}$CH$_3$OH peak. The same color scale is used in (a)-(d). The synthesized beam is shown at the bottom left of each panel.}
\end{figure*}

\clearpage

\begin{figure*}
\figurenum{1}
\epsscale{1.0}
\plotone{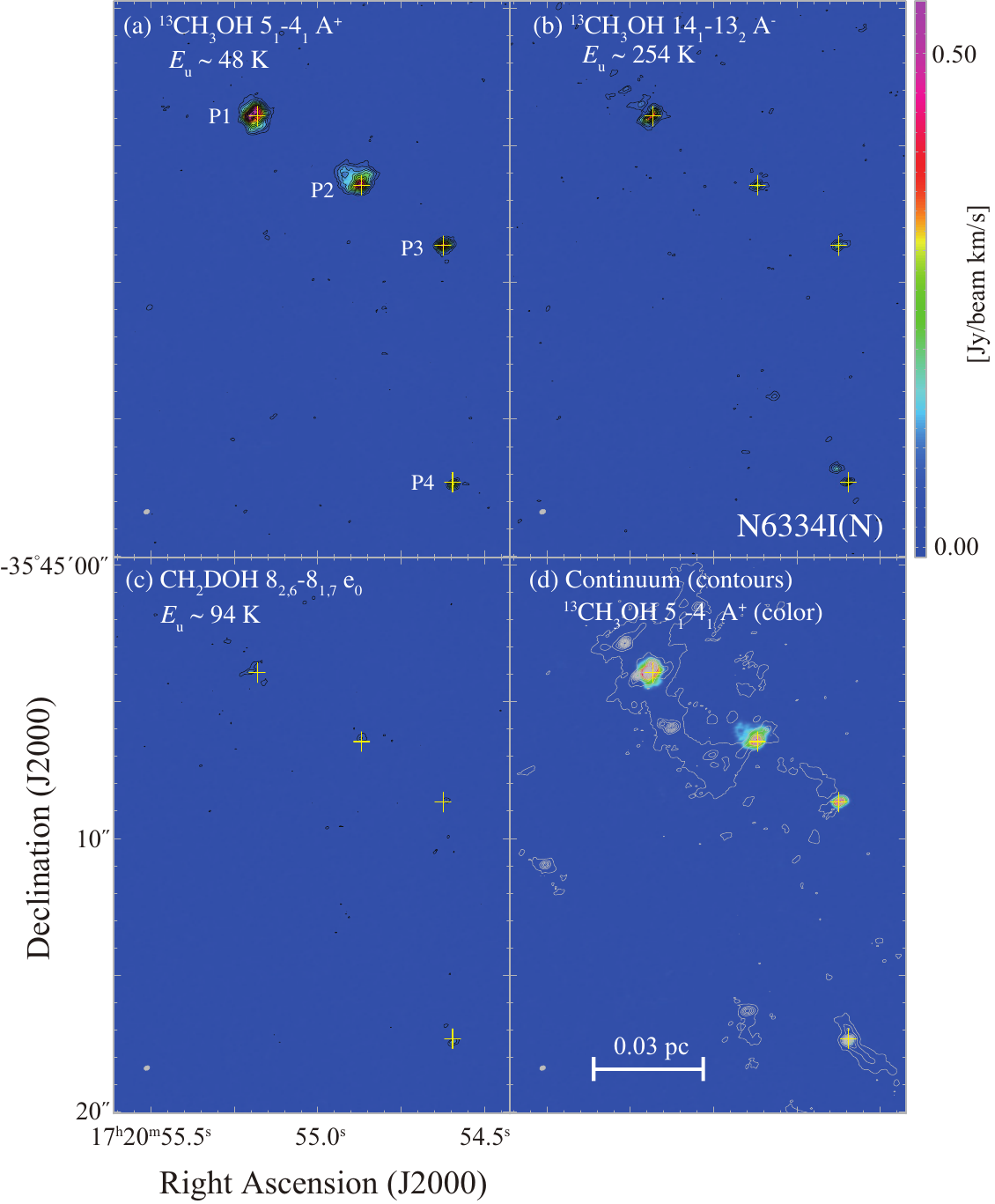}
\caption{Integrated intensity maps of $^{13}$CH$_3$OH $5_1-4_1$ A$^+$ (a), $^{13}$CH$_3$OH $14_1-13_2$ A$^-$ (b), and CH$_2$DOH $8_{2,6}-8_{1,7}$ $e_0$ (c) toward N6334IN. The integrated velocity range is from -10 km s$^{-1}$ to 0 km s$^{-1}$. Contour levels start from 3$\sigma$ and increase in steps of 1$\sigma$ for (a), (b), and (c) [(a) $1\sigma=13$ mJy beam$^{-1}$ km s$^{-1}$, (b) $1\sigma=12$ mJy beam$^{-1}$ km s$^{-1}$, (c) $1\sigma=13$ mJy beam$^{-1}$ km s$^{-1}$]. (d) the continuum image (contours) overlaid on $^{13}$CH$_3$OH $14_1-13_2$ A$^-$ integrated intensity map (color) . For (d), contour levels start from 3$\sigma$ and increase in steps of 10$\sigma$ [$1\sigma=0.4$ mJy beam$^{-1}$]. Cross marks indicate the positions of the $^{13}$CH$_3$OH peak. The same color scale is used in (a)-(d). The synthesized beam is shown at the bottom left of each panel.}
\end{figure*}

\clearpage

\begin{figure*}
\figurenum{1}
\epsscale{1.0}
\plotone{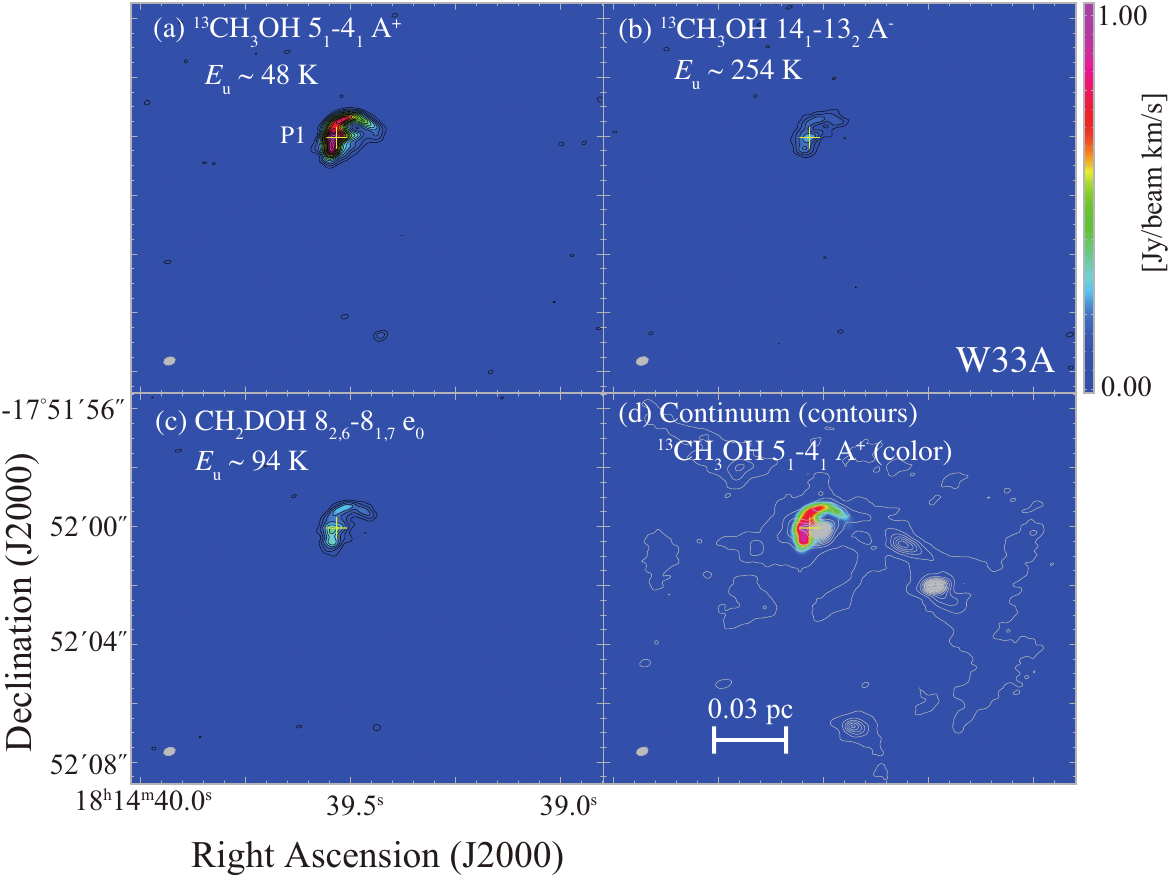}
\caption{Integrated intensity maps of $^{13}$CH$_3$OH $5_1-4_1$ A$^+$ (a), $^{13}$CH$_3$OH $14_1-13_2$ A$^-$ (b), and CH$_2$DOH $8_{2,6}-8_{1,7}$ $e_0$ (c) toward W33A. The integrated velocity range is from 32.5 km s$^{-1}$ to 42.5 km s$^{-1}$. Contour levels start from 3$\sigma$ and increase in steps of 1$\sigma$ for (a), (b), and (c) [(a) $1\sigma=18$ mJy beam$^{-1}$ km s$^{-1}$, (b) $1\sigma=17$ mJy beam$^{-1}$ km s$^{-1}$, (c) $1\sigma=18$ mJy beam$^{-1}$ km s$^{-1}$]. (d) the continuum image (contours) overlaid on $^{13}$CH$_3$OH $14_1-13_2$ A$^-$ integrated intensity map (color) . For (d), contour levels start from 3$\sigma$ and increase in steps of 10$\sigma$ [$1\sigma=0.2$ mJy beam$^{-1}$]. Cross marks indicate the positions of the $^{13}$CH$_3$OH peak. The same color scale is used in (a)-(d). The synthesized beam is shown at the bottom left of each panel.}
\end{figure*}

\clearpage

\begin{figure*}
\figurenum{2}
\epsscale{1.0}
\plotone{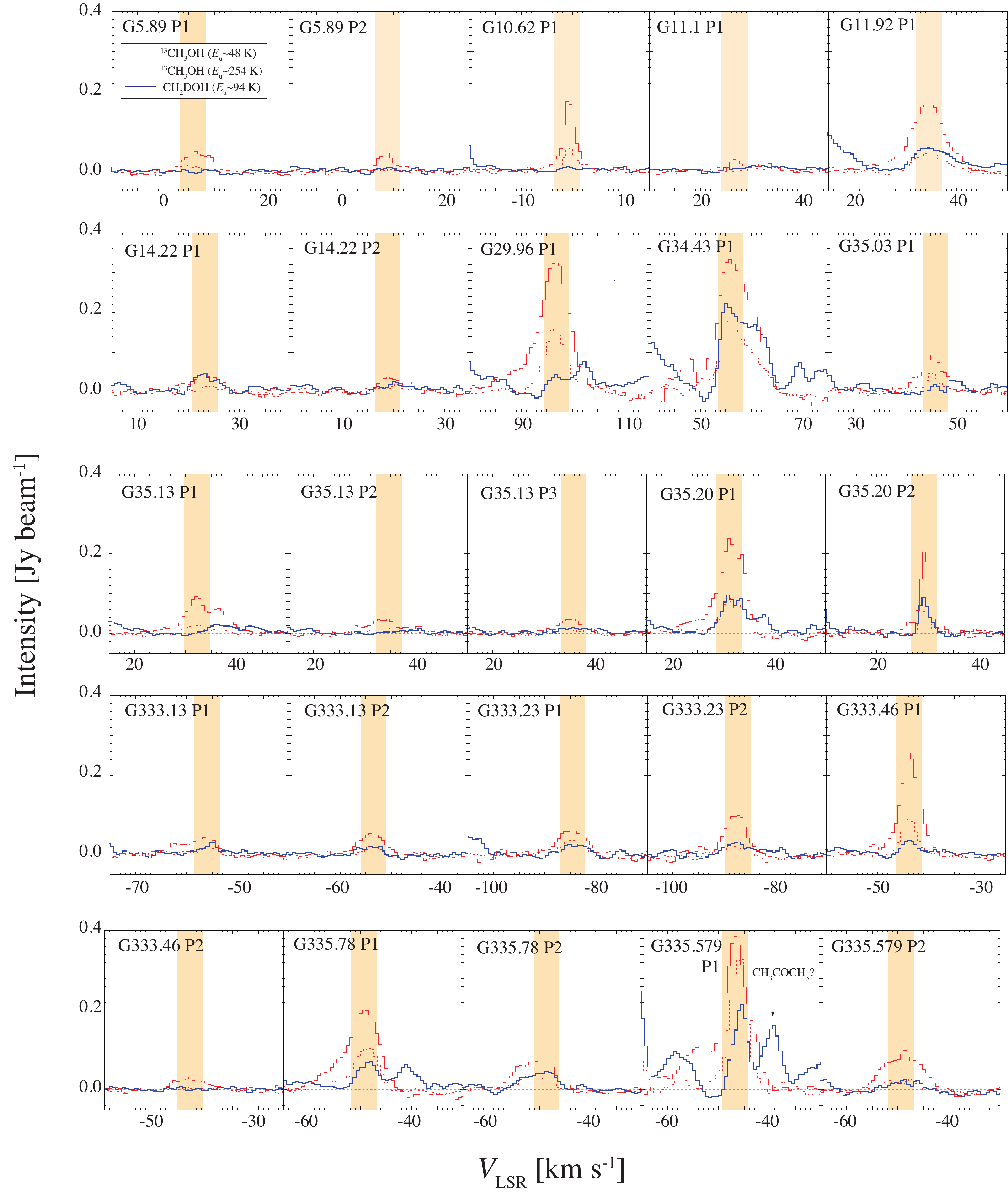}
\caption{Spectra toward the peak positions of each source. Orange regions indicate the velocity range over which the integrated intensity is calculated to derive the temperature and column densities.
\label{fig:sp}}
\end{figure*}

\begin{figure*}
\figurenum{3}
\epsscale{1.0}
\plotone{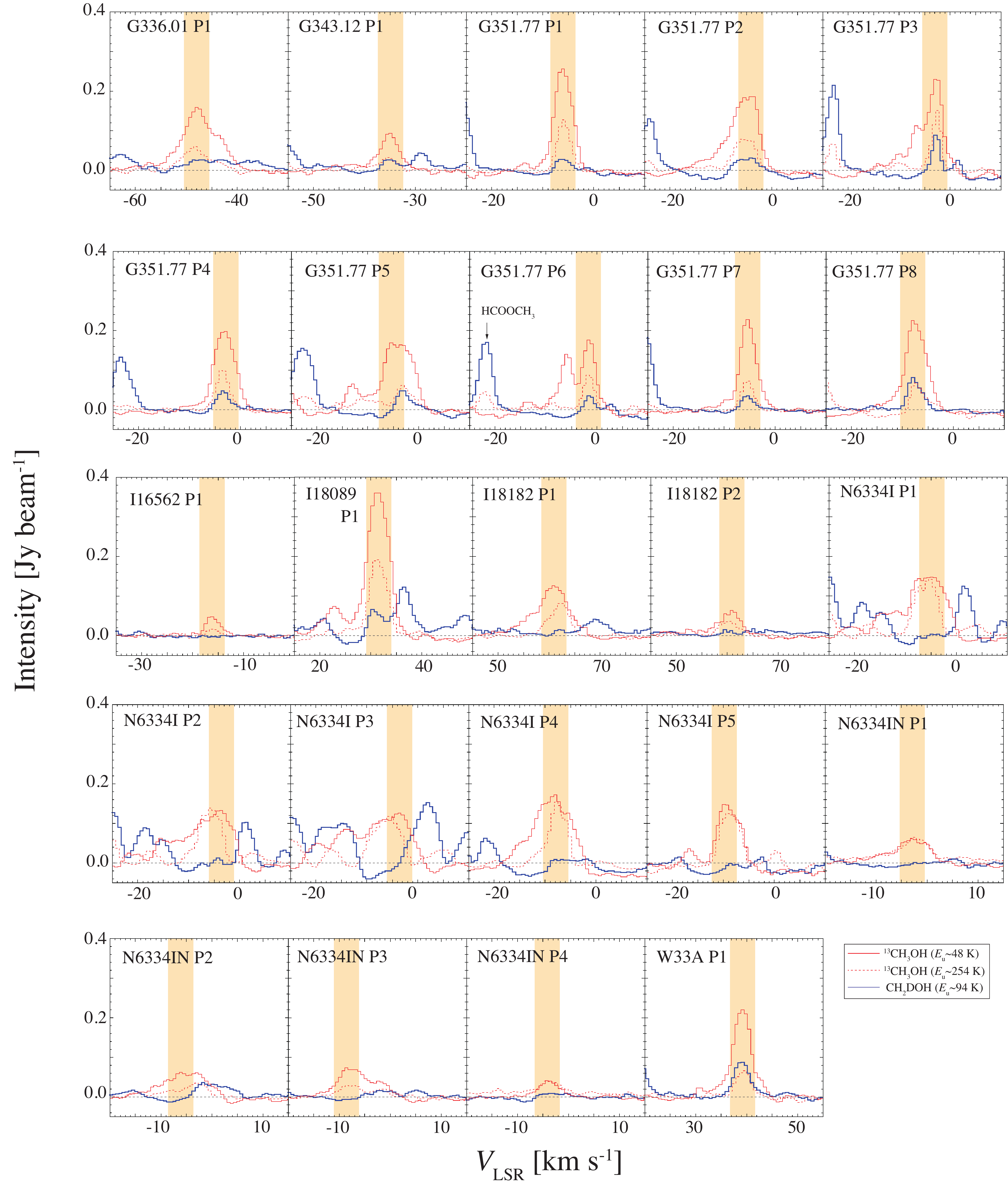}
\caption{Same as Figure 2.
\label{fig:sp2}}
\end{figure*}

\clearpage

\begin{figure*}
\figurenum{4}
\epsscale{0.7}
\plotone{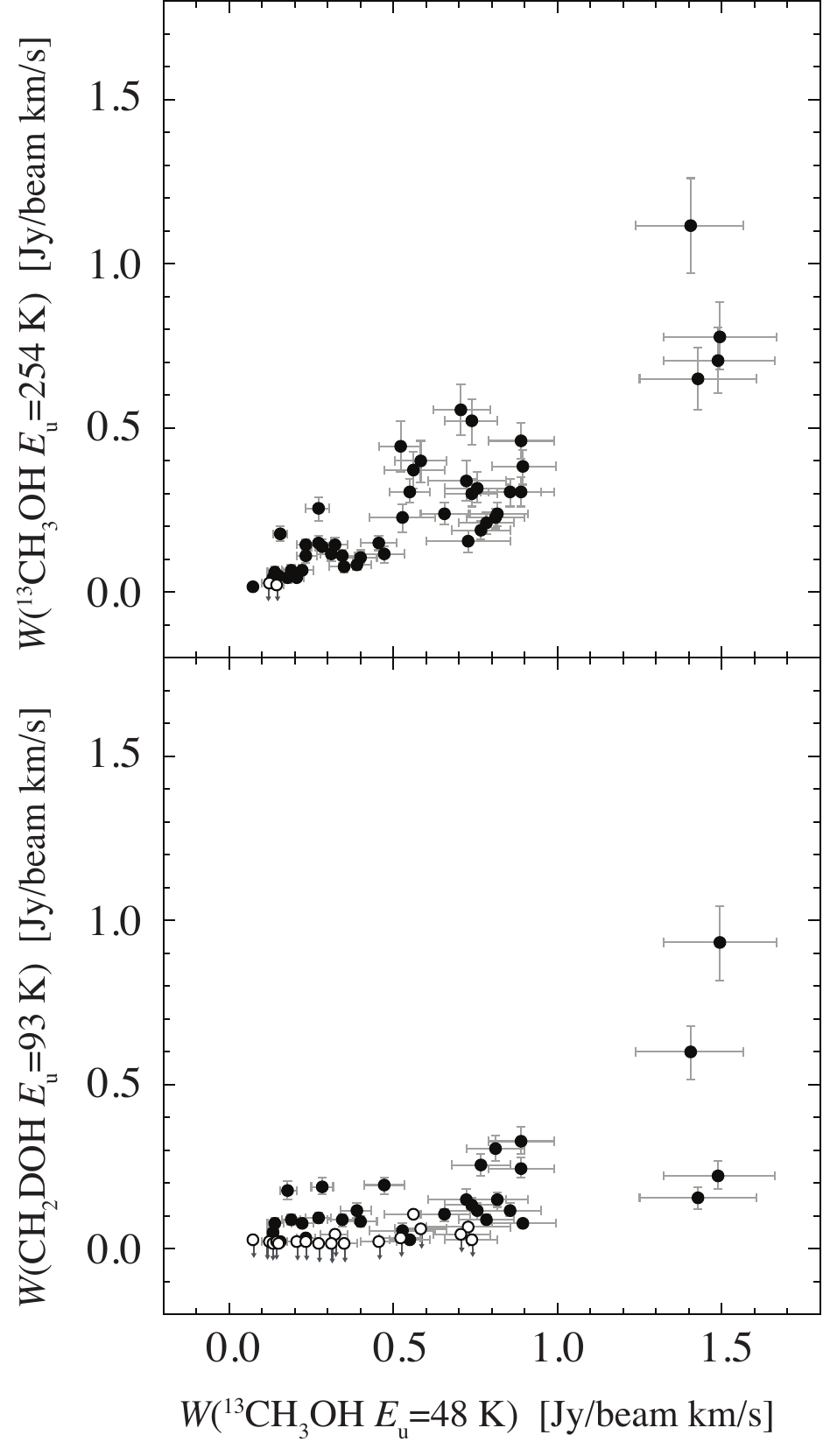}
\caption{(a) Correlation plot of the integrated intensities toward the $^{13}$CH$_3$OH peaks between $^{13}$CH$_3$OH $14_1-13_2$ A$^-$ and CH$_2$DOH $8_{2,6}-8_{1,7}$ $e_0$. (b) Same as (a), but for $^{13}$CH$_3$OH $14_1-13_2$ A$^-$ and $^{13}$CH$_3$OH $5_1-4_1$ A$^+$. Open circles indicate the upper limit values. The error bars denote the 1$\sigma$ noise level.
\label{fig:cor1}}
\end{figure*}

\clearpage

\begin{figure*}
\figurenum{5}
\epsscale{1.0}
\plotone{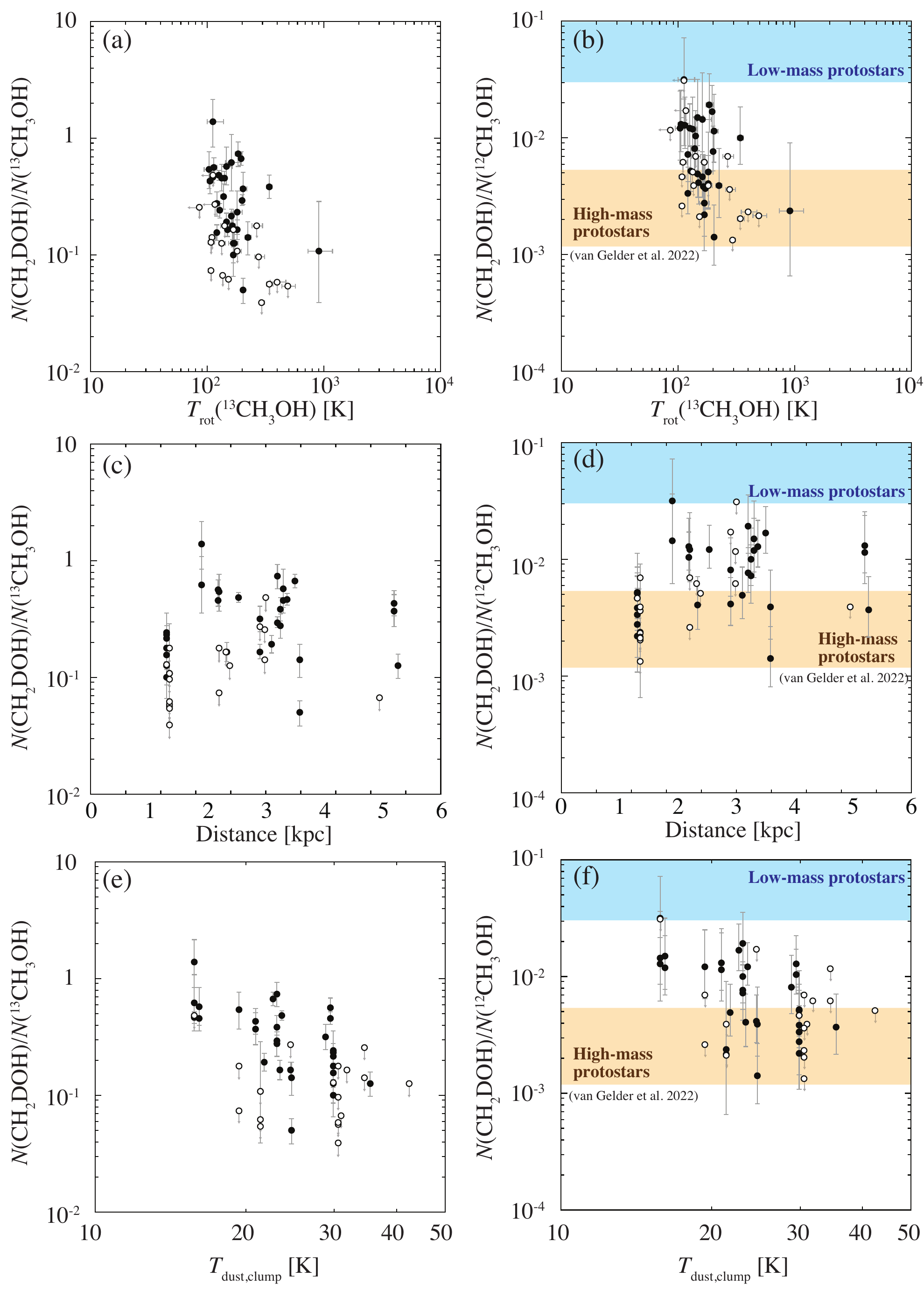}
\caption{Plots of the column density ratios against various parameters: (a) $N$(CH$_2$DOH)/$N$($^{13}$CH$_3$OH) versus $T_{\rm rot}$($^{13}$CH$_3$OH), (b) $N$(CH$_2$DOH)/$N$($^{12}$CH$_3$OH) versus $T_{\rm rot}$($^{13}$CH$_3$OH), (c) $N$(CH$_2$DOH)/$N$($^{13}$CH$_3$OH) versus distance, (d) $N$(CH$_2$DOH)/$N$($^{12}$CH$_3$OH) versus distance, (e) $N$(CH$_2$DOH)/$N$($^{13}$CH$_3$OH) versus dust temperature of clump, (f) $N$(CH$_2$DOH)/$N$($^{12}$CH$_3$OH) versus dust temperature of clump. Open circles indicate the upper limit values for $N$(CH$_2$DOH)/$N$($^{13}$CH$_3$OH) or $N$(CH$_2$DOH)/$N$($^{12}$CH$_3$OH). Orange and blue areas in (b), (d), and (e) indicate the range of the observed values for high-mass and low-mass protostars reported by van Gelder et al. (2022), respectively.
\label{fig:temp2}}
\end{figure*}

\clearpage

\begin{figure*}
\figurenum{6}
\epsscale{1.1}
\plotone{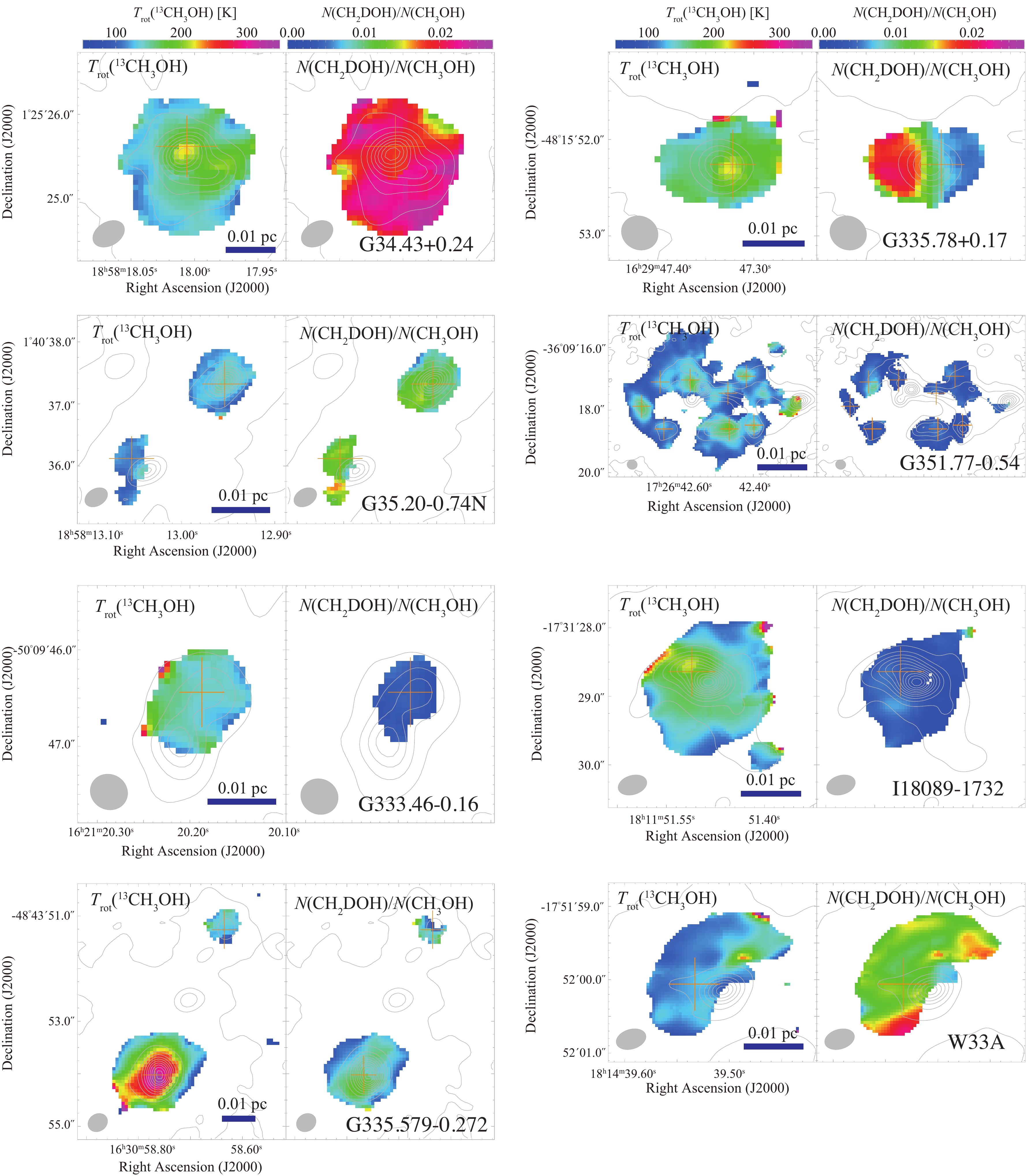}
\caption{Maps of the rotation temperature and the $N$(CH$_2$DOH)/$N$(CH$_3$OH) ratio. The same color scale is applied to all the temperature maps and separately to all the $N$(CH$_2$DOH)/$N$(CH$_3$OH) ratio maps. The orange cross marks indicate the peak positions of $^{13}$CH$_3$OH. Gray contours show the continuum emission. The synthesized beam is shown at the bottom left of each panel.
\label{fig:rmaps}}
\end{figure*}

\clearpage

\begin{figure*}
\figurenum{7}
\epsscale{1.1}
\plotone{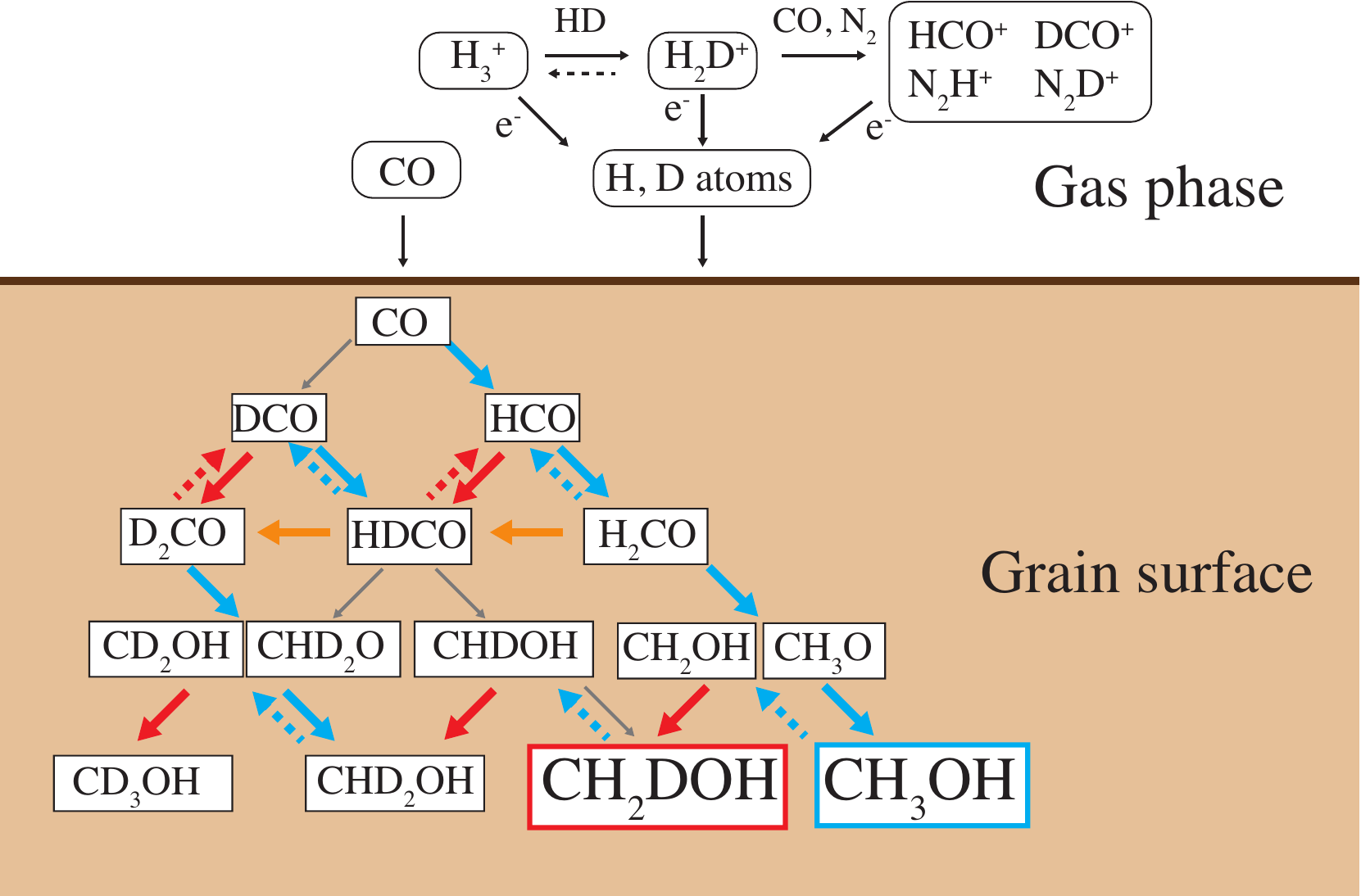}
\caption{Diagram of the formation mechanism of CH$_3$OH and CH$_2$DOH. Bold arrows represent the main reaction pathways on the grain surface, as indicated by laboratory experiments (Watanabe et al. 2006; Watanabe \& Kouchi 2008; Hidaka et al. 2009). Solid and dashed blue arrows indicate hydrogen addition reactions and hydrogen abstraction reactions, respectively. Solid and dashed red arrows represent deuterium addition reactions and deuterium abstraction reactions. Orange arrows show hydrogen-deuterium exchange reactions. Gray arrows represent the reaction pathways additionally considered in the chemical model calculation (Taquet et al. 2012).
\label{fig:form}}
\end{figure*}

\clearpage

\begin{figure*}
\figurenum{8}
\epsscale{1.1}
\plotone{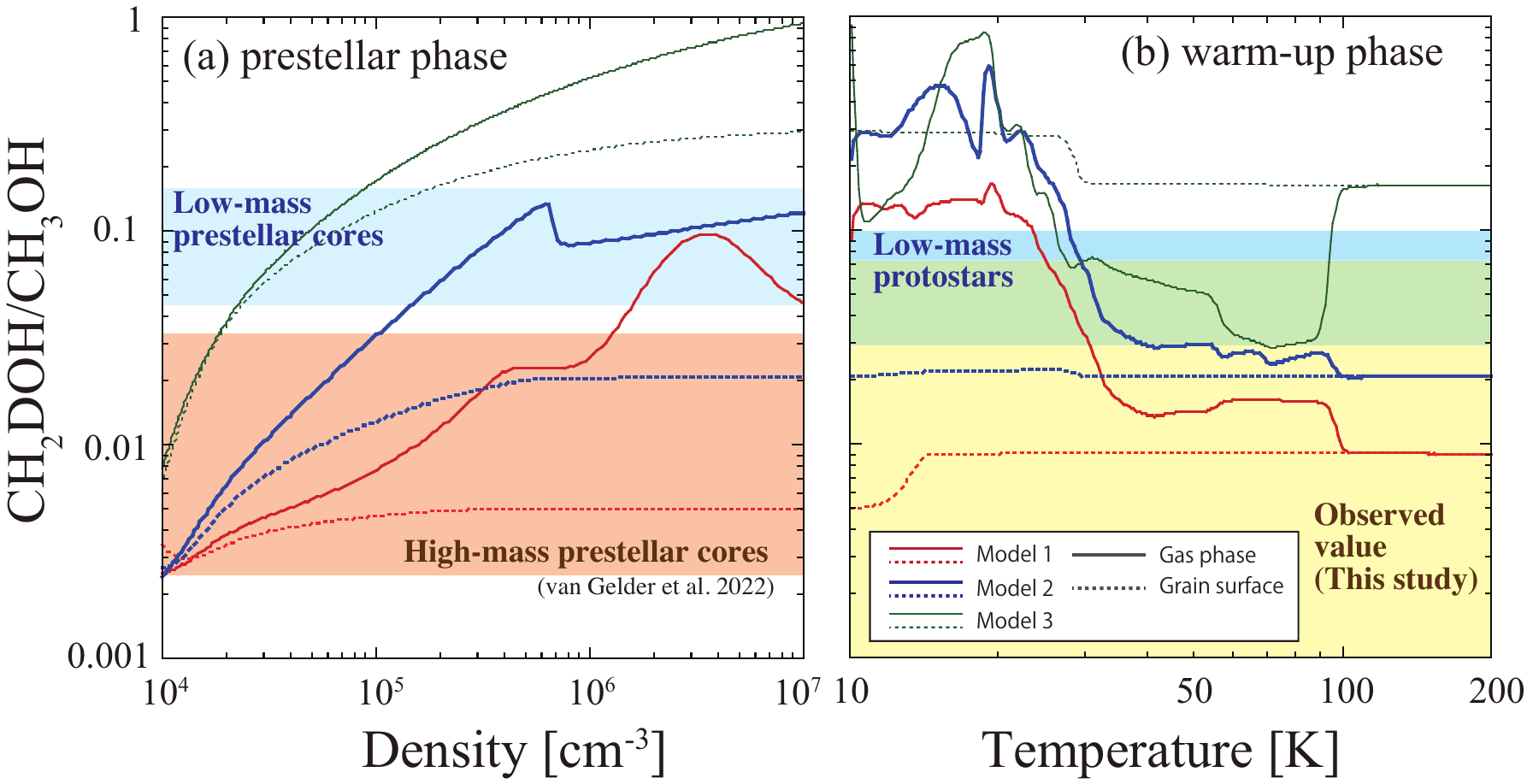}
\caption{(a) Chemical model calculation results of CH$_2$DOH/CH$_3$OH ratio in the prestellar phase. The horizontal axis is the density, which increases with time, and the vertical axis is the CH$_2$DOH/CH$_3$OH ratio. The dotted lines indicate the CH$_2$DOH/CH$_3$OH ratio on grain surface, and the solid lines indicate that in gas phase. The results of Model 1 (collapsing with one free-fall time) are shown in red, those of Model 2 (collapsing with three times the free-fall time) in blue, and those of Model 3 (collapsing with ten-times the free-fall time) in green. Red and light blue areas indicate the range of the observed values for high-mass and low-mass prestellar cores reported by van Gelder et al. (2022), respectively. (b) Chemical model calculation results of CH$_2$DOH/CH$_3$OH ratio in the warm-up phase. The horizontal axis is temperature, which increases with time, and the vertical axis is the CH$_2$DOH/CH$_3$OH ratio. Yellow area indicates the range of the observed values in this study, and blue area indicates the range of the observed values for low-mass protostars reported by van Gelder et al. (2022).
\label{fig:chmodel}}
\end{figure*}

\clearpage

\begin{figure*}
\figurenum{9}
\epsscale{1.1}
\plotone{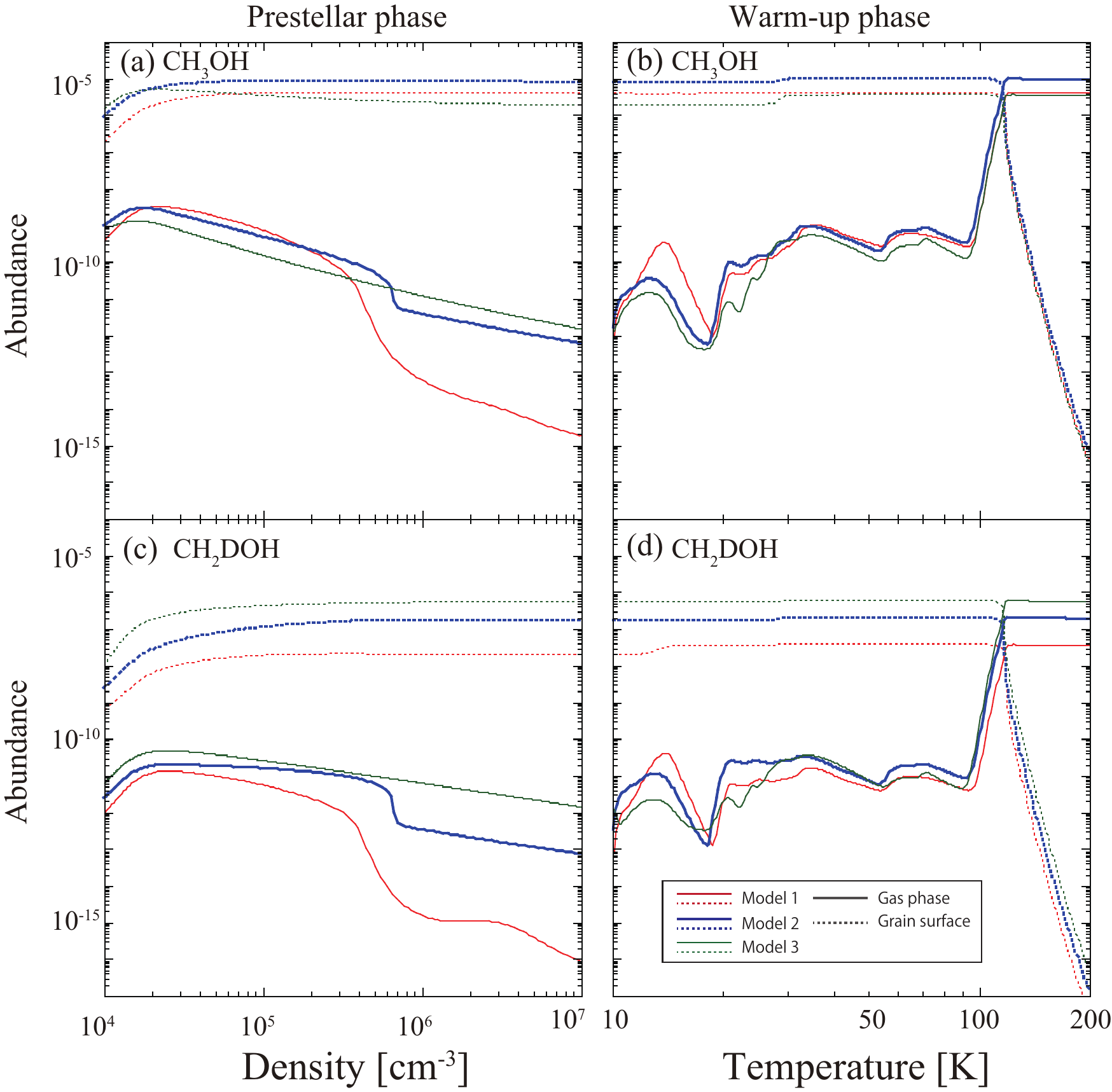}
\caption{Chemical model calculation results of the abundances of CH$_3$OH (a and b) and CH$_2$DOH (c and d). Left panels (a and c) show the results for the prestellar phase. The horizontal axis is the density, which increases with time. Right panels (b and d) show the results for the warm-up phase. The horizontal axis is temperature, which increases with time. The red line indicates the result of Model 1 (collapsing with free fall time), and the blue dashed line indicates the result of Model 2 (collapsing three times slower than the free-fall time), and the green dotted line indicates the result of Model 3 (collapsing ten times slower than the free-fall time). The dotted lines indicate the abundance on grain surface, and the solid lines indicate that in gas phase.
\label{fig:chmodel2}}
\end{figure*}

\clearpage

\begin{figure*}
\figurenum{10}
\epsscale{0.9}
\plotone{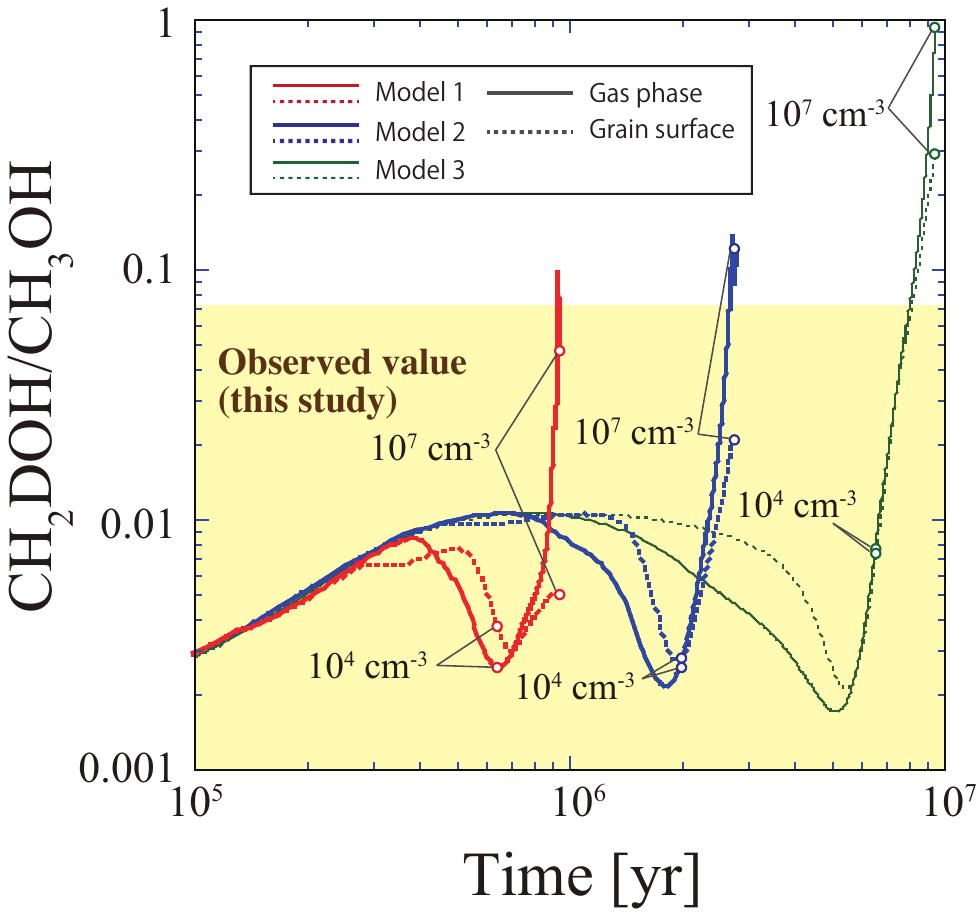}
\caption{Chemical model calculation results of CH$_2$DOH/CH$_3$OH ratio in the prestellar phase. The horizontal axis is time, and the vertical axis is the CH$_2$DOH/CH$_3$OH ratio. The red line indicates the result of Model 1 (collapsing with free fall time), and the blue dashed line indicates the result of Model 2 (collapsing three times slower than the free-fall time), and the green dotted line indicates the result of Model 3 (collapsing ten times slower than the free-fall time). The dotted lines indicate the CH$_2$DOH/CH$_3$OH ratio on grain surface, and the solid lines indicate that in gas phase. Yellow area indicates the range of the observed values in this study.
\label{fig:chmodel3}}
\end{figure*}

\clearpage

\begin{deluxetable*}{lllrrrrrlll}
\tabletypesize{\scriptsize}
\tablenum{1}
\tablecaption{Properties of the observed targets and the beam sizes of the observations\tablenotemark{a} \label{tab:freq}}
\tablewidth{0pt}
\tablehead{
\colhead{Source} & \colhead{R. A.}   & \colhead{Dec.}    &  \colhead{$D$\tablenotemark{b}}  & \colhead{$D_{\rm GC}$\tablenotemark{c}} & \colhead{Mass\tablenotemark{d}} & \colhead{$\Sigma$(H$_2$)\tablenotemark{e}} & \colhead{$T_{\rm dust}$\tablenotemark{f}} & \colhead{Beam} & \colhead{Beam} & \colhead{Beam} \\
& & & & & & & & $^{13}$CH$_3$OH & $^{13}$CH$_3$OH &  CH$_2$DOH \\
& & & & & & & & $5_1-4_1$ A$^+$ & $14_1-13_2$ A$^-$ & $8_{2,6}-8_{1,7}$ $e_0$ \\
 & \footnotesize(J2000.0) & \footnotesize(J2000.0)  & \footnotesize[kpc] &  \footnotesize[kpc] & \footnotesize[$M_\odot$] & \footnotesize[g cm$^{-2}$] & \footnotesize[K]  & \footnotesize($^{\prime\prime}\times^{\prime\prime}$, PA) &  \footnotesize($^{\prime\prime}\times^{\prime\prime}$, PA)&  \footnotesize($^{\prime\prime}\times^{\prime\prime}$, PA) \\
}
\decimalcolnumbers
\startdata
G5.89-0.37 & 18:00:30 & -24:04:01.64 & 2.99  & 5.04  & 1280 & 4.1 & 34.5& 0.437$\times$0.275(-77.2) & 0.476$\times$0.297(-76.5) & 0.436$\times$0.274(-77.1) \\
G10.62-0.38 & 18:10:29 & -19:55:49.52 & 4.95  & 3.26  & 5160 & 6.8 & 31.0 & 0.421$\times$0.275(-75.1) & 0.458$\times$0.298(-74.7) & 0.419$\times$0.275(-75.3) \\
G11.1-0.12 & 18:10:28.27 & -19:22:30.92 & 3.00  & 5.09  & 470 & 0.6 & 15.8 & 0.426$\times$0.274(-74.8) & 0.463$\times$0.297(-74.3) & 0.426$\times$0.274(-74.8) \\
G11.92-0.61 & 18:13:58 & -18:54:19.02 & 3.37  & 4.75  & 1050 & 1.0 & 22.8 & 0.416$\times$0.276(-74.9) & 0.454$\times$0.298(-74.1) & 0.415$\times$0.275(-74.7) \\
G14.22-0.50S & 18:18:13 & -16:57:21.82 & 1.90  & 6.17  & 550 & 1.2 & 15.8 &  0.546$\times$0.384(-60.5) & 0.561$\times$0.461(-55.6) & 0.543$\times$0.381(-60.4) \\
G29.96-0.02 & 18:46:04 & -02:39:22.60 & 5.26  & 4.33  & 1820 & 2.0 & 35.5 & 0.400$\times$0.269(-63.5) & 0.433$\times$0.291(-63.2) & 0.399$\times$0.268(-63.4) \\
G34.43+0.24 & 18:53:18 & 01:25:25.50 & 3.50  & 5.48  & 1400 & 3.4 & 22.7 & 0.402$\times$0.276(-59.3) & 0.434$\times$0.296(-59.9) & 0.401$\times$0.275(-59.5) \\
G35.03+0.35A & 18:54:01 & 02:01:19.30 & 2.32  & 6.24  & 170 & 1.0 & 31.8 & 0.400$\times$0.273(-59.2) & 0.434$\times$0.295(-59.1) & 0.400$\times$0.273(-59.1) \\
G35.13-0.74 & 18:58:06.30 & 01:37:05.98 & 2.20  & 6.33  & 930 & 1.7 & 19.4 & 0.364$\times$0.310(89.6) & 0.392$\times$0.333(88.2) & 0.362$\times$0.307(-86.7) \\
G35.20-0.74N & 18:58:13 & 01:40:36 & 2.19  & 6.34  & 640 & 1.7 & 29.5 & 0.392$\times$0.273(-59.7) & 0.423$\times$0.294(-59.5) & 0.392$\times$0.273(-59.8) \\
G333.12-0.56 & 16:21:36 & -50:40:50.01 & 3.30  & 5.27  & 2720 & 2.6 & 16.1 & 0.394$\times$0.382(72.2) & 0.422$\times$0.398(69.1) & 0.393$\times$0.381(75.9) \\
G333.23-0.06 & 16:19:51 & -50:15:13.00 & 5.20  & 4.09  & 2030 & 2.3 & 20.9 & 0.375$\times$0.327(-46.3) & 0.402$\times$0.355(-50.1) & 0.372$\times$0.328(-48.9) \\
G333.46-0.16 & 16:21:20 & -50:09:46.56 & 2.90  & 5.56  & 870 & 2.0 & 24.6 & 0.393$\times$0.363(62.1) & 0.422$\times$0.370(63.2) & 0.391$\times$0.363(65.5) \\
G335.78+0.17 & 16:29:47 & -48:15:52.32 & 3.20  & 5.25  & 1200 & 2.0 & 23.1 & 0.387$\times$0.346(63.5) & 0.416$\times$0.367(64.8) & 0.386$\times$0.347(64.1) \\
G335.579-0.272 & 16:30:59 & -48:43:54.01 & 3.25  & 5.22  & 1570 & 2.5 & 23.1 & 0.388$\times$0.328(-56.7) & 0.409$\times$0.352(-59.2) & 0.388$\times$0.326(-55.7) \\
G336.01-0.82 & 16:35:09 & -48:46:48.16 & 3.10  & 5.32  & 950 & 2.7 & 21.8 & 0.384$\times$0.366(78.2) & 0.415$\times$0.394(86.4) & 0.382$\times$0.366(80.9) \\
G343.12-0.06 & 16:58:17.22 & -42:52:07.49 & 2.90  & 5.29  & 1380 & 3.4 & 28.9 & 0.313$\times$0.312(80.3) & 0.336$\times$0.325(46.1) & 0.312$\times$0.311(-60.2) \\
G351.77-0.54 & 17:26:42.53 & -36:09:17.40 & 1.30  & 6.72  & 620 & 9.9 & 29.9 & 0.324$\times$0.302(80.5) & 0.348$\times$0.321(69.9) & 0.324$\times$0.300(81.3) \\
IRAS16562-3959 & 16:59:41.63 & -40:03:43.62 & 2.38  & 5.73  & 1090 & 1.5 & 42.3 & 0.225$\times$0.170(-64.5) & 0.242$\times$0.184(-60.6) & 0.224$\times$0.171(-65.2) \\
IRAS18089-1732 & 18:11:51.40 & -17:31:28.52 & 2.34  & 5.74  & 550 & 2.1 & 23.4 & 0.428$\times$0.278(-72.9) & 0.456$\times$0.297(-72.6) & 0.427$\times$0.277(-72.7) \\
IRAS18182-1433 & 18:21:09.13 & -14:31:50.58 & 3.58  & 4.68  & 650 & 1.5 & 24.7 & 0.291$\times$0.211(-62.9) & 0.316$\times$0.234(-63.0) & 0.291$\times$0.211(-63.1) \\
NGC6334I & 17:20:53.30 & -35:47:00.02 & 1.35  & 6.67  & 720 & 10.2 & 30.6 & 0.235$\times$0.175(-61.6) & 0.246$\times$0.186(-65.7) & 0.235$\times$0.174(-61.6) \\
NGC6334I(N) & 17:20:54.90 & -35:45:10.02 & 1.35  & 6.67  & 2190 & 7.6 & 21.4 & 0.227$\times$0.168(-65.6) & 0.241$\times$0.176(-66.8) & 0.226$\times$0.167(-65.4) \\
W33A & 18:14:39.40 & -17:52:01.02 & 2.53  & 5.56  & 890 & 1.9 & 23.6 & 0.415$\times$0.267(-72.9) & 0.436$\times$0.280(-73.2) & 0.416$\times$0.269(-73.0) \\
\enddata
\tablenotetext{a}{The parameters, except for beam sizes, are teken from Ishihara et al. (2024). The references for the parameters are listed in Ishihara et al. (2024).}
\tablenotetext{b}{Distance from the Sun.}
\tablenotetext{c}{Distance from the Galactic center. The distance of the Sun to the Galactic center is assumed to be 8 kpc (Eisenhauer et al. 2003)}
\tablenotetext{d}{Clump mass.}
\tablenotetext{e}{Surface density of clump.}
\tablenotetext{f}{Dust temperature of clump.}
\end{deluxetable*}

\clearpage

\begin{deluxetable*}{crrrr}
\tablenum{2}
\tablecaption{Parameters of the observed lines \label{tab:lines}}
\tablewidth{0pt}
\tablehead{
\colhead{Molecule} & \colhead{Transition}   & \colhead{Frequency\tablenotemark{a}}    &  \colhead{$\mu^2 S$}  & \colhead{$E_u/k$}  \\
 &  & \footnotesize[GHz]  & \footnotesize[D$^2$] &  \footnotesize[K]  \\
}
\decimalcolnumbers
\startdata
$^{13}$CH$_3$OH & $5_1-4_1$ A$^+$ & 234.011612 & 3.88589\tablenotemark{b} & 48.25\tablenotemark{b} \\
$^{13}$CH$_3$OH & $14_1-13_2$ A$^-$  & 217.044615 & 5.78604\tablenotemark{b} & 254.25\tablenotemark{b}\\
CH$_2$DOH & $8_{2,6}-8_{1,7}$ $e_0$ & 234.470663 & 7.78\tablenotemark{a} & 93.66\tablenotemark{c} \\
\enddata
\tablenotetext{a}{taken from the SUMIRE data (Watanabe et al. 2021; Oyama et al. 2023).}
\tablenotetext{b}{taken from CDMS (M{\"u}ller et al. 2005).}
\tablenotetext{c}{taken from JPL (Pickett et al. 1998).}
\end{deluxetable*}

\begin{deluxetable*}{lllrlll}
\tabletypesize{\scriptsize}
\tablenum{3}
\tablecaption{Peak positions and integrated intensities \label{tab:peak}}
\tablewidth{0pt}
\tablehead{
\colhead{Name} & \colhead{R. A.}   & \colhead{Dec.}  &  \colhead{$V_{\rm LSR}$\tablenotemark{a}}  &  \colhead{$\int I_{\nu} dV$}  & \colhead{$\int I_{\nu} dV$} & \colhead{$\int I_{\nu} dV$} \\
& & & & $^{13}$CH$_3$OH &  $^{13}$CH$_3$OH &  CH$_2$DOH \\
& & & & $5_1-4_1$ A$^+$ & $14_1-13_2$ A$^-$ & $8_{2,6}-8_{1,7}$ $e_0$\\
 & \footnotesize(J2000.0) & \footnotesize(J2000.0)  & \footnotesize[km s$^{-1}$] &  \footnotesize[Jy km s$^{-1}$] & \footnotesize[Jy km s$^{-1}$]  &  \footnotesize[Jy km s$^{-1}$] \\
}
\decimalcolnumbers
\startdata
G5.89-0.37 P1 & 18:030.644 & -24:4:3.16 & 5.78  & 0.204$_{\pm 0.026}$ & 0.047$_{\pm 0.010}$ & $<$0.021    \\
G5.89-0.37 P2 & 18:0:30.501 & -24:4:0.6 & 8.90  & 0.147$_{\pm 0.021}$ & $<$0.020   & $<$0.023    \\
G10.62-0.38 P1 & 18:10:28.674 & -19:55:49.36 & -1.02  & 0.456$_{\pm 0.053}$ & 0.152$_{\pm 0.023}$ & $<$0.025    \\
G11.1-0.12 P1 & 18:10:28.246 & -19:22:30.28 & 26.63  & 0.073$_{\pm 0.013}$ & $<$0.018   & $<$0.026    \\
G11.92-0.61 P1 & 18:13:58.1 & -18:54:20.3 & 34.54  & 0.766$_{\pm 0.089}$ & 0.189$_{\pm 0.030}$ & 0.257$_{\pm 0.034}$ \\
G14.22-0.50S P1 & 18:18:13.35 & -16:57:23.98 & 23.31  & 0.180$_{\pm 0.025}$ & 0.046$_{\pm 0.018}$ & 0.177$_{\pm 0.027}$ \\
G14.22-0.50S P2 & 18:18:12.86 & -16:57:20.38 & 18.89  & 0.141$_{\pm 0.024}$ & 0.063$_{\pm 0.016}$ & 0.075$_{\pm 0.017}$ \\
G29.96-0.02 P1 & 18:46:3.772 & -2:39:22.12 & 96.96  & 1.428$_{\pm 0.176}$ & 0.649$_{\pm 0.094}$ & 0.157$_{\pm 0.033}$ \\
G34.43+0.24 P1 & 18:53:18.006 & 1:25:25.62 & 55.85  & 1.496$_{\pm 0.171}$ & 0.779$_{\pm 0.104}$ & 0.933$_{\pm 0.114}$ \\
G35.03+0.35A P1 & 18:54:0.662 & 2:1:19.3 & 45.98  & 0.321$_{\pm 0.043}$ & 0.142$_{\pm 0.026}$ & $<$0.046    \\
G35.13-0.74 P1 & 18:58:6.13 & 1:37:7.48 & 32.01  & 0.349$_{\pm 0.043}$ & 0.075$_{\pm 0.014}$ & $<$0.018    \\
G35.13-0.74 P2 & 18:58:6.166 & 1:37:8.2 & 34.57  & 0.135$_{\pm 0.021}$ & 0.047$_{\pm 0.010}$ & $<$0.020    \\
G35.13-0.74 P3 & 18:58:6.278 & 1:37:7.3 & 35.82  & 0.134$_{\pm 0.018}$ & 0.027$_{\pm 0.007}$ & 0.049$_{\pm 0.011}$ \\
G35.20-0.74N P1 & 18:58:12.954 & 1:40:37.32 & 31.15  & 0.889$_{\pm 0.101}$ & 0.305$_{\pm 0.045}$ & 0.330$_{\pm 0.044}$ \\
G35.20-0.74N P2 & 18:58:13.054 & 1:40:36.12 & 29.28  & 0.471$_{\pm 0.062}$ & 0.114$_{\pm 0.023}$ & 0.192$_{\pm 0.026}$ \\
G333.12-0.56 P1 & 16:21:35.378 & -50:40:56.61 & -55.99  & 0.189$_{\pm 0.026}$ & 0.067$_{\pm 0.016}$ & 0.090$_{\pm 0.016}$ \\
G333.12-0.56 P2 & 16:21:36.243 & -50:40:47.31 & -53.48  & 0.223$_{\pm 0.031}$ & 0.068$_{\pm 0.012}$ & 0.080$_{\pm 0.015}$ \\
G333.23-0.06 P1 & 16:19:50.875 & -50:15:10.48 & -84.72  & 0.273$_{\pm 0.038}$ & 0.149$_{\pm 0.024}$ & 0.096$_{\pm 0.018}$ \\
G333.23-0.06 P2 & 16:19:51.275 & -50:15:14.56 & -87.22  & 0.386$_{\pm 0.049}$ & 0.082$_{\pm 0.018}$ & 0.115$_{\pm 0.021}$ \\
G333.46-0.16 P1 & 16:21:20.187 & -50:9:46.44 & -43.89  & 0.858$_{\pm 0.094}$ & 0.304$_{\pm 0.042}$ & 0.118$_{\pm 0.019}$ \\
G333.46-0.16 P2 & 16:21:20.168 & -50:9:48.96 & -43.26  & 0.120$_{\pm 0.019}$ & $<$0.029   & $<$0.024    \\
G335.579-0.272 P1 & 16:30:58.768 & -48:43:54.02 & -46.73  & 1.404$_{\pm 0.164}$ & 1.117$_{\pm 0.145}$ & 0.597$_{\pm 0.082}$ \\
G335.579-0.272 P2 & 16:30:58.635 & -48:43:51.26 & -48.61  & 0.400$_{\pm 0.050}$ & 0.107$_{\pm 0.019}$ & 0.082$_{\pm 0.017}$ \\
G335.78+0.17 P1 & 16:29:47.322 & -48:15:52.26 & -49.33  & 0.890$_{\pm 0.101}$ & 0.462$_{\pm 0.056}$ & 0.247$_{\pm 0.032}$ \\
G335.78+0.17 P2 & 16:29:46.138 & -48:15:49.98 & -48.71  & 0.284$_{\pm 0.035}$ & 0.137$_{\pm 0.023}$ & 0.191$_{\pm 0.026}$ \\
G336.01-0.82 P1 & 16:35:9.272 & -48:46:47.56 & -48.00  & 0.658$_{\pm 0.075}$ & 0.241$_{\pm 0.034}$ & 0.104$_{\pm 0.019}$ \\
G343.12-0.06 P1 & 16:58:17.23 & -42:52:7.55 & -34.92  & 0.343$_{\pm 0.041}$ & 0.110$_{\pm 0.019}$ & 0.086$_{\pm 0.018}$ \\
G351.77-0.54 P1 & 17:26:42.695 & -36:9:17.88 & -6.05  & 0.897$_{\pm 0.095}$ & 0.382$_{\pm 0.053}$ & 0.078$_{\pm 0.013}$ \\
G351.77-0.54 P2 & 17:26:42.567 & -36:9:18.6 & -4.17  & 0.754$_{\pm 0.099}$ & 0.318$_{\pm 0.047}$ & 0.116$_{\pm 0.025}$ \\
G351.77-0.54 P3 & 17:26:42.463 & -36:9:18.6 & -2.92  & 0.725$_{\pm 0.117}$ & 0.340$_{\pm 0.063}$ & 0.151$_{\pm 0.034}$ \\
G351.77-0.54 P4 & 17:26:42.418 & -36:9:16.92 & -2.92  & 0.738$_{\pm 0.080}$ & 0.303$_{\pm 0.043}$ & 0.135$_{\pm 0.019}$ \\
G351.77-0.54 P5 & 17:26:42.467 & -36:9:17.46 & -5.42  & 0.727$_{\pm 0.128}$ & 0.156$_{\pm 0.032}$ & $<$0.065    \\
G351.77-0.54 P6 & 17:26:42.398 & -36:9:18.48 & -1.67  & 0.528$_{\pm 0.099}$ & 0.225$_{\pm 0.040}$ & 0.058$_{\pm 0.022}$ \\
G351.77-0.54 P7 & 17:26:42.636 & -36:9:18.6 & -5.42  & 0.783$_{\pm 0.085}$ & 0.211$_{\pm 0.035}$ & 0.091$_{\pm 0.014}$ \\
G351.77-0.54 P8 & 17:26:42.646 & -36:9:17.1 & -7.97  & 0.817$_{\pm 0.092}$ & 0.237$_{\pm 0.037}$ & 0.150$_{\pm 0.023}$ \\
IRAS16562-3959 P1 & 16:59:41.562 & -40:3:42.99 & -16.19  & 0.148$_{\pm 0.021}$ & 0.048$_{\pm 0.009}$ & $<$0.015    \\
IRAS18089-1732 P1 & 18:11:51.474 & -17:31:28.64 & 31.57  & 1.491$_{\pm 0.168}$ & 0.705$_{\pm 0.098}$ & 0.224$_{\pm 0.042}$ \\
IRAS18182-1433 P1 & 18:21:9.124 & -14:31:48.52 & 60.85  & 0.551$_{\pm 0.061}$ & 0.308$_{\pm 0.038}$ & 0.026$_{\pm 0.007}$ \\
IRAS18182-1433 P2 & 18:21:8.981 & -14:31:47.61 & 60.85  & 0.231$_{\pm 0.027}$ & 0.142$_{\pm 0.021}$ & 0.032$_{\pm 0.008}$ \\
NGC6334I P1 & 17:20:53.461 & -35:46:57.23 & -4.87  & 0.707$_{\pm 0.085}$ & 0.554$_{\pm 0.077}$ & $<$0.045    \\
NGC6334I P2 & 17:20:53.472 & -35:46:57.59 & -3.62  & 0.584$_{\pm 0.076}$ & 0.398$_{\pm 0.066}$ & $<$0.059    \\
NGC6334I P3 & 17:20:53.379 & -35:46:57.18 & -3.62  & 0.563$_{\pm 0.091}$ & 0.371$_{\pm 0.056}$ & $<$0.103    \\
NGC6334I P4 & 17:20:53.405 & -35:46:59.34 & -7.99  & 0.737$_{\pm 0.082}$ & 0.521$_{\pm 0.069}$ & $<$0.031    \\
NGC6334I P5 & 17:20:53.198 & -35:46:59.48 & -9.87  & 0.520$_{\pm 0.064}$ & 0.444$_{\pm 0.076}$ & $<$0.036    \\
NGC6334I(N) P1 & 17:20:55.182 & -35:45:3.94 & -2.97  & 0.271$_{\pm 0.037}$ & 0.255$_{\pm 0.036}$ & $<$0.018    \\
NGC6334I(N) P2 & 17:20:54.868 & -35:45:6.46 & -6.10  & 0.236$_{\pm 0.032}$ & 0.109$_{\pm 0.019}$ & $<$0.023    \\
NGC6334I(N) P3 & 17:20:54.624 & -35:45:8.67 & -8.60  & 0.313$_{\pm 0.036}$ & 0.114$_{\pm 0.017}$ & $<$0.017    \\
NGC6334I(N) P4 & 17:20:54.595 & -35:45:17.31 & -4.22  & 0.156$_{\pm 0.021}$ & 0.177$_{\pm 0.024}$ & 0.024$_{\pm 0.007}$ \\
W33A P1 & 18:14:39.533 & -17:52:0.06 & 39.22  & 0.812$_{\pm 0.089}$ & 0.226$_{\pm 0.031}$ & 0.306$_{\pm 0.039}$ \\
\enddata
\tablenotetext{a}{The peak velocity of $^{13}$CH$_3$OH $5_1-4_1$ A$^+$.}
\tablecomments{The errors represent 1$\sigma$ values. Details on how the errors were derived are described in the text.}
\end{deluxetable*}

\begin{deluxetable*}{llllllll}
\tablenum{4}
\tabletypesize{\scriptsize}
\tablecaption{Temperature and column densities\label{tab:col}}
\tablewidth{0pt}
\tablehead{
\colhead{Name} & \colhead{$T_{\rm rot}$($^{13}$CH$_3$OH)}   & \colhead{$N$($^{13}$CH$_3$OH)}  &  \colhead{$N$($^{12}$CH$_3$OH)}  &  \colhead{$N$(CH$_2$DOH)}  & \colhead{$^{12}$C/$^{13}$C\tablenotemark{b}} & \colhead{$\frac{{\rm N(CH_2DOH)}}{N({\rm ^{13}CH_3OH)}}$}  & \colhead{$\frac{{\rm N(CH_2DOH)}}{N({\rm ^{12}CH_3OH)}}$}\\
& \footnotesize[K] & \footnotesize[10$^{16}$ cm$^{-2}$] & \footnotesize[10$^{18}$ cm$^{-2}$]  & \footnotesize[10$^{16}$ cm$^{-2}$] & &  \footnotesize[10$^{-1}$] &  \footnotesize[10$^{-2}$]\\
}
\decimalcolnumbers
\startdata
G5.89-0.37 P1 & 109$_{-6}^{+8}$ & 1.7$_{-0.3}^{+0.4}$ & 0.63$_{-0.31}^{+0.38}$ & $<$0.24  & 37.46$_{\pm12.14}$ & $<$1.4  & $<$0.63  \\
G5.89-0.37 P2 & $<$86  & 0.88$_{-0.19}^{+0.13}$ & 0.33$_{-0.17}^{+0.16}$ & $<$0.23  & 37.46$_{\pm12.14}$ & $<$2.6  & $<$1.2  \\
G10.62-0.38 P1 & 136$_{-4}^{+5}$ & 5.2$_{-0.9}^{+0.9}$ & 1.5$_{-0.7}^{+0.8}$ & $<$0.35  & 28.42$_{\pm10.19}$ & $<$0.67  & $<$0.39  \\
G11.1-0.12 P1 & $<$111  & 0.63$_{-0.31}^{+0.11}$ & 0.24$_{-0.16}^{+0.12}$ & $<$0.30  & 37.72$_{\pm12.20}$ & $<$4.81  & $<$3.07  \\
G11.92-0.61 P1 & 112$_{-4}^{+4}$ & 6.8$_{-1.1}^{+1.1}$ & 2.5$_{-1.2}^{+1.3}$ & 3.2$_{-0.5}^{+0.5}$ & 35.99$_{\pm11.83}$ & 4.7$_{-0.5}^{+0.6}$ & 1.3$_{-0.4}^{+0.9}$ \\
G14.22-0.50S P1 & 113$_{-14}^{+26}$ & 0.88$_{-0.26}^{+0.44}$ & 0.38$_{-0.21}^{+0.35}$ & 1.2$_{-0.3}^{+0.5}$ & 43.20$_{\pm13.39}$ & 14$_{-5}^{+8}$ & 3.2$_{-1.7}^{+4.1}$ \\
G14.22-0.50S P2 & 163$_{-18}^{+29}$ & 1.2$_{-0.4}^{+0.6}$ & 0.50$_{-0.28}^{+0.45}$ & 0.72$_{-0.24}^{+0.32}$ & 43.20$_{\pm13.39}$ & 6.2$_{-2.7}^{+4.6}$ & 1.4$_{-0.8}^{+2.2}$ \\
G29.96-0.02 P1 & 172$_{-7}^{+8}$ & 25$_{-5}^{+5}$ & 8.5$_{-4.1}^{+4.7}$ & 3.1$_{-0.8}^{+0.8}$ & 33.86$_{\pm11.36}$ & 1.3$_{-0.3}^{+0.3}$ & 0.37$_{-0.15}^{+0.33}$ \\
G34.43+0.24 P1 & 197$_{-6}^{+7}$ & 31$_{-5}^{+5}$ & 12$_{-6}^{+6}$ & 21$_{-3}^{+3}$ & 39.70$_{\pm12.63}$ & 6.7$_{-0.8}^{+0.9}$ & 1.7$_{-0.6}^{+1.1}$ \\
G35.03+0.35A P1 & 168$_{-11}^{+15}$ & 5.3$_{-1.2}^{+1.5}$ & 2.3$_{-1.2}^{+1.5}$ & $<$0.87  & 43.56$_{\pm13.46}$ & $<$1.6  & $<$0.63  \\
G35.13-0.74 P1 & 107$_{-4}^{+5}$ & 3.0$_{-0.5}^{+0.6}$ & 1.3$_{-0.6}^{+0.7}$ & $<$0.22  & 44.02$_{\pm13.56}$ & $<$0.74  & $<$0.26  \\
G35.13-0.74 P2 & 141$_{-11}^{+15}$ & 1.7$_{-0.4}^{+0.5}$ & 0.74$_{-0.38}^{+0.52}$ & $<$0.30  & 44.02$_{\pm13.56}$ & $<$1.8  & $<$0.69  \\
G35.13-0.74 P3 & 103$_{-8}^{+12}$ & 1.1$_{-0.3}^{+0.3}$ & 0.47$_{-0.24}^{+0.31}$ & 0.58$_{-0.16}^{+0.19}$ & 44.02$_{\pm13.56}$ & 5.4$_{-1.7}^{+2.2}$ & 1.2$_{-0.6}^{+1.3}$ \\
G35.20-0.74N P1 & 140$_{-4}^{+5}$ & 12$_{-2}^{+2}$ & 5.1$_{-2.3}^{+2.5}$ & 5.3$_{-0.9}^{+0.9}$ & 44.07$_{\pm13.57}$ & 4.6$_{-0.6}^{+0.6}$ & 1.0$_{-0.3}^{+0.7}$ \\
G35.20-0.74N P2 & 113$_{-6}^{+7}$ & 4.5$_{-0.9}^{+1.0}$ & 2.0$_{-1.0}^{+1.1}$ & 2.6$_{-0.5}^{+0.5}$ & 44.07$_{\pm13.57}$ & 5.6$_{-1.0}^{+1.2}$ & 1.3$_{-0.5}^{+1.0}$ \\
G333.12-0.56 P1 & 148$_{-13}^{+18}$ & 1.9$_{-0.5}^{+0.6}$ & 0.73$_{-0.38}^{+0.53}$ & 1.1$_{-0.3}^{+0.3}$ & 38.63$_{\pm12.40}$ & 5.8$_{-1.8}^{+2.6}$ & 1.5$_{-0.7}^{+1.7}$ \\
G333.12-0.56 P2 & 134$_{-7}^{+9}$ & 1.9$_{-0.4}^{+0.5}$ & 0.74$_{-0.37}^{+0.45}$ & 0.88$_{-0.20}^{+0.22}$ & 38.63$_{\pm12.40}$ & 4.6$_{-1.1}^{+1.3}$ & 1.2$_{-0.5}^{+1.1}$ \\
G333.23-0.06 P1 & 204$_{-13}^{+16}$ & 5.5$_{-1.3}^{+1.5}$ & 1.8$_{-0.9}^{+1.2}$ & 2.0$_{-0.5}^{+0.6}$ & 32.64$_{\pm11.10}$ & 3.7$_{-1.0}^{+1.4}$ & 1.1$_{-0.5}^{+1.2}$ \\
G333.23-0.06 P2 & 106$_{-6}^{+8}$ & 3.0$_{-0.6}^{+0.7}$ & 0.97$_{-0.49}^{+0.60}$ & 1.3$_{-0.3}^{+0.3}$ & 32.64$_{\pm11.10}$ & 4.3$_{-1.0}^{+1.2}$ & 1.3$_{-0.6}^{+1.2}$ \\
G333.46-0.16 P1 & 150$_{-4}^{+4}$ & 9.2$_{-1.4}^{+1.4}$ & 3.7$_{-1.7}^{+1.8}$ & 1.5$_{-0.3}^{+0.3}$ & 40.10$_{\pm12.72}$ & 1.7$_{-0.2}^{+0.3}$ & 0.41$_{-0.14}^{+0.28}$ \\
G333.46-0.16 P2 & $<$117  & 0.90$_{-0.47}^{+0.14}$ & 0.36$_{-0.25}^{+0.18}$ & $<$0.24  & 40.10$_{\pm12.72}$ & $<$2.7  & $<$1.7  \\
G335.579-0.272 P1 & 343$_{-18}^{+21}$ & 62$_{-12}^{+14}$ & 24$_{-12}^{+14}$ & 24$_{-5}^{+5}$ & 38.38$_{\pm12.34}$ & 3.8$_{-0.8}^{+1.0}$ & 1.0$_{-0.4}^{+0.8}$ \\
G335.579-0.272 P2 & 122$_{-5}^{+7}$ & 3.6$_{-0.7}^{+0.7}$ & 1.4$_{-0.7}^{+0.8}$ & 0.98$_{-0.24}^{+0.26}$ & 38.38$_{\pm12.34}$ & 2.8$_{-0.6}^{+0.7}$ & 0.72$_{-0.29}^{+0.62}$ \\
G335.78+0.17 P1 & 199$_{-5}^{+5}$ & 16$_{-2}^{+2}$ & 6.0$_{-2.7}^{+3.0}$ & 4.6$_{-0.7}^{+0.7}$ & 38.53$_{\pm12.38}$ & 3.0$_{-0.3}^{+0.3}$ & 0.77$_{-0.25}^{+0.49}$ \\
G335.78+0.17 P2 & 186$_{-11}^{+14}$ & 4.5$_{-1.0}^{+1.1}$ & 1.7$_{-0.9}^{+1.1}$ & 3.3$_{-0.7}^{+0.8}$ & 38.53$_{\pm12.38}$ & 7.3$_{-1.6}^{+2.0}$ & 1.9$_{-0.8}^{+1.7}$ \\
G336.01-0.82 P1 & 147$_{-4}^{+5}$ & 6.9$_{-1.1}^{+1.1}$ & 2.7$_{-1.2}^{+1.4}$ & 1.3$_{-0.3}^{+0.3}$ & 38.89$_{\pm12.45}$ & 1.9$_{-0.3}^{+0.4}$ & 0.49$_{-0.18}^{+0.37}$ \\
G343.12-0.06 P1 & 137$_{-6}^{+7}$ & 4.7$_{-0.9}^{+0.9}$ & 1.8$_{-0.9}^{+1.0}$ & 1.5$_{-0.4}^{+0.4}$ & 38.73$_{\pm12.42}$ & 3.2$_{-0.7}^{+0.9}$ & 0.82$_{-0.34}^{+0.71}$ \\
G351.77-0.54 P1 & 167$_{-5}^{+6}$ & 16$_{-2}^{+3}$ & 7.6$_{-3.3}^{+3.6}$ & 1.7$_{-0.3}^{+0.3}$ & 46.00$_{\pm13.99}$ & 1.0$_{-0.2}^{+0.2}$ & 0.22$_{-0.08}^{+0.15}$ \\
G351.77-0.54 P2 & 166$_{-7}^{+8}$ & 14$_{-3}^{+3}$ & 6.3$_{-2.9}^{+3.4}$ & 2.4$_{-0.6}^{+0.7}$ & 46.00$_{\pm13.99}$ & 1.8$_{-0.4}^{+0.5}$ & 0.39$_{-0.16}^{+0.33}$ \\
G351.77-0.54 P3 & 181$_{-15}^{+20}$ & 15$_{-4}^{+5}$ & 6.9$_{-3.6}^{+5.0}$ & 3.5$_{-1.1}^{+1.3}$ & 46.00$_{\pm13.99}$ & 2.3$_{-0.8}^{+1.2}$ & 0.51$_{-0.26}^{+0.61}$ \\
G351.77-0.54 P4 & 162$_{-5}^{+6}$ & 12.9$_{-2.0}^{+2.1}$ & 6.0$_{-2.6}^{+2.9}$ & 2.8$_{-0.5}^{+0.5}$ & 46.00$_{\pm13.99}$ & 2.2$_{-0.3}^{+0.3}$ & 0.47$_{-0.15}^{+0.30}$ \\
G351.77-0.54 P5 & 107$_{-6}^{+8}$ & 7.1$_{-1.8}^{+2.1}$ & 3.3$_{-1.7}^{+2.1}$ & $<$0.91  & 46.00$_{\pm13.99}$ & $<$1.3  & $<$0.47  \\
G351.77-0.54 P6 & 167$_{-14}^{+19}$ & 9.7$_{-2.9}^{+3.7}$ & 4.5$_{-2.4}^{+3.4}$ & 1.2$_{-0.6}^{+0.7}$ & 46.00$_{\pm13.99}$ & 1.3$_{-0.6}^{+1.0}$ & 0.28$_{-0.17}^{+0.44}$ \\
G351.77-0.54 P7 & 122$_{-4}^{+5}$ & 9.1$_{-1.4}^{+1.5}$ & 4.2$_{-1.9}^{+2.1}$ & 1.4$_{-0.3}^{+0.3}$ & 46.00$_{\pm13.99}$ & 1.6$_{-0.2}^{+0.3}$ & 0.34$_{-0.12}^{+0.23}$ \\
G351.77-0.54 P8 & 127$_{-4}^{+5}$ & 10$_{-2}^{+2}$ & 4.7$_{-2}^{+2}$ & 2.4$_{-0.4}^{+0.5}$ & 46.00$_{\pm13.99}$ & 2.4$_{-0.3}^{+0.4}$ & 0.52$_{-0.18}^{+0.34}$ \\
IRAS16562-3959 P1 & 135$_{-8}^{+10}$ & 5.1$_{-1.1}^{+1.3}$ & 2.1$_{-1.0}^{+1.3}$ & $<$0.64  & 40.97$_{\pm12.90}$ & $<$1.3  & $<$0.52  \\
IRAS18089-1732 P1 & 183$_{-6}^{+7}$ & 26$_{-4}^{+4}$ & 11$_{-5}^{+5}$ & 4.3$_{-0.9}^{+1.0}$ & 41.02$_{\pm12.91}$ & 1.7$_{-0.3}^{+0.4}$ & 0.40$_{-0.15}^{+0.31}$ \\
IRAS18182-1433 P1 & 203$_{-5}^{+6}$ & 22$_{-3}^{+3}$ & 7.7$_{-3.6}^{+3.9}$ & 1.1$_{-0.3}^{+0.3}$ & 35.63$_{\pm11.75}$ & 0.50$_{-0.12}^{+0.13}$ & 0.14$_{-0.06}^{+0.13}$ \\
IRAS18182-1433 P2 & 223$_{-12}^{+14}$ & 11$_{-2}^{+2}$ & 3.8$_{-1.9}^{+2.2}$ & 1.5$_{-0.5}^{+0.5}$ & 35.63$_{\pm11.75}$ & 1.4$_{-0.4}^{+0.5}$ & 0.39$_{-0.18}^{+0.42}$ \\
NGC6334I P1 & 343$_{-23}^{+28}$ & 96$_{-22}^{+25}$ & 44$_{-22}^{+27}$ & $<$5.5  & 45.74$_{\pm13.94}$ & $<$0.57  & $<$0.21  \\
NGC6334I P2 & 278$_{-24}^{+31}$ & 56$_{-15}^{+18}$ & 26$_{-13}^{+18}$ & $<$5.4  & 45.74$_{\pm13.94}$ & $<$0.96  & $<$0.36  \\
NGC6334I P3 & 267$_{-24}^{+33}$ & 51$_{-15}^{+19}$ & 23$_{-13}^{+18}$ & $<$9.0  & 45.74$_{\pm13.94}$ & $<$1.8  & $<$0.69  \\
NGC6334I P4 & 293$_{-13}^{+15}$ & 78$_{-14}^{+15}$ & 35$_{-16}^{+19}$ & $<$3.0  & 45.74$_{\pm13.94}$ & $<$0.39  & $<$0.14  \\
NGC6334I P5 & 400$_{-49}^{+71}$ & 91.21$_{-28.50}^{+40.33}$ & 41.72$_{-23.05}^{+35.51}$ & $<$5.36  & 45.74$_{\pm13.94}$ & $<$0.59  & $<$0.23  \\
NGC6334I(N) P1 & 494$_{-55}^{+76}$ & 73$_{-23}^{+31}$ & 33$_{-18}^{+27}$ & $<$4.0  & 45.74$_{\pm13.94}$ & $<$0.54  & $<$0.22  \\
NGC6334I(N) P2 & 182$_{-12}^{+15}$ & 13$_{-3}^{+3}$ & 5.8$_{-2.9}^{+3.6}$ & $<$1.4  & 45.74$_{\pm13.94}$ & $<$1.1  & $<$0.39  \\
NGC6334I(N) P3 & 151$_{-6}^{+7}$ & 13$_{-2}^{+2}$ & 5.8$_{-2.6}^{+3.0}$ & $<$0.79  & 45.74$_{\pm13.94}$ & $<$0.62  & $<$0.21  \\
NGC6334I(N) P4 & 905$_{-165}^{+277}$ & 118$_{-49}^{+89}$ & 54$_{-34}^{+68}$ & 13$_{-6}^{+11}$ & 45.74$_{\pm13.94}$ & 1.1$_{-0.7}^{+1.8}$ & 0.24$_{-0.17}^{+0.66}$ \\
W33A P1 & 127$_{-3}^{+3}$ & 8.8$_{-1.2}^{+1.3}$ & 3.5$_{-1.6}^{+1.7}$ & 4.3$_{-0.6}^{+0.6}$ & 40.10$_{\pm12.72}$ & 4.9$_{-0.4}^{+0.5}$ & 1.2$_{-0.4}^{+0.7}$ \\
\enddata
\tablenotetext{a}{The lowest temperature is assumed to be 80 K for the derivation of the column densities.}
\tablenotetext{b}{derived from the relationship between $^{12}$C/$^{13}$C and the distance from the galactic center reported by Yan et al. (2019).}
\tablecomments{The errors represent 1$\sigma$ values. Details on how the errors were derived are described in the text.}
\end{deluxetable*}

\end{document}